\documentclass[twocolumn]{aastex631}
\usepackage{threeparttable}

\usepackage{float}
\usepackage{hyperref}
\usepackage{graphicx}
\usepackage{booktabs}
\usepackage{amsthm}
\usepackage{amsmath}
\usepackage{soul}

\newcommand{\Fvar}{F_{\rm var}}

\newcommand{\fermi}{\textit{Fermi}-LAT}

\newcommand{\AUC}{\text{AUC}}
\newcommand{\BSS}{\text{BSS}}

\begin{document}

\title[Blazar flare forecasting]{Advance warning of $\gamma$-ray blazar flares from \textit{Fermi}-LAT light curves:\\
a strictly causal machine-learning backtest}
\author[0000-0003-1458-4396]{Zahir Shah}
\affiliation{Manipal Centre for Natural Sciences, Manipal Academy of Higher Education, Manipal Karnataka 576104, India}
\email{zahir.shah@manipal.edu}

\author[0009-0005-3498-1104]{Sikandar Akbar}
\affiliation{Department of Physics, University of Kashmir, Srinagar 190006, India}
\correspondingauthor{Sikandar Akbar}
\email{darprince46@gmail.com}

\begin{abstract}

Long-term \textit{Fermi}-LAT monitoring makes it possible to ask whether a
blazar light curve shows signs of an upcoming flare before it becomes
obvious in the $\gamma$-ray emission. We present a strictly causal
machine-learning framework for forecasting $\gamma$-ray blazar flares
from 3-d binned LAT light curves. Flare intervals are identified with
Bayesian Blocks, and 42 variability features are measured from 365-d
trailing windows. We train separate WATCH and TRIGGER models: WATCH
predicts whether flare activity will appear within the next 90 d, while
TRIGGER predicts whether a new flare onset will occur within the next
45 d. To avoid temporal leakage, all scaling, calibration, threshold
selection, and validation use only pre-cutoff data (before MJD 60000).
We apply the method to the FSRQ 4FGL\,J1048.4$+$7143, using 13 bright
blazars as auxiliary training sources. Among logistic regression,
polynomial logistic regression, and random forest classifiers,
polynomial logistic regression gives the strongest held-out WATCH
performance, with ROC AUC $=0.884$, average precision $=0.392$, and a
block-permutation probability $p_{\rm perm}=0.009$. At the selected
threshold it recovers 21 of the 21 positive WATCH windows.
 For TRIGGER, the same model gives AUC $=0.751$, but with just nine
positive test windows its AP $=0.121$ is only about twice the random
baseline (0.052), and no reliable pre-onset TRIGGER alert is obtained.
The WATCH state appears before both held-out flares, with alerts 4.5 and
2.5 d before onset. These results suggest that the {\fermi} light curve
of this source contains useful predictive information about its flare
build-up during the period considered.

\end{abstract}


\section{Introduction}
\label{sec:intro}

Blazars are a subclass of radio-loud active galactic nuclei (AGN) whose  relativistic jets are oriented at small angles  to our line of sight \citep{Urry1995}.  This geometric alignment causes the jet emission to be strongly Doppler                                          
  boosted, making blazars among the most powerful non-thermal                                      
  sources in the Universe. They emit across the  electromagnetic spectrum---from                                      
  radio waves to GeV/TeV energies                                                             
  \citep{Blandford1978, Ghisellini1993}.                                                                           
  Blazars are broadly classified into BL Lacertae objects (BL Lacs) and                                            
  flat-spectrum radio quasars (FSRQs) based on the rest-frame equivalent                                           
  width of their optical emission lines \citep{Stickel1991, Padovani1995}.                                         
  BL Lacs are further subdivided according to the synchrotron peak frequency                                       
  $\nu_{\rm peak}^{\rm syn}$ into high-synchrotron-peaked (HBL),                                                   
  intermediate-synchrotron-peaked (IBL), and low-synchrotron-peaked (LBL)                                          
  subclasses \citep{Padovani1995, Ackermann2011_2LAC}. The broadband spectral energy distribution (SED) of blazars is 
characterized by two broad humps. The low-energy hump typically peaks between infrared and X-ray energies (depending on source class), while the high-energy hump peaks between MeV and TeV energies. 
The low-energy hump  is attributed to 
synchrotron emission from relativistic electrons in the jet, while the high-energy hump is generally explained within the one-zone leptonic framework as inverse Compton (IC) scattering of either the synchrotron photon field 
(synchrotron self-Compton, SSC; \citealt{Maraschi1992, Bloom1996}) or 
external photon fields such as the broad-line region (BLR) or dusty torus 
(external Compton, EC; \citealt{Dermer1993, Sikora1994, 2017MNRAS.470.3283S, 2024ApJ...977..111A}). Hadronic models, invoking proton synchrotron or photo-meson cascades,  provide alternative explanations for the high-energy emission  \citep{Mannheim1993, Mucke2003, Bottcher2013}.

The \textit{Fermi} Large Area Telescope \citep[LAT;][]{Atwood2009} has
revolutionized the study of blazar variability since its launch in 2008.
Operating as an all-sky monitor in the 0.1--300\,GeV band, the LAT has now
provided nearly continuous multi-year light curves for hundreds of
blazars.  These sources exhibit extreme flux variability, with timescales as short as minutes in VHE-emitting HBLs such as Mrk~501 \citep{Albert2007} and PKS~2155$-$304 \citep{Aharonian2007}. Such rapid variability places tight constraints on the size and physical conditions of the emitting region.  The \fermi\ mission has revealed a wide range of variability phenomena in blazars, including quasi-periodic oscillations (QPOs) \citep{Ackermann2015, 2025PhRvD.112f3061A, AKBAR2026100608}, log-normal flux behaviour \citep{Shah2018_lognormal}, orphan flares \citep{Macdonald2015}.
Spectral hysteresis was first established in X-ray/TeV observations \citep{Katarzynski2005}, and this type of behavior has also been reported at GeV energies from Fermi-LAT light
curves, e.g. during bright flares of PKS 1222+216 \citep{2014ApJ...796...61K}. Major $\gamma-$ray outbursts in FSRQs such as 3C\,454.3, 3C\,279, PKS\,1510$-$089, and CTA\,102 further highlight the broad dynamical range accessible to \fermi\ observations.

Statistical studies of \textit{Fermi}-LAT blazar light curves show that the
$\gamma$-ray flux distributions of blazars are better described
by a log-normal  probability density function
\citep{Shah2018_lognormal,2025MNRAS.539.2185M}. Such behavior favors
multiplicative variability processes and  point to fluctuations that
couple the accretion flow to the jet, although the physical origin is not
yet uniquely established \citep{Biteau2012}. The variability amplitude is
commonly quantified through the fractional variability, $F_{\rm var}$
\citep{Vaughan2003}. Applied to large LAT blazar samples, these diagnostics
show that FSRQs are, on average, more variable than BL Lacs at
$\gamma$-ray energies \citep{Abdo2010_var, Nalewajko2013, 61tz-jk8c}; this trend also
remains evident in multi-timescale analyses of 3-, 7-, and 30-d
\textit{Fermi}-LAT light curves.

These statistical properties are not only useful for describing blazar
variability, but also for asking whether the light curve contains
measurable signatures of an approaching flare. Understanding the temporal evolution of blazar flares has direct implications for jet physics, particle acceleration, and the underlying emission processes \citep{2020ApJ...891..120B}. Recent study has addressed the challenge of early detection
of blazar high states using machine-learning frameworks \citep{2025ApJ...980..141S}. The authors develop an
unsupervised deep-learning anomaly-detection framework  to automatically detect unusual behavior (especially flares) in blazar light curves, with the goal of triggering timely VHE observations.
From an
observational perspective, $\gamma$-ray flares are often accompanied by
activity at other wavelengths and therefore provide a strong motivation
for coordinated radio, optical, X-ray, and VHE campaigns. This has led
to continuous public monitoring through the \textit{Fermi}-LAT
monitored-source light curves and optical surveys such as ASAS-SN, as
well as rapid follow-up with pointed facilities such as \textit{Swift},
\textit{NuSTAR}, and \textit{AstroSat}. It has therefore become
increasingly important to explore whether blazar flares can be
anticipated, rather than identified only after they are already in
progress.

Despite these extensive monitoring and follow-up
efforts, operational flare-response strategies remain
largely reactive. They typically identify a source only
after it has already entered a bright state, or they rely on
a small number of multi-wavelength indicators to trigger further observations. A small number of recent
machine-learning studies have begun to address parts
of this problem: \citet{2025ApJ...980..141S} use real-time anomaly
detection to trigger VHE follow-up of an ongoing high
state, while \citet{2021APh...13202625F} use a random forest
to predict one-day-ahead Fermi-LAT flux for Cherenkov target-of-opportunity
selection. In the later work, the authors find that the short-term prediction is
driven mainly by typical source properties rather than
by the most recent measurements. There remains limited work on statistically controlled forecasts of flare occurrence within specified future time windows. This motivates a machine-learning
approach that searches the recent LAT light curve for combinations of
variability features that may appear before a flare. At the same time,
flare forecasting is especially vulnerable to temporal data leakage,
because labels, feature scaling, probability calibration, or model
validation can inadvertently use information from after the prediction
time. The central question is whether the recent variability history of a blazar
contains enough information to provide advance warning of an upcoming flare.
 In this paper we therefore develop a strictly causal machine-learning
framework for forecasting $\gamma$-ray blazar flares from 3-d binned
\textit{Fermi}-LAT light curves. We apply this framework to
4FGL\,J1048.4$+$7143 as the primary target source, while using 13
additional bright blazars to provide auxiliary training data and improve
the statistical basis of the model.

The paper is organized as follows. In Section~\ref{sec:data}, we describe
the \textit{Fermi}-LAT data and the preparation of the light curves.
Section~\ref{sec:methods} presents the methodology, including Bayesian-Blocks
flare identification, feature construction, model training, calibration,
and evaluation. Section~\ref{sec:results} gives the held-out backtest
results and the inferred warning times. In Section~\ref{sec:discussion},
we discuss these results in the broader context of blazar variability and
flare physics. Section~\ref{sec:conclusions} summarizes our main
conclusions.                                                            
                                                                                   
 \section{Data}                                                                                                   
  \label{sec:data}                                                                                                 
                                         
To test whether recent $\gamma$-ray variability can provide advance warning
of a flare, we require long, uniformly processed light curves that sample
multiple activity cycles for each source. The \textit{Fermi}-LAT
\citep{Atwood2009} is well suited to this purpose. It is a pair-conversion
$\gamma$-ray telescope covering the energy range from $\sim$20\,MeV to
$>$300\,GeV, with a peak effective area of $>$8000\,cm$^2$ at normal
incidence, an angular resolution (68\% containment radius) of
$\sim$0\fdg6 at 1\,GeV, and an energy resolution of $\sim$10\%.
Operating in continuous all-sky survey mode, the LAT scans the full sky
every $\sim$3\,hr and provides nearly uninterrupted long-term monitoring
of individual blazars. Over the mission lifetime, this has yielded more
than 17 years of continuous $\gamma$-ray data, encompassing multiple
distinct active episodes for each source in our sample
(Table~\ref{tab:sources}).

We use light curves from the \textit{Fermi}-LAT Light Curve Repository
\citep[LCR;][]{Abdollahi2023}, a public archive that provides
processed 3-d, 7-d, and 30-d binned light curves for 1525 sources variable in the 4FGL Data Release 2 catalogue (4FGL-DR2). Each light
curve is derived from an unbinned maximum-likelihood analysis of Pass~8
SOURCE-class events in the energy range 100\,MeV--100\,GeV within a
$12\degr$ region of interest. In this work, we use the 3-d binned light
curves of the 14 bright $\gamma$-ray blazars listed in
Table~\ref{tab:sources}. For each time bin, we extract the integrated
photon flux $F$ and its uncertainty $\sigma_F$, the Test Statistic
$\mathrm{TS}=2\Delta\ln\mathcal{L}$, and the photon spectral index
$\Gamma$ with its uncertainty $\sigma_\Gamma$ when available. The 3-d
binning provides a  balance between temporal resolution and
photon statistics: shorter bins often do not contain enough counts for
stable likelihood fits except during bright flares, whereas longer bins
tend to smooth out sub-week variability that may carry predictive
information about an approaching flare. All subsequent quality filtering
and feature extraction are applied to these 3-d binned light curves, as
described in Section~\ref{sec:cuts}.

  \subsection{Sample selection and quality cuts}
  \label{sec:sources}
  \label{sec:cuts}
  \label{sec:data_cuts}
We analyse a sample of 14 bright $\gamma$-ray blazars selected from the
fourth \textit{Fermi}-LAT source catalogue and included in the
\textit{Fermi}-LAT Light Curve Repository \citep[LCR;][]{Abdollahi2023}.
The light curves span 2008 August to 2026 March, and in this work we use
 3-d binned light curves. Before Bayesian-Blocks segmentation
and feature extraction, each light curve is subjected to a uniform set of
quality cuts. We retain only bins with $\mathrm{TS} \geq 4$, positive
reported flux, and a finite flux uncertainty. None of the
retained bins are flagged by the LCR as non-convergent.
A further cut is applied to the photon index,
$0 < \Gamma < 5$, to ensure that all spectral values lie
within the allowed fit range. In addition, a hard upper
limit of $F_{\max}=10^{-4}$\,ph\,cm$^{-2}$\,s$^{-1}$ is applied to remove
isolated outlier bins, which are likely to arise from processing artifacts
or solar contamination. 
As an additional check, we verified that our sample is not affected by the LCR-documented artifact of erroneously high flux values during periods of low exposure, by inspecting reported flux against TS for all quality-cut-passing bins in each light curve. We identify candidate artifacts as bins with flux exceeding ten times the source's median flux at only marginal detection significance ($\mathrm{TS} < 15$). Two of the 20{,}657 retained bins meet this criterion ($\mathrm{TS}\approx4$--5, flux 13--88 times the source's typical level, with a reported uncertainty comparable to the flux itself), both in auxiliary training sources. We exclude these two bins as an additional quality cut, leaving 20{,}655 bins in the full sample. Very small reported flux uncertainties can make
the Bayesian-Blocks segmentation too sensitive to a few individual bins.
To avoid this, we apply a minimum uncertainty of
$\sigma_F \geq 0.10\,\widetilde{|F|}$ in each segmentation run, where
$\widetilde{|F|}$ is the median absolute flux of the corresponding
light-curve segment. This floor does not replace the reported
uncertainties altogether; it only raises those bins whose reported errors
are smaller than the adopted minimum level. This prevents a small number
of formally high-precision bins from driving spurious change points. The
original measurement uncertainties are retained without modification for
the subsequent rolling-window feature extraction. All 14 sources remain
persistently detected over multi-year baselines and exhibit
peak-to-quiescent flux ratios $R_{pk/q}\equiv F_{\rm peak}/F_q$ large
enough to support flare-classification training
(Table~\ref{tab:sources}).


\subsubsection{Target source}
\label{sec:target}

The primary prediction target is 4FGL\,J1048.4$+$7143, the flat-spectrum
radio quasar (FSRQ) S5\,1044+71, at redshift $z=1.15$
\citep{Polatidis1995}. The source is included in the \textit{Fermi}-LAT
monitored-source list and has shown repeated LAT-reported episodes of
enhanced $\gamma$-ray activity \citep{Ojha2013_ATel4941, Ojha2017_ATel9928}.
It is a bright and strongly variable FSRQ with pronounced multi-epoch
$\gamma$-ray activity over the full \textit{Fermi} mission baseline.
During a major flare reported in early 2017, the daily flux above
100\,MeV reached $(1.1\pm0.2)\times10^{-6}$~ph\,cm$^{-2}$\,s$^{-1}$
\citep{Ojha2017_ATel9928}. Its 3-d binned light curve yields 1443
bins after quality cuts (Section~\ref{sec:data_cuts}). The quiescent flux
level, defined as the 30th percentile of the
training-period flux distribution, is
$F_q \simeq 7.68\times10^{-8}$~ph\,cm$^{-2}$\,s$^{-1}$, and the
peak-to-quiescent ratio is $R_{pk/q}=14.4$
(Table~\ref{tab:sources}). The long light curve of this source, spanning more than 17 years and containing multiple activity cycles, makes it well suited for training and held-out evaluation of predictive models.

 \subsubsection{Auxiliary training sources}
\label{sec:aux}

The remaining 13 sources listed in Table~\ref{tab:sources} are used exclusively as auxiliary training data; no forecasts are issued for these sources. The sample includes mostly FSRQs spanning a redshift range $0.1 \lesssim z \lesssim 1.6$. The auxiliary sample spans a range of blazar classes, redshifts, and brightness levels. This allows the classifier to learn from a broader variety of variability patterns and spectral behavior. At the same time, all of the selected sources show strong and sustained $\gamma$-ray flaring activity over the full \textit{Fermi} mission baseline. All auxiliary sources are drawn from the \textit{Fermi}-LAT LCR and 
satisfy the same quality criteria as the target source 
(Section~\ref{sec:data_cuts}). Each source exhibits substantial 
peak-to-quiescent flux variability, with $R_{pk/q} \equiv 
F_{\rm peak}/F_q$ values ranging from 7.0 to 150.0
(Table~\ref{tab:sources}), ensuring that elevated flux states are 
well represented in the training data. The selection deliberately 
excludes sources with very low variability amplitudes, to avoid 
biasing the classifier towards quiescent-dominated behavior that 
would be uninformative for flare prediction.

The sources in our sample span more than an order of magnitude in brightness. We therefore normalize all flux-dependent features by the quiescent flux level of each source, $F_q$, defined as the 30th percentile of its training-epoch flux distribution (Section~\ref{sec:features}). This places the features on a common relative scale and allows the classifier to learn shared variability patterns rather than source-specific flux levels. The auxiliary sources also enlarge the training set and increase its statistical diversity, providing many more examples of pre-flare behavior than the target light curve alone. The strictly causal train--test split used for the rolling windows is described in Section~\ref{sec:split}.

\begin{table*}
\centering
\small
\setlength{\tabcolsep}{4pt}
\caption{Properties of the 14 \fermi\ blazar light curves used in this study.
$N_{\rm bins}$ is the number of 3-d bins surviving quality cuts;
$N_{\rm flare}$ is the number of training-epoch bins classified as flaring
by Bayesian Blocks; $F_q$ is the 30th percentile of the
training-epoch flux distribution; and
$R_{pk/q}\equiv F_{\rm peak}/F_q$. The target source is highlighted in bold.}
\label{tab:sources}
\begin{tabular}{llcclrrrl}
\hline\hline
4FGL name & Common name & Class & $z$ & Role
  & $N_{\rm bins}$ & $N_{\rm flare}$
  & $F_q\;(10^{-8}$\,ph\,cm$^{-2}$\,s$^{-1})$
  & $R_{pk/q}$ \\
\hline
\textbf{J1048.4$+$7143} & \textbf{S5\,1044$+$71} & \textbf{FSRQ} & \textbf{1.15} & Target+Train
  & \textbf{1443} & \textbf{186} & \textbf{7.68} & \textbf{14.4} \\
J1512.8$-$0906 & PKS\,1510$-$089     & FSRQ & 0.361 & Train & 1839 &  79 & 28.90 & 31.8 \\
J1224.9$+$2122 & 4C +21.35    & FSRQ  & 0.43 & Train & 1167 &  33 & 13.10 &  57.3 \\
J0904.9$-$5734 & PKS\,0903$-$573     & FSRQ & 0.2621   & Train & 1238 &  28 & 9.94 & 64.3 \\
J1256.1$-$0547 & 3C\,279             & FSRQ & 0.536 & Train & 1943 &  36 & 29.10 &  53.6 \\
J2253.9$+$1609 & 3C\,454.3           & FSRQ & 0.859 & Train & 1989 &  50 & 38.20 & 150.0 \\
J0538.8$-$4405 & PKS\,0537$-$441     & BL LAC & 0.892 & Train & 1956 & 263 & 11.20 &   8.9 \\
J0739.2$+$0137 & PKS\,0736$+$017     & FSRQ & 0.189 & Train & 1310 & 116 & 11.70 & 20.0 \\
J1159.5$+$2914 & B2\,1156$+$295      & FSRQ & 0.729 & Train & 1667 & 117 & 11.15 & 28.7 \\
J0730.3$-$1141 & PKS\,0727$-$115     & FSRQ & 1.591 & Train & 1425 & 240 & 13.10 &   7.0 \\
J1310.5$+$3221 & B2\,1308$+$326      & FSRQ  & 0.997 & Train & 1076 & 102 &  7.08 &  44.9 \\
J1443.9$+$2501 & PKS\,1441$+$25      & FSRQ & 0.939 & Train &  788 &  79 & 4.74 & 24.1 \\
J1522.1$+$3144 & B2\,1520$+$31       & FSRQ & 1.487 & Train & 1405 & 217 & 14.30 &  11.2 \\
J0403.9$-$3605 & PKS\,0402$-$362     & FSRQ & 1.417 & Train & 1409 &  73 & 10.90 &  45.1 \\
\hline
\end{tabular}
\end{table*}

\section{Methodology}
\label{sec:methods}

\subsection{Overview}
\label{sec:methods_overview}
\begin{figure}
\centering
\includegraphics[width=\columnwidth]{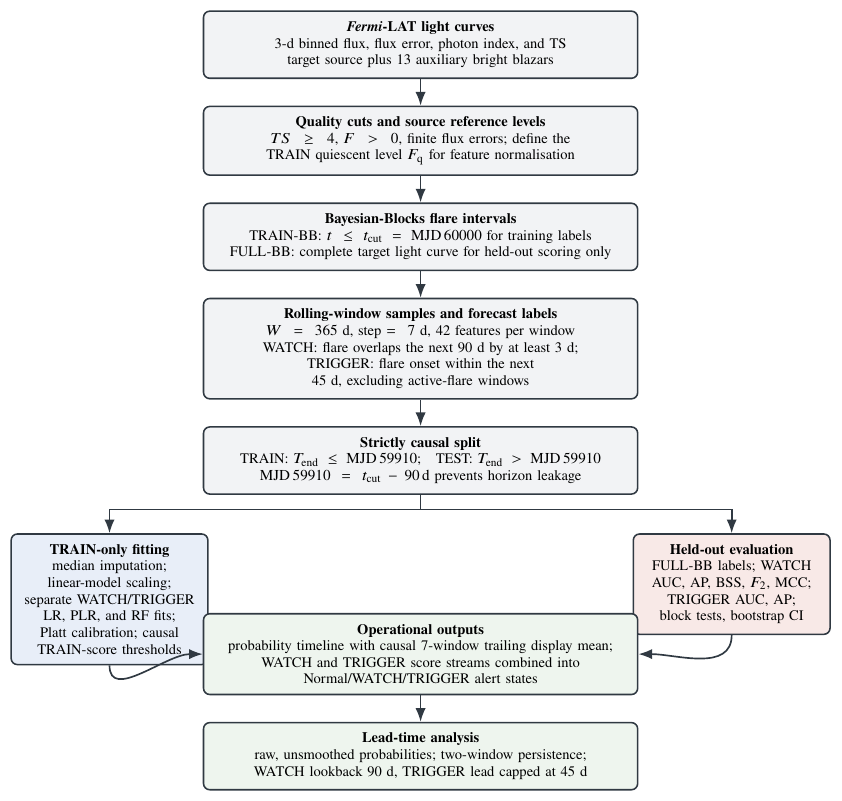}
\caption{Schematic of the strictly causal blazar flare-forecasting
pipeline. The light curves first pass through quality cuts, and the
TRAIN-only quiescent level $F_{\rm q}$ is defined for later
normalization of flux-dependent features. Bayesian Blocks are then
computed in two ways:
TRAIN-BB uses only data up to $t_{\rm cut}=\mathrm{MJD}\,60000$ for
training labels, while FULL-BB is used only to score the held-out target
stream. Rolling 365-d windows with a 7-d step are assigned two future
targets: a WATCH label based on flare overlap within the next 90 d and a
TRIGGER label based on flare onset within the next 45 d. The train/test
boundary is placed at
$T_{\rm boundary}=t_{\rm cut}-90\,{\rm d}=\mathrm{MJD}\,59910$, so that no
training WATCH horizon reaches into the held-out period. Each classifier
family is then fitted separately for WATCH and TRIGGER, calibrated on a
trailing TRAIN tail, and thresholded using causal TRAIN-score sequences
before the fixed pipeline is evaluated on the held-out windows.}
\label{fig:pipeline}
\end{figure}

With the source sample, light curves, and quality cuts now defined, we
next describe how the forecasting dataset is constructed and evaluated.
The full workflow is summarized in Figure~\ref{fig:pipeline}. We first
identify $\gamma$-ray flare intervals using the Bayesian-Blocks analysis
on the training and full light curves. We then extract variability
features from rolling windows to build the supervised learning dataset.
This is followed by a strictly causal train--test split, model training,
and probability calibration. The WATCH and TRIGGER thresholds are then
determined using only the causal TRAIN-score sequences. Finally, the
models are evaluated on the held-out target stream, including the
warning-time analysis. This causal separation is maintained throughout the
pipeline so that no information from the future enters the training or
evaluation steps.

\subsection{Bayesian-Blocks flare identification}
\label{sec:bb}
\label{sec:bayesian_blocks}

We identify flare intervals using the Bayesian-Blocks (BB) method
\citep{Scargle2013}, as implemented in \textsc{astropy}
\citep{Astropy2022}.  Specifically, we use astropy.stats.bayesian$_-$blocks with fitness $=$ `measures', the mode appropriate for a time series of point measurements with Gaussian uncertainties. The inputs are the observation times (MJD), the 3-d binned photon fluxes, and their associated uncertainties. As described in Section~\ref{sec:data_cuts}, a minimum uncertainty floor of $0.10\,\widetilde{|F|}$ is applied before this step to prevent a few formally high-precision bins from driving spurious block boundaries.
This method divides the light curve into a set of
piecewise-constant blocks while taking the reported flux uncertainties
into account. For
the BB segmentation we adopt a false-alarm probability
of $p_0 = 0.05$, which provides a conservative balance between detecting
real flux changes and avoiding spurious change points.
A block is classified as a flare only if it satisfies both a flux and a
duration condition. Its mean flux must satisfy
\begin{equation}
\hat{\mu}_b \geq \max\!\left(\hat{\mu}_{70\rm th\%}, \;
\mu_{\rm q} + 3\sigma_{\rm q}\right),
\label{eq:flare_criterion}
\end{equation}
where $\hat{\mu}_b$ is the mean flux of the $b$th block. The quantity
$\hat{\mu}_{70\rm th\%}$ is the 70th percentile of the mean fluxes of all
blocks returned by that BB run. To define the quiescent level, we first
identify the blocks whose mean fluxes lie in the lowest 40 per cent of
that same set of block means. The quantities $\mu_{\rm q}$ and
$\sigma_{\rm q}$ are then taken as the mean and standard deviation of
those low-flux blocks. This combines two ingredients used
separately in the literature: a quiescent-plus-uncertainty threshold,
similar to \citet{Yoshida2023}, and a percentile-of-block-flux threshold,
similar to \citet{Kouch2026}. Blocks spanning large gaps in the
light curve are not classified as flare blocks. A gap is defined as
more than 3$\times$ the median time spacing between consecutive bins.
This is computed independently of the duration condition below; it
happens to also evaluate to 9~d for all 14 sources in our sample because
each light curve is 3-d binned ($3\times3\,{\rm d}=9\,{\rm d}$), not
because the two quantities are the same rule.
In addition, the block duration must satisfy
$\Delta t_b \geq 9$ d. Adjacent flare blocks separated by $\leq 9$ d are
merged into a single flare interval, so that multi-peaked outbursts are
not split into several smaller events. Applying only the flux
condition of Equation~\ref{eq:flare_criterion}, 463 blocks qualify
across the 14 training light curves. Excluding blocks that span a large
gap reduces this to 287. Applying the 9-d duration cut on top of both
leaves 193 flare blocks, excluding 94 segments (33 per cent). Among the
287 gap-free segments, none is shorter than 6 d. In
terms of the underlying binning, this corresponds to 1987
flux-qualifying bins, reduced to 1808 after gap exclusion, and 1619
after the duration cut removes 189 bins (10.5 per cent) -- these 1619
bins are the flare-labelled bins summed as $N_{\rm flare}$ in
Table~\ref{tab:sources}. We do not lower the duration threshold to 6 d:
our rolling window advances in 7-day steps (Section~\ref{sec:features}),
and a 6-day flare is shorter than a single window update, so labeling it
as a flare would give the model a label it cannot resolve at its own
temporal scale. This duration cut means the backtest targets sustained
flares of duration $\gtrsim$9 d, rather than all flare-like activity.
To preserve temporal causality, we perform two separate BB runs for each
source. The TRAIN-BB run uses only data up to the cutoff time,
$t_{\rm cut}=\mathrm{MJD}\,60000$, and its flare intervals are used to
define the training labels. The FULL-BB run uses the complete light curve
and is used only for the held-out evaluation. This separation ensures
that information from after the cutoff does not enter the training step.

\subsection{Feature engineering}
\label{sec:features}
Using the cleaned light curves, we construct the machine-learning dataset
with rolling windows. For each window ending at $T_{\rm end}$, we use the
preceding 365 d of data, i.e. the interval
$[T_{\rm end}-W,\,T_{\rm end}]$ with $W=365$ d, and shift the window
forward in steps of 7 d. Only windows containing at least 30 valid data
points are retained. From each window we compute a 42-dimensional feature
vector that summarizes the variability behavior of the source. Because
the blazars in our sample span a wide range of flux levels, all
flux-dependent features are normalized by the quiescent flux of that
source, $F_{\rm q}$. Here $F_{\rm q}$ is defined from the training epoch
as the 30th percentile of the flux distribution. This
places the flux-dependent features on a common relative scale and allows
direct comparison across sources. The 42 features are grouped into the
following categories.

\paragraph{Flux distribution and tail statistics.}
The first group contains 16 features and describes the overall shape of
the flux distribution within each window. We first measure the average
spacing between valid data points, together with the mean flux and the
standard deviation of the flux, both normalized by $F_{\rm q}$. We also
include the two parameters of a log-normal fit,
$(\mu_{\ln}, \sigma_{\ln})$, along with the skewness and kurtosis, so
that the width and asymmetry of the distribution are captured. To
describe the bright end of the distribution, we add several tail
measures. One feature gives the fraction of flux values in the current
window that lie above the 95th-percentile flux level of the same source
during its pre-cutoff training epoch. We also measure the fraction of
points in the current window that lie more than $2\sigma$ above the
window mean. In addition, we include the ratios
$p_{95}/\mathrm{median}$ and $p_{99}/\langle F\rangle$, where
$p_{95}$ and $p_{99}$ are the 95th and 99th percentiles of the flux
distribution within the current window. We further include the
interquartile range normalized by the median, the peak-to-mean ratio, the
Gini coefficient, and the Kolmogorov--Smirnov statistic and $p$-value for
the log-normal fit. Together, these features describe the typical flux
level, the spread of the distribution, and the strength of the high-flux
tail. Most windows have far more than the 30-point minimum (median
99 points; only 6.3 per cent have 30--40). We checked the stability of
the log-normal fit and the KS statistic near this floor by repeatedly
resampling 30 points from larger windows and refitting: both are
essentially unbiased, with modest sampling noise (typical spread 0.10 in
$\mu_{\ln}$, 0.08 in $\sigma_{\ln}$, and 0.03 in the KS statistic), not
large enough to distract training.

\paragraph{Variability amplitude and temporal evolution.}
The second group contains 15 features and is designed to describe how
strongly the source varies, and whether that variability becomes stronger
towards the end of the window. We first use two standard measures of the
overall variability amplitude: the fractional variability $\Fvar$
\citep{Vaughan2003} and the normalised excess variance
$\sigma^2_{\rm NXS}$ \citep{Edelson2002}. We then measure the long-term behaviour of the light curve through the
linear slope of the normalised flux $F/F_{\rm q}$ and the slope of the
log-flux $\ln F$. The latter is useful for capturing multiplicative
changes in the flux. To check
whether the source is brightening towards the end of the window, we
compare the mean flux in the first and last 90, 45, and 30 d of the same
window. We also measure the flux slope over the final 30 and 45 d, and
the fraction of bins above $1.5F_{\rm q}$ over these same recent
intervals. To follow slower changes across the full 365-d window, we
divide the window into four equal parts and measure how the mean flux,
$\Fvar$, and the log-normal width $\sigma_{\ln}$ change across these
sub-intervals. Finally, we include the time since the most recent
TRAIN-BB flare onset as a simple measure of possible recurrence
behavior. Together, these features describe both the strength of the
variability and its temporal build-up within the window.

\paragraph{Temporal timing and periodicity.}
The third group contains 4 features and describes the time structure of
the variability. We first use the Lomb--Scargle periodogram to search for
any preferred variability timescale within the window. We use
\textsc{astropy}'s implementation on the flux and flux error in their
original units, without normalizing by the quiescent flux. The frequency grid has 500 points, spaced evenly in frequency from
 from $1/180$ to $1/10\,{\rm d}^{-1}$. We use Astropy's default normalization `standard' and default mean-fitting
(\texttt{fit\_mean=True}), with no further detrending. From this, we take
three quantities: the highest periodogram power, the period at which this
maximum occurs, and the corresponding false-alarm probability. 
Unlike the log-normal fit, we find the periodogram is not stable near
the 30-point floor: resampling 30 points from larger windows recovers a
peak period more than 30 d from the well-sampled value in all the cases
tested. At 30 points, the peak power
also comes out high, by about 0.15 on average. This affects the same $\sim$6 per cent of windows
with 30--40 points, and is a genuine limitation for a source with a
short monitoring history. We then
use the first-order structure function to measure how the flux variations
grow with time lag, and include its slope as an additional feature.
Together, these quantities describe whether the variability shows a
characteristic timescale and how strongly the flux changes are correlated
across time.

\paragraph{Spectral evolution.}
The final group contains 7 features and describes how the photon spectral
index, $\Gamma$, evolves within each window. We first measure the mean
value of $\Gamma$, its standard deviation, and its linear slope with
time. We also include the difference between the mean spectral index in
the current window and the quiescent spectral index of the same source,
where the quiescent value is defined from the training epoch. To examine
whether the spectrum changes towards the end of the window, we further
compare the mean $\Gamma$ values in the first and last 90, 45, and 30 d
of the same window. Together, these features are intended to capture
spectral hardening or softening before major $\gamma$-ray activity.

\subsection{Label definition}
\label{sec:label_def}
Once the 42 features have been computed for each rolling window, the next
step is to assign the future label that the model is asked to predict.
Each sample is based only on the past 365-d interval ending at
$T_{\rm end}$, but its label is defined from what happens after
$T_{\rm end}$. In this work, we assign two future labels to each window:
a WATCH label and a TRIGGER label. The WATCH label, $Y_{\rm W}$, is designed to identify windows that are
followed by flare activity within the next 90 d. For this purpose, we
define the future horizon as $(T_{\rm end},\,T_{\rm end}+90\,{\rm d}]$.\footnote{Half-open interval notation: the round bracket ``('' excludes $T_{\rm end}$ itself, and the square bracket ``]'' includes the endpoint $T_{\rm end}+90$~d.} A
window is assigned $Y_{\rm W}=1$ if any Bayesian-Blocks flare interval
overlaps this future horizon by at least $\Delta_{\rm min}=3$ d; otherwise
it is assigned $Y_{\rm W}=0$. The minimum-overlap condition avoids
labelling a window as positive when only a very small decaying part of a
flare enters the horizon. Since every flare lasts at least 9 d
(Section~\ref{sec:bb}), this rule never rejects a flare for being too
short; it only affects the small number of windows right at the start
or end of a flare's presence in the horizon. Using a looser rule that
counts any nonzero overlap as positive changes only 52 of 9{,}004
training windows (0.6 per cent), all at these edges, confirming the
effect is small and does not remove any flare from the training data.
The TRIGGER label, $Y_{\rm T}$, is more restrictive and is intended to
capture the start of a flare rather than flare activity in general. Here
we use a shorter, onset-focused horizon of 45 d. A window is assigned
$Y_{\rm T}=1$ if a flare onset occurs within
$(T_{\rm end},\,T_{\rm end}+45\,{\rm d}]$, and $Y_{\rm T}=0$ otherwise.
To keep this task focused on advance warning, windows that already lie
inside an ongoing flare interval are not counted as trigger-positive,
even if a new flare onset falls within the next 45 d. The 90-d
WATCH and 45-d TRIGGER horizons were fixed a priori and were not tuned
on the data. WATCH is deliberately broad, trading precise timing for a
longer warning window, while TRIGGER is deliberately narrow, trading
warning length for a more specific onset window. This tradeoff is
reflected in the held-out results (Section~\ref{sec:results}): WATCH
identifies two of the held-out flares, whereas TRIGGER identifies none
at its fixed alert threshold, and its held-out metrics are too
noisy to establish onset-level skill (Section~\ref{sec:discussion}). Given the small number of
held-out flares, we did not attempt to optimize these horizon lengths or
draw conclusions about whether 45 and 90 d are the best choices. For
model training, both WATCH and TRIGGER labels are derived from the
TRAIN-BB flare intervals and flare onsets, constructed using only data up
to the cutoff time $t_{\rm cut}=\mathrm{MJD}\,60000$. For the held-out
evaluation of the target source, the same two labels are recomputed
retrospectively from the FULL-BB segmentation of the complete light
curve. These held-out labels are used only for performance evaluation. They are not used for model training, probability calibration, or threshold selection.

\subsection{Strictly causal train--test split}
\label{sec:split}
Once the WATCH and TRIGGER labels have been defined, the next step is to
separate the rolling-window samples into training and held-out sets. This
step is especially important in a forecasting problem, because leakage can
arise even when the input window itself uses only past data: the label is
still defined by what happens after the window end time. To keep the
analysis causal, we must consider not only the past data used as input,
but also the next 90 d used to assign the WATCH label.
The temporal cutoff is set at $t_{\rm cut}=\mathrm{MJD}\,60000$. The
WATCH label is based on whether a flare occurs within the next $H=90$ d.
To ensure that no training label depends on information from after the
cutoff, we exclude any window whose 90-d WATCH horizon extends beyond
$t_{\rm cut}$. The corresponding boundary in window end time is therefore
\begin{equation}
T_{\rm boundary} = t_{\rm cut} - H
= \mathrm{MJD}\,60000 - 90
= \mathrm{MJD}\,59910.
\label{eq:boundary}
\end{equation}

Training windows satisfy $T_{\rm end} \leq T_{\rm boundary}$, so the full
90-d WATCH horizon of every training sample remains before the cutoff.
All windows with $T_{\rm end} > T_{\rm boundary}$ are placed in the
held-out set. This implies that some held-out windows can still end before
$\mathrm{MJD}\,60000$, but they are excluded from training because their
future WATCH horizon extends beyond the cutoff. The shorter 45-d TRIGGER
horizon is automatically contained within the same split, so no TRIGGER
training label depends on post-cutoff information either.
In the code, this split is applied directly through the window end time
$T_{\rm end}$. The training set is built from all sources using windows
with $T_{\rm end} \leq \mathrm{MJD}\,59910$, whereas the post-boundary
windows of the target source are kept separate for the final causal
backtest.

 \subsection{Machine learning classification}
     \label{sec:models}
We now train the machine-learning models that use the 42-dimensional
feature vectors to estimate the probability of a future flare. Because
WATCH and TRIGGER represent two different forecasting tasks, each model
family is trained separately for the two labels. The WATCH model returns
the probability $\hat{p}_{\rm W}=P(Y_{\rm W}=1\mid\mathbf{x})$, while the
TRIGGER model returns $\hat{p}_{\rm T}=P(Y_{\rm T}=1\mid\mathbf{x})$.
We use a small set of supervised classifiers that span a range of model
complexity, from simple linear models to a non-linear tree-based model.
For the linear classifiers, the input features are standardized to zero
mean and unit variance before fitting. The mean and variance used
here are computed from the TRAIN windows only. Held-out windows are
never used to compute them; they are only rescaled afterward using the
values already computed from TRAIN. Both WATCH and TRIGGER are
class-imbalanced problems, with many more negative windows than positive
ones. To account for this, we use balanced class weights in the logistic
regression, polynomial logistic regression, and random-forest models.

\paragraph{Logistic Regression (LR).}
We use logistic regression as the baseline linear classifier. In this
model, the flare probability is obtained from a weighted sum of the 42
input features:
\begin{equation}
P(Y=1|\mathbf{x}) = \sigma(\mathbf{w}^\top\mathbf{x} + b),
\quad \sigma(z) = \frac{1}{1+e^{-z}}.
\label{eq:logreg}
\end{equation}
Here $\mathbf{x}$ is the 42-dimensional feature vector of a given
window, $\mathbf{w}$ is the set of learned weights, and $b$ is the
intercept term. The quantity $\mathbf{w}^\top\mathbf{x}+b$ gives the
linear model score, and the sigmoid function $\sigma(z)$ converts this
score into a value between 0 and 1. This value is interpreted as the
predicted probability that the window belongs to the positive class,
i.e. either WATCH-positive or TRIGGER-positive, depending on the task
being fitted. The model is fitted with L2 regularization
($C=1$) using the \texttt{lbfgs} solver.

\paragraph{Polynomial Logistic Regression (PLR).}
A purely linear model can miss cases where two or more features become
important only when they vary together. To account for this, we expand
the original 42 input features by adding all second-order terms,
including squared terms ($x_i^2$) and pairwise products ($x_i x_j$).
This polynomial expansion retains the original linear terms as well. The
total number of features therefore increases from 42 to
$42 + 42(43)/2 = 945$. Because this expanded space is much larger, we
retain only the 50 most informative features using ANOVA F-score
ranking (\texttt{SelectKBest}; \citealt{Pedregosa2011}). This
ranking is also computed from the TRAIN windows only, using only the
TRAIN WATCH/TRIGGER labels to decide which features to keep; held-out
windows play no part in the ranking and are only ever restricted to the
resulting fixed set of 50 features. The choice to retain 50
features is based on two checks. First, a common rule of thumb for
logistic regression models is at least 10 positive training examples
per retained feature \citep{Peduzzi1996}. Our WATCH task has 1500
positive TRAIN windows, giving 30.0 examples per feature when 50 are
retained; our TRIGGER task has 563 positive TRAIN windows, giving 11.3
examples per feature, just within this guideline. Second, we test 25,
50, and 75 retained features directly (Table~\ref{tab:k_sensitivity}).
WATCH performance changes only marginally between 25 and 50 retained
features, and drops more sharply as more are retained, up to 75, while
TRIGGER shows the opposite but noisier trend, given its much smaller
positive sample. We choose 50 as a compromise between these two
trends: it keeps WATCH performance close to its best value at 25
features (a drop of only 0.010 in AUC) while giving TRIGGER, which has
far fewer positive examples to learn from, a larger gain than the
smaller additional improvement available at 75 features, which comes
at a much greater cost to WATCH performance. We also checked what kind
of terms are actually selected at 50 features: only 4 of the 50 are
plain linear terms, and the remaining 46 are interaction or squared
terms, showing that the feature-selection step strongly favours
combinations of features over individual ones, consistent with our
finding that most of the useful pre-flare signal is carried by feature
interactions rather than by single variability measures
(Section~\ref{sec:results}).

\begin{table*}
\centering
\caption{PLR held-out performance as a function of the number of
retained features. LR and RF are unaffected by this choice and are
omitted.}
\label{tab:k_sensitivity}
\begin{tabular}{lccc}
\hline\hline
Retained features & WATCH AUC & WATCH BSS & TRIGGER AUC \\
\hline
25          & 0.894 & 0.267 & 0.716 \\
50 (used)   & 0.884 & 0.197 & 0.751 \\
75          & 0.845 & 0.022 & 0.776 \\
\hline
\end{tabular}
\end{table*}
 A logistic-
regression model is then fitted in this reduced feature space with
stronger L2 regularization ($C=0.05$) to limit overfitting. This model
can capture simple feature interactions that are missed by standard
logistic regression, for example when increased variability and spectral
change occur together before a flare.

\paragraph{Random Forest (RF).}
Random forest is a non-linear ensemble model made up of 200 decision
trees, with at least 5 samples required in every terminal leaf
\citep{Breiman2001}. Neither value was chosen with a grid search.
More trees do not make a random forest overfit, however, adding more trees only
helps up to a point. After that, it just takes longer to run without
improvig results. For example \citet{Oshiro2012} found that performance saturated at around 128 trees, while \citet{Probst2018} found that most performance gains occurred within the first 250 trees. Our choice
of 200 trees lies within this range. For the minimum leaf size, we
adopted a value of 5, above the \texttt{scikit-learn} default of 1. This
stops a leaf from being built out of fewer than five training windows,
so no single window can drive the fitted rule on its own. This matters
especially for TRIGGER, which has only 563 positive training windows,
compared to 1500 for WATCH. Each tree
is trained on a bootstrap resample of the training windows. At each
split, the tree examines only a random subset of the input features,
rather than all 42 features at once. This reduces the similarity
between the trees and makes the final model more robust. The predicted
flare probability is then obtained by averaging the probabilities
returned by all trees. To account for the smaller number of positive
windows, we use balanced class weights during training. Unlike the
linear models, RF does not need an explicit expansion of the feature
space. It can learn non-linear relations and interactions between
features directly from the data. We also compute the out-of-bag score
during training. This uses the samples left out of the bootstrap
resampling for each tree and provides an internal check of model
consistency \citep{Breiman1996}.

\subsection{Probability calibration and threshold selection}
\label{sec:calibration}
\label{sec:threshold}

Before the models are evaluated on the held-out target stream, their raw
scores must be converted into calibrated probabilities and then into
practical alert decisions. Both steps are performed using only the
pre-cutoff TRAIN data, so that the full procedure remains causal.
The raw output of a classifier is not always a reliable probability. For
this reason, we calibrate the WATCH and TRIGGER models before choosing
the final alert thresholds. We use the sigmoid calibration of
\citet{Platt1999}, in which a raw model score $s$ is converted into a
calibrated probability
\begin{equation}
\hat{p} = \sigma(As + B),
\end{equation}
where $\sigma$ is the logistic sigmoid function, and $A$ and $B$ are
determined from part of the TRAIN sequence. This step does not change the
time ordering of the samples; it only adjusts the probability scale so
that the model output is more consistent with the observed fraction of
positive windows. To keep this step causal, the calibration uses only the most recent part
of TRAIN. We first keep the last 20 per cent of the TRAIN sequence as a
calibration tail. If this tail contains too few positive windows, it is
expanded in steps of 5 per cent, up to a maximum of 60 per cent, until
it contains both positive and negative windows. The earlier part of
TRAIN is then used to fit the base model, while the trailing tail is used
only to adjust its probability scale. In this way, the calibration is
based entirely on pre-cutoff data and is anchored to the most recent
behavior before the cutoff. If the calibration split still does not
contain both classes, the model is left uncalibrated. Across our 14
sources, most already reach enough flares within 20 per cent of TRAIN.
A few need more, and two need up to 50 per cent. We set the limit at 60
per cent to leave these two sources a margin, without taking too much
data away from the base model. None of our 14 sources ever needed more
than 50 per cent, so we had no reason to test a limit above 60 per
cent. The tail is taken out of TRAIN rather than added to it, so a
larger tail leaves less data to fit the base model. We also compared
the 20 per cent baseline
against a smaller 10 per cent tail. At 20 per cent, the held-out Brier
score  is 0.0861 and the Brier Skill Score (BSS) is 0.197. At 10 per cent, the Brier score
is worse (0.0924) and the BSS drops to 0.138, and the reliability curve
sits further from the diagonal (Figure~\ref{fig:plr_diagnostics}c). This
confirms 20 per cent as the better choice for our data.

For threshold selection, we then need one probability for each TRAIN
window, kept in time order. For the earlier TRAIN windows, we mimic real
forecasting: the model is first trained on the oldest part of the TRAIN
sequence and used to predict the next set of windows. The training range
is then extended forward in time, and the same procedure is repeated. For
the final calibration tail, we use the calibrated probabilities
directly. Joining these two parts gives a single causal TRAIN
probability sequence for each task. These sequences are used only for
threshold selection.
The WATCH and TRIGGER models return a probability for every window, but
in practice a probability alone is not enough: we also need a rule for
deciding when it is high enough to issue an alert. We therefore determine
separate thresholds for the WATCH and TRIGGER tasks from their own causal
TRAIN probability sequences. For WATCH, these causal TRAIN
probabilities are first constructed separately within each source and are
then pooled, so that each blazar contributes its own time-ordered
pre-cutoff WATCH behavior before a common threshold is chosen. For
TRIGGER, the threshold is still determined from the combined causal TRAIN
trigger sequence.  The two tasks are treated separately because they are
designed for different purposes: WATCH is intended to identify an
elevated-risk state before flare activity, whereas TRIGGER is intended to
identify a more selective pre-onset alert.

\paragraph{WATCH threshold.}
To turn the WATCH probability into a practical alert, we must choose a
single probability threshold, $\tau_{\rm W}$. We test candidate
thresholds from 0.05 to 0.95 in steps of 0.01. At a trial threshold
$\tau$, any window with $\hat{p}_{\rm W}(T_{\rm end}) \geq \tau$ is
classified as WATCH-positive. These predictions are then compared with
the true WATCH labels of the TRAIN windows.
For each trial threshold, we compute the precision $P(\tau)$ and recall
$R(\tau)$. Here precision is the fraction of predicted WATCH-positive
windows that are truly WATCH-positive, while recall is the fraction of
true WATCH-positive windows that are successfully recovered. We then
evaluate the recall-weighted $F_2$ score,
\begin{equation}
F_2(\tau) = \frac{5\,P(\tau)\,R(\tau)}{4\,P(\tau)+R(\tau)},
\end{equation}
and choose the threshold that maximises it:
\begin{equation}
\tau_{\rm W} = \arg\max_\tau F_2(\tau).
\label{eq:f2_threshold}
\end{equation}
We use the $F_2$ score because it gives more weight to recall than to
precision. This is appropriate for WATCH, where missing a genuine
pre-flare window is usually more costly than issuing an extra early
warning.

\paragraph{TRIGGER threshold and alert states.}
The TRIGGER threshold is selected in a similar way, but with a more
conservative goal. Whereas WATCH is intended to identify an elevated-risk
state, TRIGGER is meant to provide a smaller number of more selective
pre-onset alerts. For this reason, we again test thresholds from 0.05 to
0.95 in steps of 0.01, but now use the onset-based TRIGGER probabilities
and labels.
At each trial threshold, windows with
$\hat{p}_{\rm T}(T_{\rm end}) \geq \tau$ are classified as
TRIGGER-positive and compared with the true TRIGGER labels of the TRAIN
windows. For every trial threshold, we compute the precision, recall, the
precision-weighted $F_{0.5}$ score, and the alert fraction. Here the
alert fraction means the fraction of TRAIN windows that would be placed in
the TRIGGER state at that threshold. We use $F_{0.5}$ rather than $F_2$
because TRIGGER is intended to be more selective, so precision is more
important here than recall.
Not every threshold is accepted. We require the precision to be at least
0.25 and the alert fraction to remain below 0.20. Among the thresholds
that satisfy these conditions, we choose the one that gives the best
overall TRIGGER behavior. We selected it by a score defined as

\begin{equation}
S(\tau) = F_{0.5}(\tau) + 0.10\,\bigl[\operatorname{precision}(\tau)
             - \operatorname{alert\ fraction}(\tau)\bigr],
\label{eq:trigger_score}
\end{equation}
The score is evaluated only among thresholds that already pass the precision and
alert-fraction limits above. The single 0.10 weight is small next to the
$F_{0.5}$ term, so $F_{0.5}$ still sets the ranking; it only adds a small,
equally-sized bonus for higher precision and penalty for placing too
many windows in TRIGGER. If no threshold satisfies both constraints, the
code falls back to the best available threshold, giving highest priority
to precision and then to the remaining ranking measures.

Once the WATCH and TRIGGER thresholds, $\tau_{\rm W}$ and $\tau_{\rm T}$,
have been fixed, each window is assigned to one of three operational
states. A window is first checked against the TRIGGER model. If
$\hat{p}_{\rm T}(T_{\rm end}) \geq \tau_{\rm T}$, it is assigned to the
\textit{TRIGGER} state. If this condition is not met, the WATCH model is
then checked. A window is assigned to \textit{WATCH} when
$\hat{p}_{\rm W}(T_{\rm end}) \geq \tau_{\rm W}$, and to
\textit{Normal} otherwise. In this way, TRIGGER takes priority over
WATCH.
For the timeline figures, we show a causal seven-window trailing mean of
the WATCH and TRIGGER probabilities in order to make the visual trends
easier to see. However, the threshold selection, performance metrics, and
lead-time calculations are all based on the raw calibrated
probabilities.

\subsection{Model evaluation}
\label{sec:eval}

After calibrating the WATCH and TRIGGER probabilities and fixing the
final thresholds from the TRAIN data, we evaluate the models on the
held-out target windows. This tells us how well the method performs on
data that were not used in model fitting, calibration, or threshold
selection. We examine the results from four related viewpoints: how well
the model ranks flare-related windows above non-flare windows, how well
the selected WATCH threshold works as a binary forecast, how reliable the
WATCH probability scale is, and how much advance warning the WATCH and
TRIGGER alerts provide before flare onset. Because the onset-based
TRIGGER task contains fewer positive examples, we summarize it mainly
through held-out area under the receiver-operating characteristic curve (ROC AUC) and the Average
Precision (AP) and the observed warning behavior.

To evaluate ranking performance, we use the ROC AUC and the AP,  ROC AUC measures how well the model assigns higher scores
to true positive windows than to true negative windows, without assuming
any particular probability threshold. AP summarizes the corresponding
precision--recall curve, and is especially useful here because positive
windows are much rarer than negative ones \citep{Saito2015}. We report
these two ranking metrics for both the WATCH and TRIGGER probabilities.

To check whether the WATCH probabilities are meaningful as probabilities,
and not only as ranking scores, we use the BSS. This compares the average probability error of the model with that of a simple baseline forecast that assigns the same probability to every window, equal to the fraction of WATCH-positive windows in the evaluation set, 
$\bar{y}$. The Brier score of the model is
$\mathrm{BS}_{\rm model}=\overline{(\hat{p}-y)^2}$, while the reference
forecast has $\mathrm{BS}_{\rm clim}=\bar{y}(1-\bar{y})$. The BSS is then
\begin{equation}
\BSS = 1 - \frac{\mathrm{BS}_{\rm model}}{\mathrm{BS}_{\rm clim}}
     = 1 - \frac{\overline{(\hat{p} - y)^2}}{\bar{y}(1-\bar{y})}.
\label{eq:bss}
\end{equation}
Here $y$ is the true binary WATCH label, $\hat{p}$ is the predicted WATCH
probability, and $\bar{y}$ is the fraction of WATCH-positive windows in
the evaluation set. A positive BSS means that the model probabilities are better than a simple baseline that assigns the same WATCH probability to every window, whereas a negative BSS means that the predicted probabilities should be interpreted with caution, because their numerical values are less reliable than a simple constant baseline forecast.

We also evaluate the WATCH model  (classifier trained using the WATCH labels) after converting the probabilities into
binary WATCH alerts at the selected threshold. This conversion is done by comparing the WATCH probability of each window with the selected WATCH threshold, and assigning the window to the WATCH state when the probability exceeds that threshold.
For this purpose, we
report the $F_2$ score and the Matthews Correlation Coefficient (MCC).
The $F_2$ score is useful here because it gives more weight to recall
than to precision, which is appropriate when missing a flare-related
window is more costly than issuing an extra alert. We also report MCC,
which uses all four entries of the confusion matrix and therefore gives a
balanced summary of binary classification performance:
\begin{equation}
\mathrm{MCC} =
\frac{TP \cdot TN - FP \cdot FN}
{\sqrt{(TP+FP)(TP+FN)(TN+FP)(TN+FN)}},
\label{eq:mcc}
\end{equation}
where $TP$, $TN$, $FP$, and $FN$ are the numbers of true positives, true
negatives, false positives, and false negatives, respectively. MCC is
especially useful because it remains informative even when the numbers of
positive and negative windows are very different.

The TEST windows are not independent, because each 365-d rolling window
is separated from the next by only 7 d and therefore overlaps heavily
with it. As a result, neighboring TEST windows share about 98 per cent
of their input data, so standard significance tests that assume
independent samples are not suitable. To test whether the held-out WATCH
ROC AUC could arise by chance, we use a block-shuffle permutation test
\citep{Politis1994, Good2013}.
In this test, the WATCH labels in the TEST sequence are divided into
contiguous blocks of length $L_B=9$ windows, corresponding to about
63 d. These blocks are then randomly reordered, while the model scores
are kept fixed. This preserves the short-range grouping of positive
labels, but removes their true association with the predicted
probabilities. For each shuffled realization, we recompute the WATCH AUC
and compare it with the observed held-out value. The empirical
$p$-value is
\begin{equation}
p = \frac{\#\{\mathrm{AUC}_{\rm perm} \geq \mathrm{AUC}_{\rm obs}\} + 1}
         {N_{\rm perm,eff} + 1},
\label{eq:pvalue}
\end{equation}
where $\mathrm{AUC}_{\rm obs}$ is the observed held-out WATCH AUC,
$\mathrm{AUC}_{\rm perm}$ is the AUC obtained after block shuffling, and
$N_{\rm perm,eff}$ is the number of valid shuffled realizations. In this
work we use 1000 permutation trials. The Phipson--Smyth correction is
included so that the $p$-value remains well defined even if none of the
shuffled realizations exceeds the observed AUC \citep{Phipson2010}.

We estimate the uncertainty of the held-out WATCH AUC with a circular
block bootstrap \citep{Kunsch1989, Politis1992}. This is again needed
because the TEST windows are not independent. In each bootstrap
realization, we rebuild a TEST-like sequence by drawing blocks of length
$L_B$ with replacement, while allowing the blocks to wrap around at the
ends of the time series. We then recompute the WATCH AUC for each
resampled sequence. The 95 per cent confidence interval is taken from the
2.5th and 97.5th percentiles of the bootstrap AUC distribution, using the
baseline block length $L_B=9$ windows and 500 bootstrap resamples. As a
sensitivity check, we repeat the same calculation for block lengths of
5, 7, 9, 10, and 15 windows, using 200 resamples for each case. We
choose $L_B=9$ windows as a baseline value that is short enough to give
a workable number of quasi-independent blocks from the 172-window TEST
sequence, while still exceeding the short-range grouping of individual
pre- and post-flare windows. The results are consistent across all of
these block lengths: the confidence interval on the held-out WATCH AUC
barely changes, and it always excludes chance performance
(AUC$=0.5$).

In addition to the held-out test, we also assess model behavior within
the pre-cutoff data by performing a purged walk-forward validation on the
combined TRAIN sample \citep{dePrado2018}. In each fold, the model is
trained on an earlier segment of the TRAIN sequence and tested on a later
segment, so that the time ordering is preserved. A purge gap of
$L_{\rm purge}=6$ windows (about 42 d) is left between the training and
validation segments to reduce leakage from neighboring windows. In the
present implementation, the initial training segment contains at least
120 windows, each validation chunk contains 90 windows, and the
walk-forward procedure advances in 90-window steps. The validation
predictions from all folds are then combined to obtain a pooled
out-of-fold AUC, which provides a leakage-resistant summary of
training-period performance. We chose $L_{\rm purge}=6$ windows
because it is close to the 45-d TRIGGER horizon (about 7 windows). It
is, however, shorter than the 90-d WATCH horizon (about 13 windows). To
check whether this matters, we recomputed the pooled
out-of-fold WATCH AUC with larger purge gaps, up to $L_{\rm purge}=26$
windows (182~d, the lag at which the WATCH label autocorrelation
function first drops below $1/e$).
For PLR, our best-performing model, the pooled OOF AUC is
0.841 at $L_{\rm purge}=6$, 0.836 at $L_{\rm purge}=13$ (matching the
WATCH horizon), and 0.834 at $L_{\rm purge}=26$: a decrease of only
0.007 as the purge gap more than quadruples. Since enlarging the purge
gap changes the pooled OOF AUC by at most about 1 per cent, our
baseline choice of $L_{\rm purge}=6$ windows does not materially
inflate this internal diagnostic.

Finally, we examine whether the WATCH and TRIGGER alerts provide useful
advance warning before a flare. For each flare onset associated with the
held-out target period, we search the raw probability sequence for valid
WATCH or TRIGGER threshold crossings before the onset. To avoid counting
isolated one-window fluctuations, a crossing is accepted only if the
threshold remains exceeded for at least two consecutive windows.

For WATCH, the main search is carried out within 90 d before the flare
onset. For TRIGGER, the same procedure is used, but the maximum allowed
lead time is limited to 45 d, making TRIGGER a more conservative
near-term alert. For each detected case, we record three lead-time
measures: the earliest valid crossing, the latest valid crossing, and the
start of the alert episode containing the final valid crossing. In the
last case, nearby alert episodes are linked if they are separated by
fewer than five consecutive non-alert windows. As a sensitivity check, we
repeat the search with 30-d and 180-d lookback windows. Because the
number of held-out flare onsets is small, these lead times are
interpreted as an operational summary rather than as a precise population
measurement.

\section{Results}
\label{sec:results}

After defining the WATCH and TRIGGER labels, training the models, calibrating the probability outputs, and fixing the alert thresholds using only the TRAIN data, we now turn to the held-out target-source results. We first examine the Bayesian-Blocks flare segmentation used to define the labels, then compare the classifiers on the held-out WATCH and TRIGGER tasks, and finally study how the best-performing model behaves as a practical alert system.

\begin{figure*}
\gridline{
  \fig{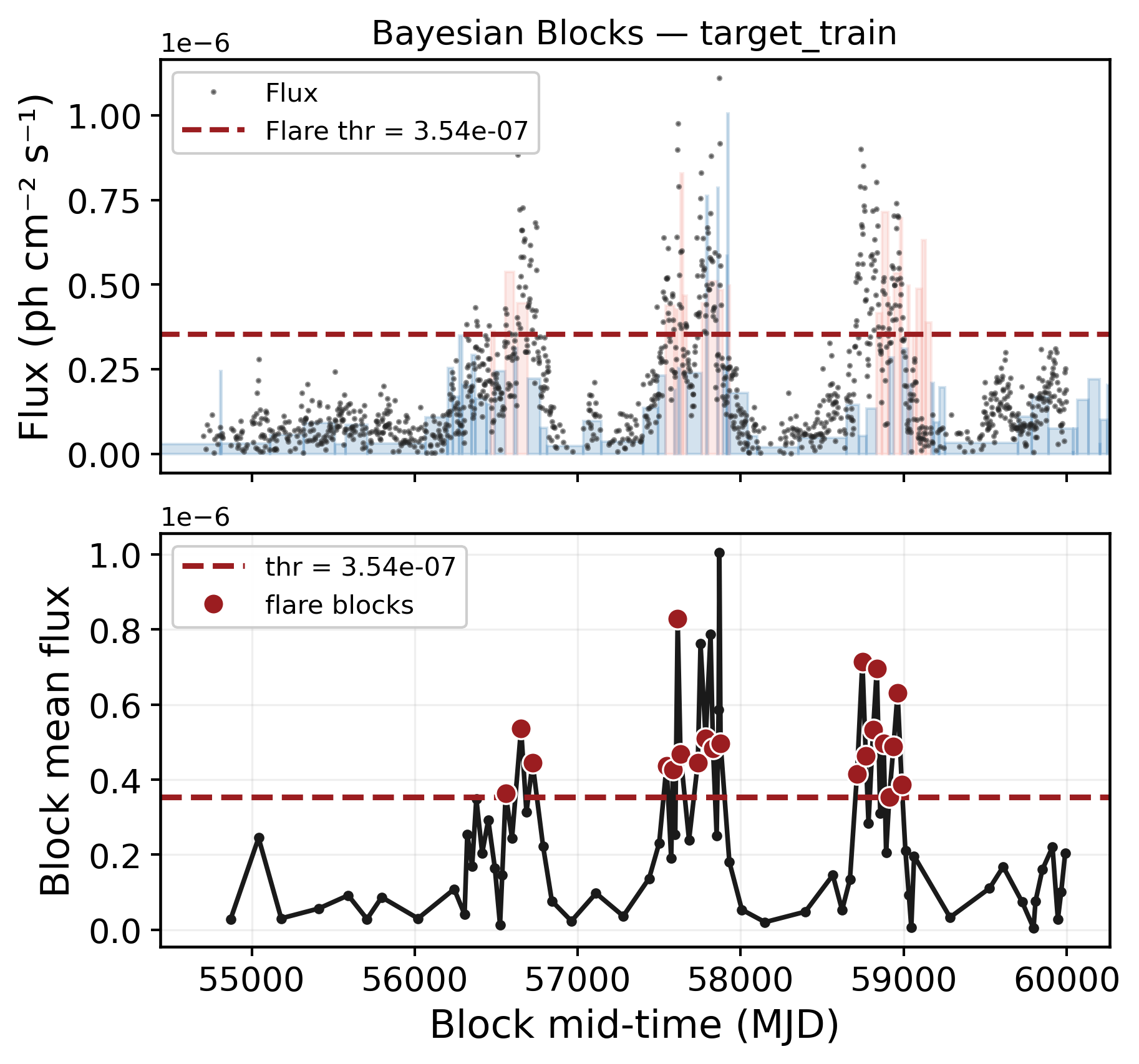}{0.48\textwidth}{(a) TRAIN-BB segmentation using only data with $t \leq \mathrm{MJD}\,60000$.}
  \fig{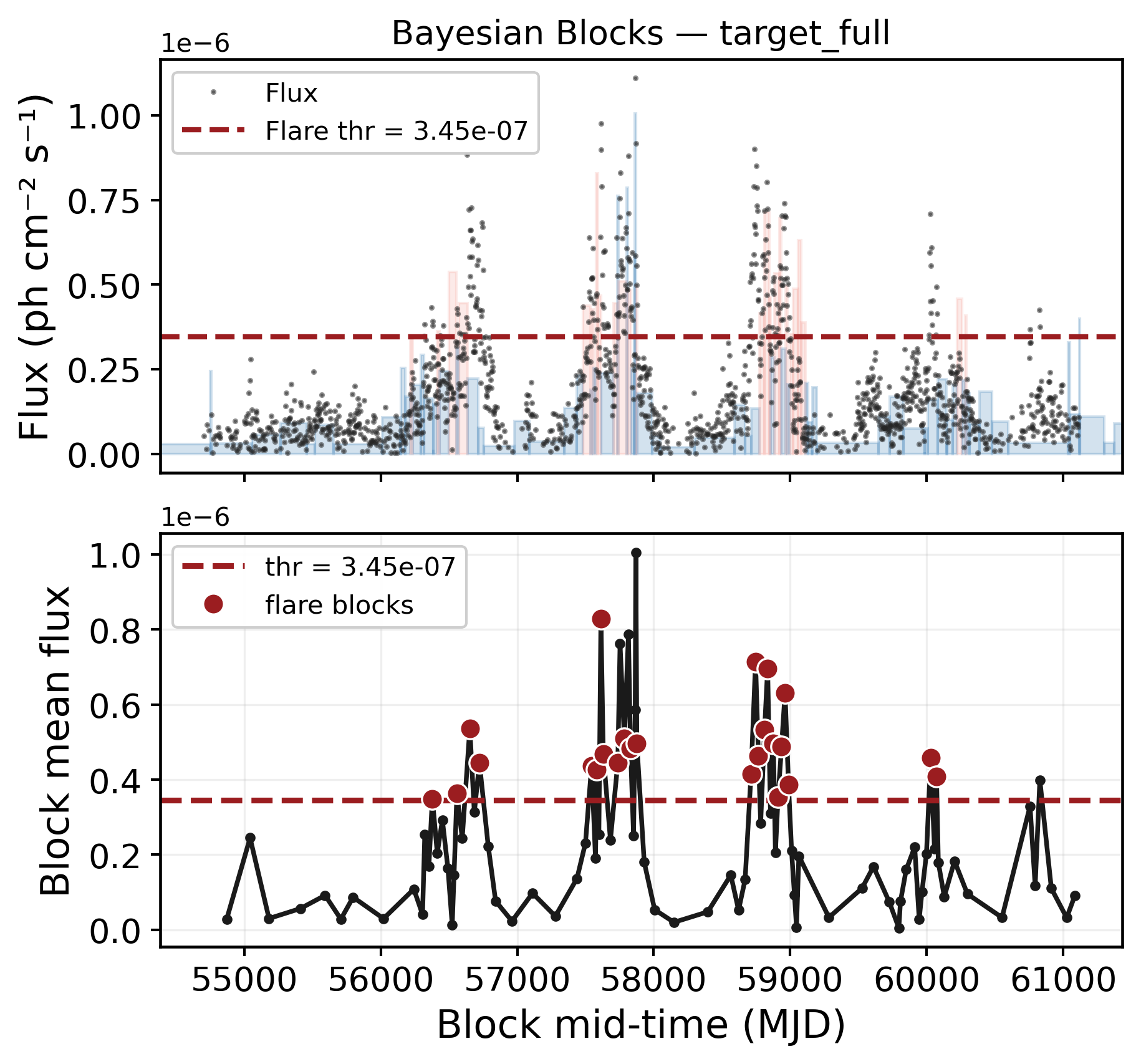}{0.48\textwidth}{(b) FULL-BB segmentation of the complete light curve.}
}
\caption{Bayesian-Blocks diagnostics for the target source. In each
panel, the upper plot shows the 3-d \fermi\ light curve together
with the Bayesian-Blocks representation. Each semi-transparent shaded
strip spans one Bayesian block in time and extends from zero flux to the
block-mean flux: red/pink strips mark blocks classified as flaring,
whereas light-blue strips mark non-flaring blocks. The red dashed line is
the flare-selection threshold from equation~(\ref{eq:flare_criterion});
its numerical value is listed in the panel legend and differs slightly
between TRAIN-BB and FULL-BB because the threshold is computed separately
for each segmentation. The lower plot shows the block-mean flux against
block mid-time, and the red points identify the flare blocks used to
define flare intervals.}
\label{fig:bb_diagnostics}
\end{figure*}

Figure~\ref{fig:bb_diagnostics} compares the Bayesian-Blocks flare segmentation of the target source obtained in the TRAIN-BB and FULL-BB runs. The TRAIN-BB segmentation is derived using only the data before the cutoff at $t_{\rm cut}=\mathrm{MJD}\,60000$, and is used to assign the training labels. The FULL-BB segmentation is derived from the complete light curve and is used only to evaluate the held-out forecasts. In this way, the training step remains causal, while the held-out predictions can still be compared with the flare activity that actually occurred.

With the training and scoring segmentations defined, we next evaluate the
forecasting performance on the held-out target-source windows. The held-out set
consists of rolling input windows from the target light curve. Each
window spans 365 d, is shifted forward in steps of 7 d, and is identified
by its end time $T_{\rm end}$.
 The held-out set
contains all target-source windows with
$T_{\rm end}>T_{\rm boundary}$, giving 172 rolling-window samples with
end times spanning MJD 59911.5--61108.5. These windows are not used in
model fitting. They are labelled only for evaluation, using the FULL-BB
flare intervals and flare onsets defined in
Section~\ref{sec:label_def}. Under the WATCH definition, 21 held-out
windows are positive and 151 are negative. Under the onset-based TRIGGER
definition, 9 windows are positive and 163 are negative.
Table~\ref{tab:test_metrics} summarizes the held-out performance of the
three classifiers. For each model, the WATCH AUC and WATCH AP are
computed from the 90-d WATCH probabilities and measure how well the 21
WATCH-positive windows are ranked above the 151 WATCH-negative windows.
The TRIGGER AUC and TRIGGER AP are computed separately from the 45-d
onset-based TRIGGER probabilities. The WATCH AUC confidence interval, the block-permutation $p$-value, and the BSS are then obtained with the procedures described in Section~\ref{sec:eval}. These procedures are chosen to account for the strong overlap between neighboring rolling windows.

\begin{figure*}
\gridline{
  \fig{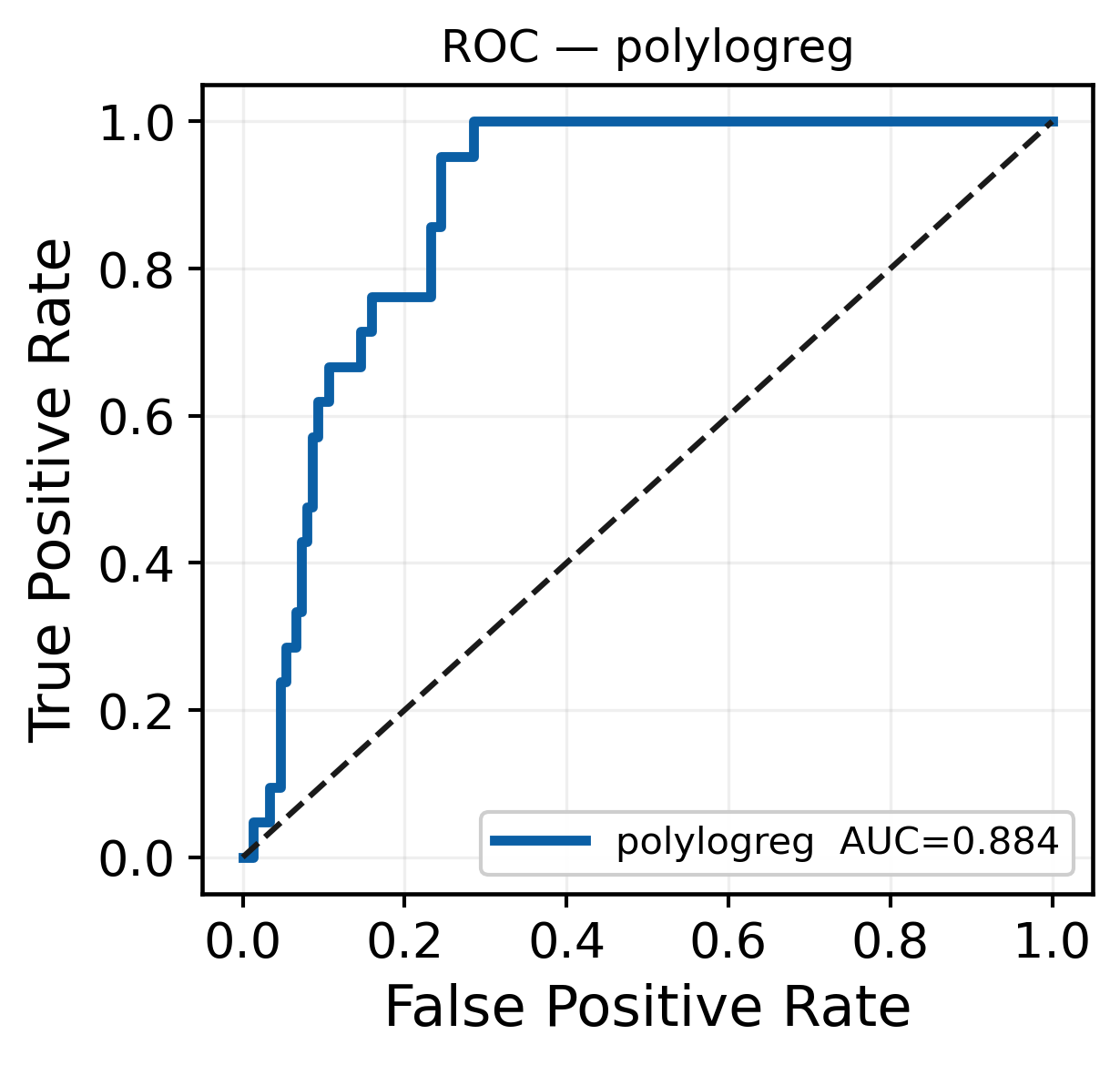}{0.31\textwidth}{(a) ROC curve.}
  \fig{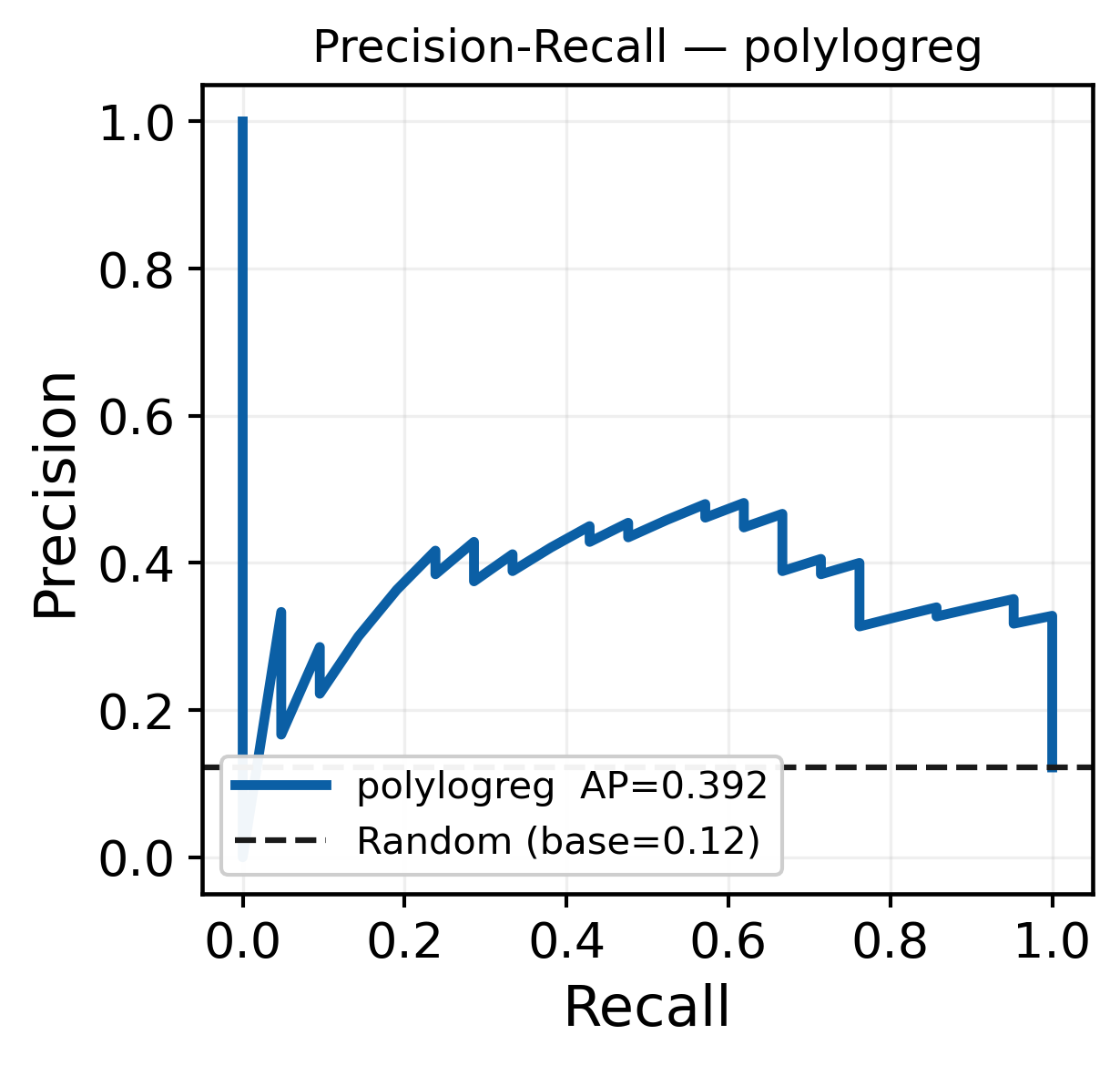}{0.31\textwidth}{(b) Precision--recall curve.}
  \fig{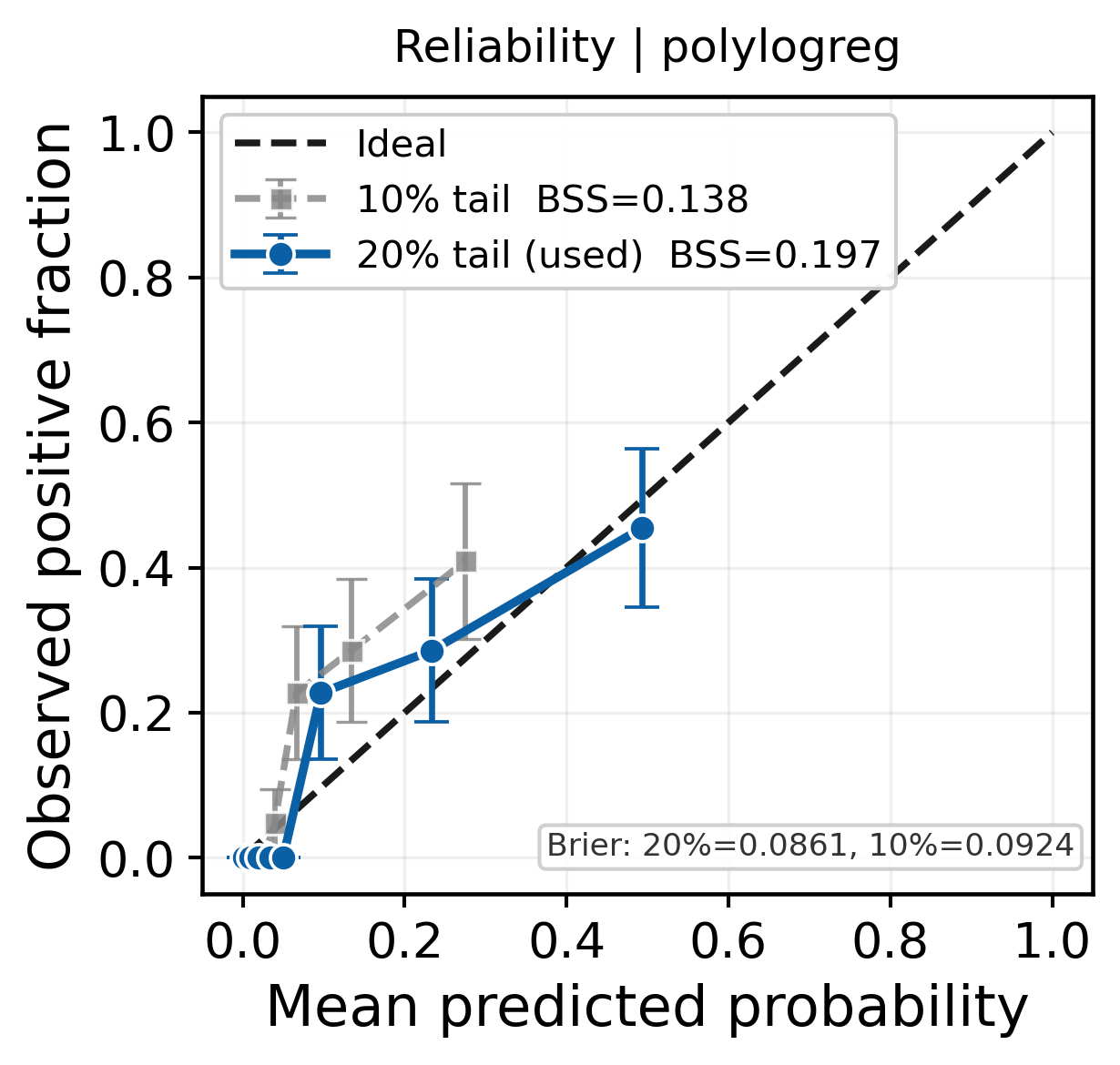}{0.31\textwidth}{(c) Reliability curve, 20 per cent tail (used) vs.\ 10 per cent tail.}
}
\caption{Held-out WATCH diagnostic curves for the PLR model. In the ROC
panel, the blue step-like curve is obtained by varying the WATCH
probability threshold across the held-out windows and plotting the
true-positive rate against the false-positive rate at each step. The
black dashed diagonal marks the random-ranking case, and the legend gives
the area under the curve ($\mathrm{AUC}=0.884$). In the
precision--recall panel, the blue curve shows the same threshold sweep in
precision--recall space. The dashed horizontal line marks the held-out
WATCH positive fraction, $21/172\simeq0.12$, which is the precision
expected from random alerting; the legend gives the average precision
($\mathrm{AP}=0.392$). In the reliability panel, the held-out windows are
grouped into bins with similar numbers of samples according to their
predicted WATCH probability. For each bin, the horizontal coordinate
shows the mean predicted probability and the vertical coordinate shows
the fraction of truly WATCH-positive windows; the error bars show the
binomial uncertainty. The black dashed diagonal marks perfect agreement
between predicted probability and observed positive fraction. The
blue curve uses the 20 per cent calibration tail adopted in this paper;
the gray curve shows the same model calibrated on a 10 per cent tail
instead, for comparison. The legend gives the BSS for
each curve ($\mathrm{BSS}=0.197$ at 20 per cent, $\mathrm{BSS}=0.138$ at
10 per cent), and the panel annotation gives the corresponding Brier
scores ($\mathrm{Brier}=0.0861$ and $0.0924$).}
\label{fig:plr_diagnostics}
\end{figure*}

Polynomial logistic regression (PLR) gives the strongest held-out WATCH
performance (AUC$=0.884$, AP$=0.392$; $p_{\rm perm}=0.009$, so this
ranking is unlikely to arise by chance),the highest held-out
TRIGGER AUC and AP of the three models (AUC$=0.751$, AP$=0.121$), and is the only model with a
positive held-out BSS ($\mathrm{BSS}=0.197$), meaning its
probabilities are not just well-ranked but also modestly more accurate
than a constant-probability baseline (Figure~\ref{fig:plr_diagnostics}c).
Unlike LR, PLR expands the input features with pairwise products and
squared terms before fitting, so it can represent how pairs of
features move together rather than only their individual effects. When
its ANOVA feature-selection step is left free to choose the 50 most
informative terms out of 945 candidates, 46 turn out to be interaction
or squared terms, and only 4 are plain linear terms -- direct evidence
that most of the useful pre-flare signal in our feature set is carried
by feature interactions. By comparison, RF gives the highest pooled
out-of-fold WATCH AUC on TRAIN (0.856; Table~\ref{tab:test_metrics})
but a negative held-out BSS ($-0.067$), and LR gives the lowest pooled
out-of-fold AUC (0.822) and is the weakest model on every held-out
metric (WATCH AUC$=0.687$, TRIGGER AUC$=0.539$,
$\mathrm{BSS}=-0.075$). PLR therefore
remains the best overall model for the main forecasting task in the
present backtest;

\begin{figure}
\centering
\includegraphics[width=0.48\columnwidth]{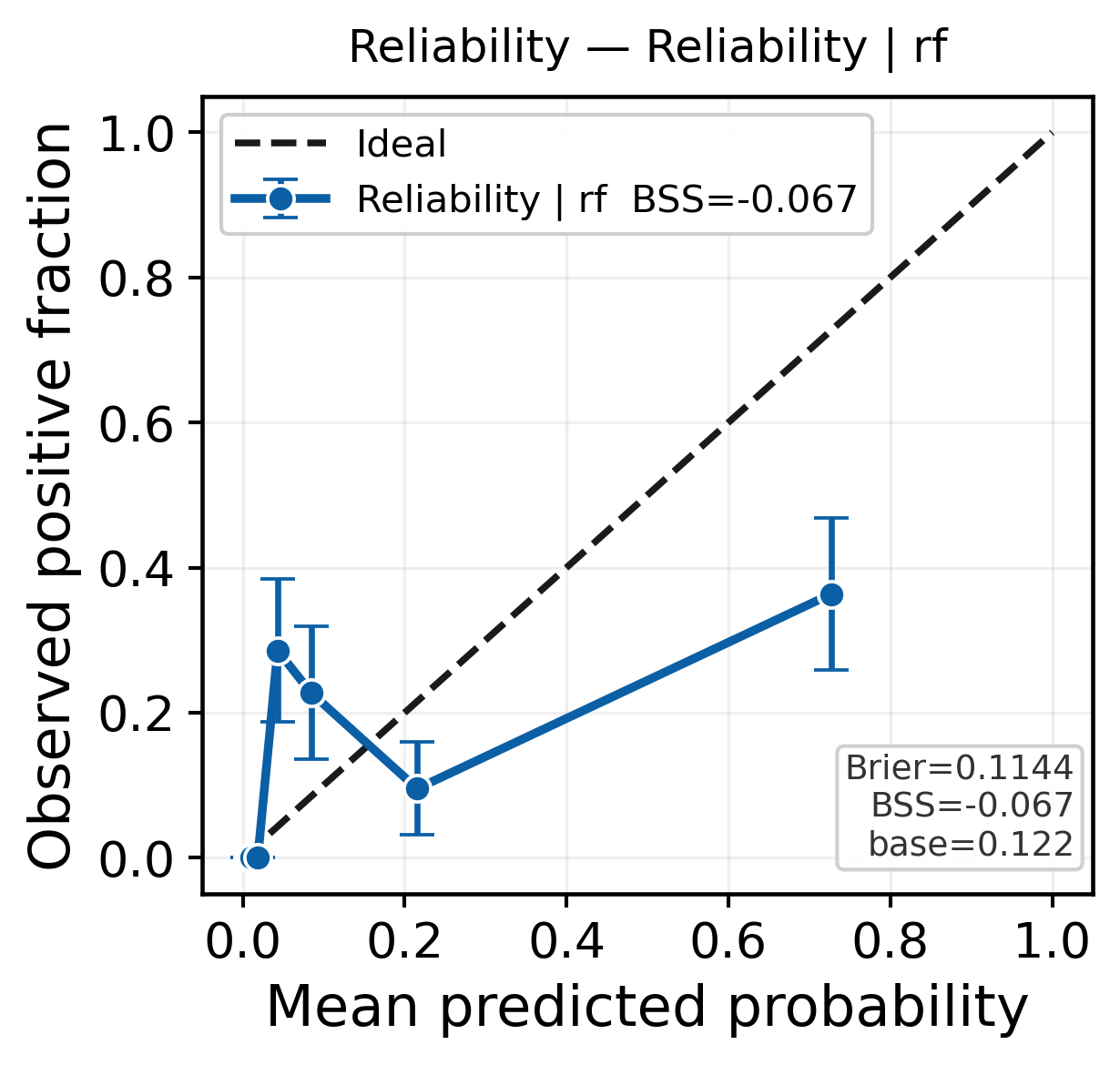}
\includegraphics[width=0.48\columnwidth]{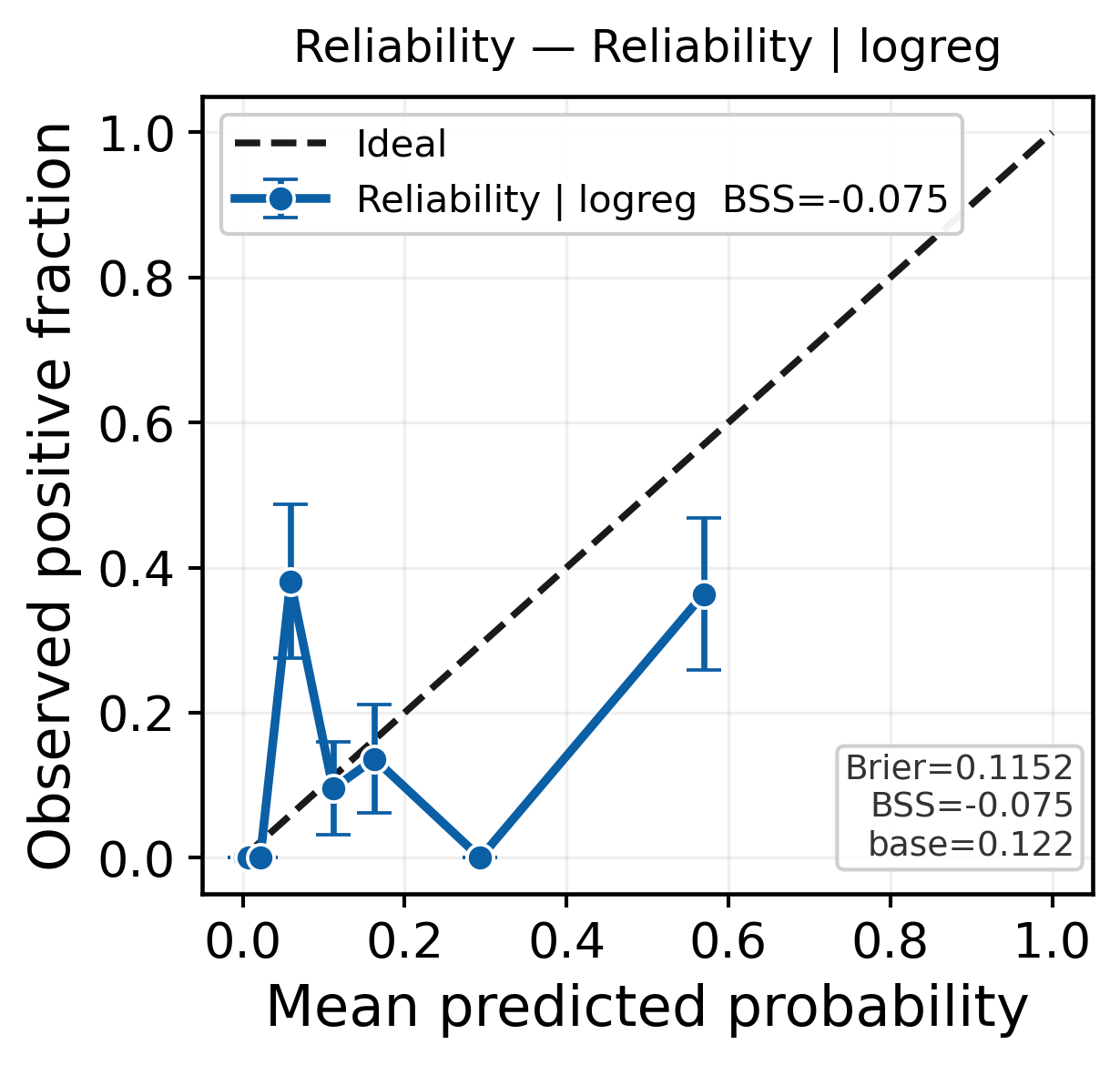}
\caption{Held-out WATCH reliability curves for (a) RF and (b) LR,
in the same format as Figure~\ref{fig:plr_diagnostics}c. RF's predicted
probabilities sit below the diagonal at higher probability bins
(over-prediction), while LR's curve is noisy and non-monotonic,
consistent with its weaker ranking performance. Both give a negative
BSS, in contrast to PLR (Figure~\ref{fig:plr_diagnostics}c).}
\label{fig:rf_lr_reliability}
\end{figure}

RF provides strong TRAIN performance but a weaker held-out
result. This is caused by a structural difference from
PLR and by overfitting. Structurally, PLR fits a single, constrained
equation using only 50 selected terms, whereas RF builds 200 decision
trees, each free to split the training windows using whatever rule
best separates WATCH-positive from WATCH-negative windows at each
step. This extra flexibility leads to overfitting: fit on the full
combined TRAIN sample, RF reaches a training accuracy of 0.991, well
above PLR's 0.828, and gives the best in-TRAIN ranking of the three
models, but this advantage does not transfer to the held-out period,
where its AUC drops to 0.792 and its BSS turns negative.
This is visible directly in RF's reliability curve
(Figure~\ref{fig:rf_lr_reliability}a), where predicted probabilities
sit below the diagonal at higher probability bins, i.e., RF
over-predicts the true flare risk there: predicted probabilities of
$\sim$0.73 correspond to an observed flare frequency of only
$\sim$0.36, even though its ranking (AUC)
is still reasonable. In practice, RF's ranking of windows by risk
remains usable, but its raw probabilities are not reliable without
further recalibration.

LR shows a different pattern. Unlike PLR, it does not account for
feature interactions: it is purely linear, combining each feature with
a fixed weight, and cannot represent the interactions identified above
as the main source of PLR's advantage. This is a structural
limitation, not overfitting: LR's training accuracy (0.840) is close
to PLR's (0.828) and far below RF's (0.991), so it remains the weakest
model on every held-out metric without ever fitting the training data
too closely. LR's own reliability curve
(Figure~\ref{fig:rf_lr_reliability}b) is correspondingly noisy and
non-monotonic. In practice, LR is unlikely to improve without adding
feature interactions, as PLR does.

\begin{figure}
\centering
\includegraphics[width=\columnwidth]{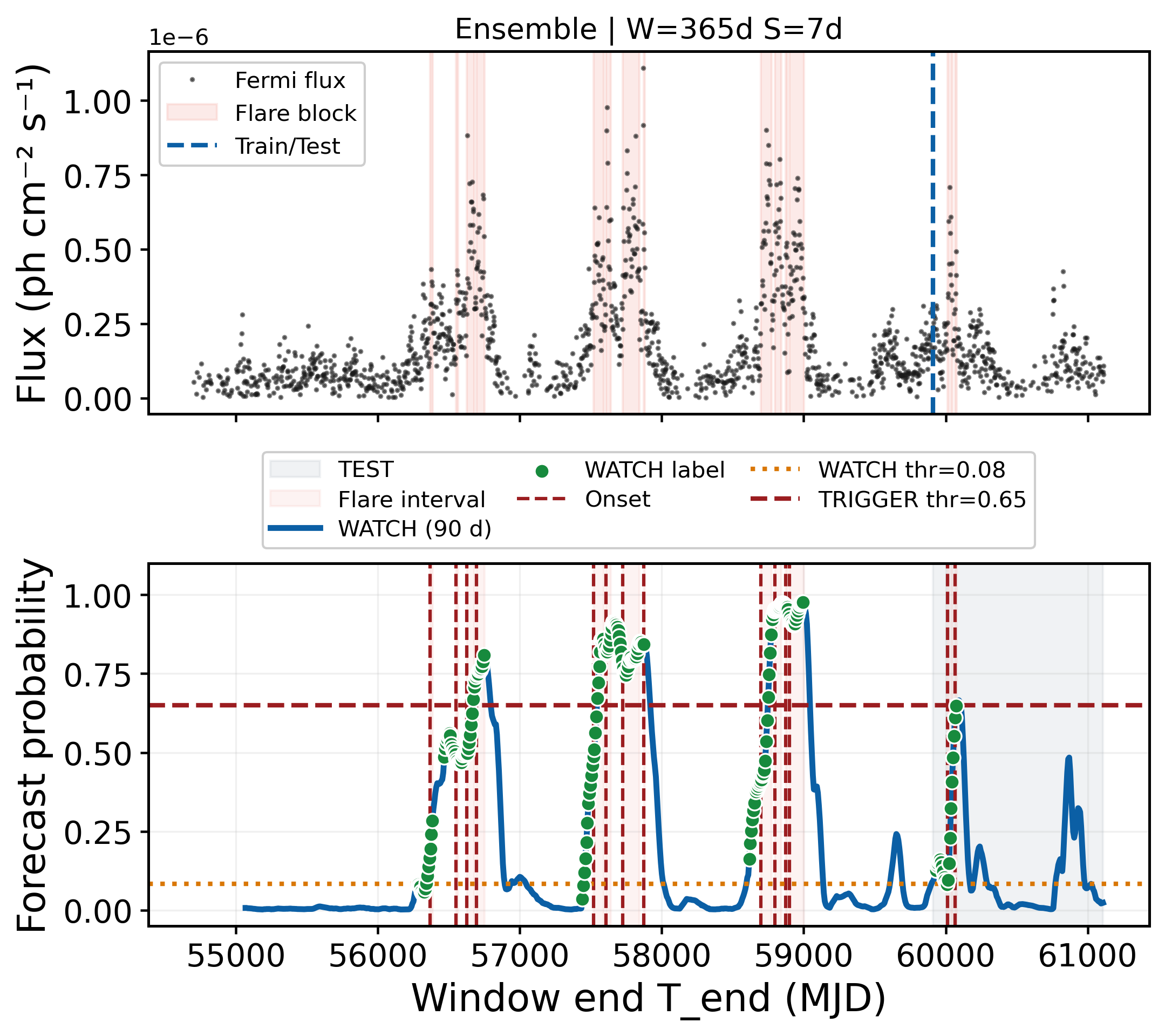}
\caption{WATCH-only soft-voting ensemble for the target source. For each
rolling window, the raw calibrated WATCH probabilities from logistic
regression, polynomial logistic regression, and random forest are
averaged to form the ensemble WATCH probability. The top panel shows the
3-d binned \fermi\ flux; the salmon-shaded bands mark the FULL-BB flare
intervals, and the navy dashed vertical line marks the train/test
boundary at $T_{\rm boundary}=\mathrm{MJD}\,59910$. The lower panel
shows the ensemble WATCH probability after applying a causal seven-window
trailing mean for visual clarity. Filled green circles mark the
WATCH-positive FULL-BB scoring windows, dark-red dashed vertical lines
mark FULL-BB flare onsets, and the pale shaded region marks the held-out
test interval. The orange dotted horizontal line shows the mean WATCH
threshold of the three component models,
$\langle\tau_{\rm W}\rangle=(0.06+0.06+0.13)/3=0.083$. The red dashed
horizontal line shows the corresponding mean of the three individual
TRIGGER thresholds,
$\langle\tau_{\rm T}\rangle=(0.87+0.77+0.31)/3=0.65$, and is included
only as a visual guide; no separate ensemble TRIGGER probability is
constructed in this work.}
\label{fig:timeline_ensemble}
\end{figure}

For completeness, we also construct a soft-voting WATCH ensemble by
averaging the raw calibrated WATCH probabilities from logistic
regression, polynomial logistic regression, and random forest on each
held-out window. The WATCH AUC and WATCH AP of the ensemble are then
computed from this averaged probability sequence, rather than by averaging
the individual model AUC values. On the held-out WATCH task, the ensemble
reaches AUC $=0.801$ and AP $=0.362$, so it does not outperform
PLR.
Nevertheless, the ensemble timeline remains useful as a comparison,
because it highlights the same broad elevated-risk intervals before the
two held-out flare episodes (Fig.~\ref{fig:timeline_ensemble}). In our analysis, the ensemble is constructed only for the WATCH probabilities; no corresponding ensemble is defined for the TRIGGER task.

\begin{table*}
\centering
\caption{Summary of model performance on the combined TRAIN sample and on
the held-out target-source TEST set.}
\label{tab:test_metrics}
\resizebox{\textwidth}{!}{%
\begin{tabular}{lcccccccc}
\toprule
Model & WATCH pooled OOF AUC & WATCH AUC & WATCH AUC 95\% CI & $p_{\rm perm}$ & WATCH AP & WATCH BSS & TRIGGER AUC & TRIGGER AP \\\midrule
Logistic Regression & 0.822 & 0.687 & 0.418--0.938 & 0.157 & 0.259 & $-0.075$ & 0.539 & 0.069 \\
Polynomial Logistic Regression & 0.841 & 0.884 & 0.792--0.978 &  0.009 & 0.392 & $0.197$ & 0.751 & 0.121 \\
Random Forest & 0.856 & 0.792 & 0.564--0.971 & 0.059 & 0.375 & $-0.067$ & 0.746 & 0.105 \\
\bottomrule
\end{tabular}
}
\vspace{2pt}

\begin{minipage}{0.96\textwidth}
\footnotesize
\emph{Notes.} Each row reports the held-out performance of one classifier.
The TEST set contains 172 rolling windows from the target source with
$T_{\rm end}>T_{\rm boundary}$. Under the WATCH definition, 21 windows are
positive and 151 are negative; under the TRIGGER definition, 9 windows are
positive and 163 are negative. The WATCH pooled OOF AUC is the pooled
out-of-fold ROC AUC from purged walk-forward validation on the combined
TRAIN sample. WATCH AUC and WATCH AP are computed from the held-out WATCH
probabilities, while TRIGGER AUC and TRIGGER AP are computed from the
separate onset-based TRIGGER probabilities. The WATCH AUC 95\% confidence
interval is obtained from a circular block bootstrap with 500 trials and a
9-window block length. The permutation probability $p_{\rm perm}$ is the
block-shuffle probability, based on 1000 trials, of obtaining a WATCH AUC
at least as large as the observed value by chance. WATCH BSS is the Brier
Skill Score relative to a simple baseline forecast that assigns the same
WATCH probability to every held-out window, equal to the held-out
WATCH-positive fraction.
\end{minipage}
\end{table*}

Since PLR gives the strongest held-out WATCH performance, we examine its
diagnostic results in more detail. For WATCH, the threshold is
fixed in advance from the pooled source-local causal TRAIN WATCH scores
by maximising the $F_2$ score, as described in
Section~\ref{sec:threshold}.  For PLR, this gives
$\tau_{\rm W}=0.06$. We then apply this fixed threshold to the
raw calibrated WATCH probabilities of the held-out target windows, so
that each window with $\hat{p}_{\rm W}\geq\tau_{\rm W}$ is counted as a
WATCH-positive forecast.
With this threshold, the held-out confusion matrix is
$TP=21$, $FP=49$, $TN=102$, and
$FN=0$. These counts correspond to a
precision of $21/70=0.300$ and a recall of $21/21=1.000$, giving
$F_2=0.682$ and $\mathrm{MCC}=0.450$. In other words, the model
recovers 21 of the 21 WATCH-positive held-out windows, while 49
of the 151 WATCH-negative windows are incorrectly assigned to WATCH.
These threshold-specific values are obtained directly from the confusion
matrix at $\tau_{\rm W}=0.06$.
Figure~\ref{fig:plr_diagnostics} shows the held-out ROC and
precision--recall curves of the PLR WATCH probabilities, together with
the reliability curve. The ROC AUC and AP summarize the ranking
performance across all thresholds, whereas the precision, recall, $F_2$,
and MCC values above describe the binary WATCH behavior only at the
selected threshold $\tau_{\rm W}=0.06$. The improvement of PLR over the
purely linear model suggests that useful pre-flare information is carried
by combinations of variability features, rather than by single features
alone.

Good ranking performance does not necessarily mean that the WATCH
probabilities are perfectly calibrated. Figure~\ref{fig:plr_diagnostics}
shows the reliability curve of the PLR WATCH probabilities. In this
panel, the horizontal axis gives the mean predicted WATCH probability in
each bin, while the vertical axis gives the observed fraction of
WATCH-positive windows in the same bin. The overall trend is upward,
which means that bins with higher predicted WATCH probabilities generally
contain a larger fraction of true WATCH-positive windows. In the
present backtest, the held-out WATCH BSS is positive
($\mathrm{BSS}=0.197$). This means that the numerical WATCH
probabilities are modestly better than a simple baseline forecast that
assigns the same WATCH probability to every held-out window. The WATCH
output can therefore be interpreted as a meaningful probabilistic
indicator of flare risk, although its clearest practical use is still to
identify elevated-risk intervals before flare onset.

\begin{figure}
\centering
\includegraphics[width=\columnwidth]{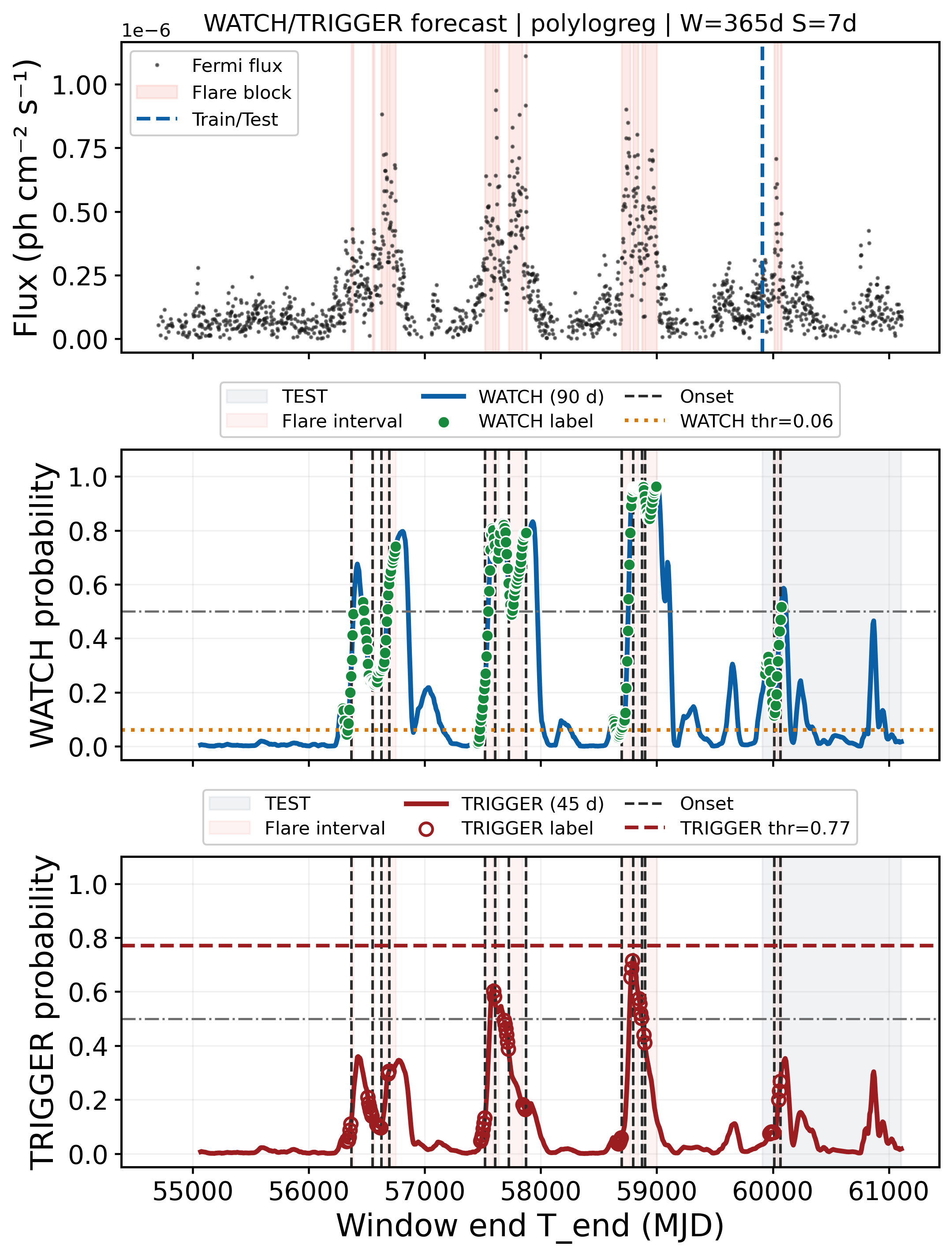}
\caption{PLR alert-state timeline. The top panel shows the 3-d binned \fermi\ flux as black
points; salmon shading marks FULL-BB flare intervals, and the navy dashed
vertical line marks the train/test boundary at
$T_{\rm boundary}=\mathrm{MJD}\,59910$. The middle panel shows the
causal seven-window trailing mean of the WATCH score (solid blue);
filled green circles mark windows that are WATCH-positive under the
FULL-BB scoring labels, and the orange dotted horizontal line is the
WATCH threshold ($\tau_{\rm W}=0.06$). The bottom panel shows the
TRIGGER score (solid dark red); open dark-red circles mark windows that
are TRIGGER-positive under the onset-based scoring labels, and the
dark-red dashed horizontal line is the TRIGGER threshold
($\tau_{\rm T}=0.77$). In both lower panels, the pale grey background
marks the held-out test stream, the grey dash-dotted horizontal line is
a visual probability guide at $\hat{p}=0.50$, and black dashed vertical
lines mark FULL-BB flare onsets. Lead-time values quoted in the text
are computed separately from the raw, unsmoothed probabilities using
the primary 90-d look-back window and the two-window persistence
requirement.}
\label{fig:timeline_lead}
\end{figure}

With this probabilistic WATCH interpretation in mind, we next examine how the PLR
WATCH and TRIGGER probabilities evolve in time. Figure~\ref{fig:timeline_lead}
shows the target-source probability timeline. In the middle panel, the
solid blue curve is the WATCH probability; in the bottom panel, the
solid dark-red curve is the TRIGGER probability. For visual clarity,
both curves are shown after applying a causal seven-window trailing
mean. The orange dotted horizontal line marks the WATCH threshold,
$\tau_{\rm W}=0.06$, and the dark-red dashed horizontal line marks the
TRIGGER threshold, $\tau_{\rm T}=0.77$.
In this smoothed display, both held-out flare onsets near MJD 60014 and
MJD 60068 are preceded by WATCH-level probabilities. By contrast, the
smoothed TRIGGER curve does not cross the TRIGGER threshold before either
flare. Before the second flare it rises somewhat, peaking at
$\simeq$0.27 (smoothed) against $\tau_{\rm T}=0.77$, but remains well
below the threshold throughout. The open red circles mark the true onset-based
TRIGGER-positive windows; they are evaluation labels, not threshold
crossings of the model. The raw held-out
TRIGGER probabilities also remain below the TRIGGER threshold, so no
valid pre-onset TRIGGER crossing is obtained in either the smoothed
display or the raw scoring sequence.

To make the warning behaviour more explicit, we next measure when the
PLR model enters the WATCH state before the two held-out flare onsets.
Here a window is counted as WATCH-active when the raw WATCH probability
exceeds the WATCH threshold, $\tau_{\rm W}=0.06$, for at least two
consecutive windows. The lead time is then the time difference between
that accepted WATCH window and the flare onset. We use the raw WATCH
probabilities for this calculation, rather than the smoothed curves shown
in Fig.~\ref{fig:timeline_lead}, so that the timing is not shifted by the
visual smoothing.
We report two related timing measures. The first is the last accepted
WATCH window before flare onset. This means the final window in the last
valid WATCH sequence before the flare begins, and it gives a
conservative estimate of the final warning time. For the two held-out
flares, these last accepted WATCH windows occur at MJD 60009.5 and
MJD 60065.5. The corresponding lead times, defined as the time between
the accepted WATCH window and the flare onset, are 4.5 and 2.5 d, with
a median of 3.5 d.
The second measure is the start of the underlying WATCH episode. Here a
WATCH episode means the broader period during which the source stays in
the WATCH state; nearby WATCH runs are linked into the same episode if
they are separated by fewer than five consecutive non-WATCH windows.
Using this definition, the WATCH state switches on at MJD 59855.5 and
remains continuously active through both onsets, without a single
WATCH-negative window in between: this is one extended elevated-risk
episode covering both flares, not two separate shorter episodes. This
corresponds to lead times of 158.5 d before the first onset and 212.5 d
before the second. In simple terms, the model identifies one broad,
sustained period of elevated flare risk that covers both flares, and
each flare is additionally preceded by its own shorter final WATCH
warning close to onset. No comparable pre-onset TRIGGER detection is
found. The TRIGGER alert threshold is set high on purpose, to avoid
false alarms (Section~\ref{sec:threshold}). Although the TRIGGER AUC
(0.751) is above random, the TRIGGER AP (0.121) is only about twice the
random baseline of $9/172=0.052$, so the TRIGGER ranking is not strong
enough to support reliable onset alerts in this held-out sample.

\section{Discussion}
\label{sec:discussion}

The possibility of predicting blazar flares is closely connected to the
unique time-domain role of the \fermi\ . Since the start of the mission, the
LAT has monitored the $\gamma$-ray sky with a high duty cycle, producing
long and nearly uniform light curves for a large population of variable
sources \citep{Atwood2009, Abdollahi2023}.  The \fermi\ Light Curve
Repository is especially important in this context: it provides
multi-cadence light curves for more than 1500 variable LAT sources, most of
which are blazars, and therefore turns blazar variability from a collection
of isolated events into a statistical forecasting problem.  Instead of
asking only whether a source is flaring now, one can ask whether its recent
light curve resembles source states that have preceded flares in the past.

This is the motivation behind the present work.  We use a sample of bright
\fermi\ blazars to learn source-normalized variability patterns, and then
test whether those patterns carry predictive information for the target
source 4FGL\,J1048.4$+$7143.  The source normalization is important because
different blazars have different average fluxes, noise levels, and flare
amplitudes.  The aim is therefore not to learn that a particular absolute
flux is high, but to learn whether the recent variability state of a source
has become flare-like relative to its own long-term behavior.

The physical basis for this approach is that a blazar flare, although it
may appear as a rapid $\gamma$-ray outburst, need not be a completely
sudden event.  In shock-in-jet, turbulent, and magnetic-reconnection
pictures, the final radiative peak is produced by changes in the jet
environment, particle acceleration, magnetic field structure, Doppler
factor, or external photon field \citep{Marscher1985, Spada2001,
Marscher2014, Sironi2015}.  These changes can leave statistical traces in
the light curve before the peak itself.  Long-term studies have shown that
blazar variability is structured, often non-Gaussian or log-normal, and
strongly dependent on the source state \citep{Giebels2009,
Shah2018_lognormal}.  Thus the 365-d rolling window used here is not meant
to predict a flare from a single data point.  It asks whether the source
has entered a broader elevated-risk state before the flare becomes obvious.

This forecasting view is also valuable observationally.  Existing
survey-based alerts are extremely useful, but they are mostly reactive:
the wide-field instrument first detects that a source has already become
active, and pointed optical, X-ray, radio, or TeV facilities are then
asked to follow it.  A predictive WATCH state would move part of this
process earlier.  Even a modest warning can help observers increase
cadence, request Target-of-Opportunity observations, prepare
multi-wavelength coverage, and avoid missing the rise of a flare.  This is
particularly important because major blazar outbursts often require
simultaneous information from several bands to locate the emitting region
and distinguish between competing jet scenarios \citep{Marscher2010,
Agudo2011, Hayashida2015, Raiteri2017}.

This places the present work between two familiar approaches to blazar
variability. On the observational side, flare monitoring often relies on
simple flux-threshold alerts in the $\gamma$-ray light curve, or on
tracking individual indicators such as optical polarization degree, and EVPA rotation
\citep{Blinov2015, Kiehlmann2016}. These indicators are physically
informative, but they are usually followed one at a time, or in a small
number of combinations. For that reason, they do not easily capture the
broader, non-linear variability patterns that may develop across several
time-scales before a flare. On the machine-learning side, most work on $\gamma$-ray blazars has
focused on retrospective classification rather than future-time
forecasting. This includes the classification of blazar candidates of
uncertain type in \fermi\ catalogues \citep{Chiaro2016, Kovacevic2020,
Sahakyan2023, Tolamatti2023}, as well as $\gamma$-ray blazar
classification using self-supervised learning \citep{Bhatta2024}. The
problem studied here is different. For a source that is already being
monitored, we ask whether the recent light curve indicates elevated flare
risk within a specified future horizon. This forecasting setting also
requires special care, because future information can leak into the
labels, preprocessing, calibration, or validation. Such temporal leakage
is a well-known problem in time-series forecasting
\citep{Hyndman2018}, and it is the reason for the strict causal design
adopted in this work.

Within this framework, the main result of the present backtest is that a
strictly causal pipeline does recover non-trivial predictive information
from the \fermi\ light curve. Among the tested models, polynomial logistic
regression gives the strongest held-out WATCH performance, with
$\AUC=0.884$, AP $=0.392$, and a block-permutation probability
$p_{\rm perm}=0.009$. This shows that the WATCH ranking skill is unlikely
to arise by chance, even after accounting for the strong overlap between
neighboring rolling windows. The same model also gives the
highest held-out TRIGGER AUC and AP of the three models, with TRIGGER AUC $=0.751$ and TRIGGER
AP $=0.121$. However, because the TRIGGER task has only a small number of
positive held-out windows and no separate permutation significance test is
applied, this result should be interpreted more cautiously. 
WATCH's longer 90-d window admits 21 held-out positive windows, while
TRIGGER's shorter, onset-only 45-d window admits only 9; this small
positive count is a direct result of the horizon choice. A
shorter horizon is a stricter condition by construction, so it always
yields fewer positive examples to both train and test on, which is why
the TRIGGER metrics
above are noisier and harder to validate than the WATCH ones. This
also explains why the TRIGGER AUC is high while its AP is low: AUC
averages over all 163 negative windows, whereas AP is dominated by the
top-ranked windows that an alert actually uses, and the TRIGGER-positive
windows are not concentrated there. The present data are therefore not
sufficient to establish the effectiveness of the onset-based TRIGGER
alerts. This limitation is specific to the TRIGGER task; the main
conclusions of this work are based on the WATCH results. Overall, the
held-out results suggest that the recent \fermi\ light curve does contain
useful information about future flare risk.

The success of PLR is also physically suggestive.  A purely linear model
uses each variability feature separately, whereas PLR can also respond to
pairwise combinations of features.  Consistent with this, when PLR's
feature-selection step is left free to choose the 50 most informative
terms out of 945 candidates, 46 turn out to be interaction or squared
terms, and only 4 are plain linear terms, showing that most of the useful
pre-flare signal is carried by combinations of features rather than by
any single feature alone.  We do not identify which specific combinations
are most informative here; disentangling this is left for future
feature-importance work.  This general picture, that flare risk depends
on how features change together rather than on one extreme quantity
alone, is broadly consistent with multi-wavelength studies in which
$\gamma$-ray activity is connected to optical polarisation rotations,
radio-core activity, spectral changes, and complex, sometimes multi-zone,
jet behaviour \citep{Marscher2010, Agudo2011, Hayashida2015,
Raiteri2017}.

The WATCH/TRIGGER formulation is important for interpreting the
alert behavour. WATCH asks whether any flare interval will overlap the
next 90 d, whereas TRIGGER asks whether a flare onset will occur within the
next 45 d while excluding windows already inside a flare. For PLR, the
causal TRAIN optimization selects $\tau_{\rm W}=0.06$ and
$\tau_{\rm T}=0.77$. The WATCH threshold is learned from source-local
causal TRAIN WATCH scores pooled across the training blazars, so that
each source contributes its own pre-cutoff WATCH history before a common
threshold is chosen. This lower WATCH threshold is consistent with the
purpose of WATCH as a broad elevated-risk state, whereas the higher
TRIGGER threshold still reflects the more conservative precision-weighted
trigger objective and the smaller number of positive trigger windows
available for training. The correct interpretation is therefore that the
present data support a broad WATCH state more robustly than a sparse
high-confidence trigger.

The alert behavior should be interpreted in the same way. In the held-out
interval, the PLR WATCH state appears before both flare onsets, whereas no
valid pre-onset TRIGGER crossing is obtained. This does not make the model
useless; rather, it shows that the present system is better suited to
identifying elevated-risk intervals than to issuing rare high-confidence
triggers. The last accepted WATCH windows occur 4.5 and 2.5 d before the
two held-out flare onsets, while the underlying WATCH-active episode
itself begins 158.5 d before the first onset and continues, without
interruption, through the second onset 212.5 d later -- both flares
occur within a single extended elevated-risk episode rather than two
separate ones. These two timescales describe
different kinds of warning: the shorter one is the final warning close to
flare onset, whereas the longer one marks how early the source
already appears more flare-prone than usual.

The WATCH state is not free of false alarms, however. Grouped
into episodes, the model raises WATCH four separate times over the
held-out period, but only one is followed by a real flare (defined by
the flux and 9-d duration conditions of
Section~\ref{sec:bb}, Equation~\ref{eq:flare_criterion}); the other
three are false positives. The source spends about 41 per cent of the
held-out period in the WATCH state. This sounds high on its own, but
two of the three false-positive episodes show real, elevated flux that
simply missed our strict flare definition; only one shows no real flux
increase at all. So WATCH is better understood as a first-pass
screening signal than a precise, standalone flare predictor. These
percentages come from only two real flares, so a larger held-out
sample would sharpen them; a softer, graded flare definition and
training on more sources should also help reduce false positives.

The WATCH probability scale also deserves a careful interpretation. For
the adopted PLR model, the held-out BSS is positive
($\BSS=0.197$): a modest value, meaning the WATCH probabilities reduce
the mean-squared probability error by about 19.7 per cent relative to
a simple constant baseline forecast. For forecasting, this means the probabilities are useful for ranking windows by risk and deciding when to raise
an alert. This is how we use them throughout
this work, but the exact numbers should not be
taken as precise. Because the BSS is computed entirely on the held-out target-source period, its positive value provides evidence of out-of-sample probability skill and argues against severe overfitting: an
overfit model would typically show near-zero or negative skill on
genuinely unseen data. The modest size instead reflects a harder
standard than ranking: PLR's held-out AUC is high ($=0.884$), showing
it separates WATCH-positive from WATCH-negative windows well, whereas BSS additionally requires the absolute probability values to be accurate. This is a more demanding criterion, particularly given the small held-out sample (21 positive windows). The reliability curve (Figure~\ref{fig:plr_diagnostics}c)
tracks the diagonal reasonably closely and shows no obvious systematic bias, suggesting that the modest BSS may partly reflect the limited held-out sample rather than an evident calibration failure. Taken
together with the confusion matrix at the operational threshold
(precision $=0.300$, recall $=1.000$; Section~\ref{sec:results}), this
indicates that WATCH is best treated as a ranking and state indicator --
flagging which windows carry elevated risk and when to raise an alert --
rather than as a precisely calibrated forecast suitable for high-stakes
decisions.
By contrast, the random forest has a negative held-out BSS
($\BSS=-0.067$), consistent with the overfitting discussed in
Section~\ref{sec:results}, showing again that discrimination skill and
probability calibration do not necessarily select the same model. In the
present work, the main scientific aim is advance warning before flare
onset, so held-out ranking skill, threshold behavior, and lead-time
performance provide the most relevant basis for adopting PLR as the
reference model.

These results also depend on the strict temporal design of the analysis,
not only on the choice of classifier. The separation between TRAIN-BB and
FULL-BB, the use of TRAIN-only preprocessing, the causal probability
calibration, and the split at $t_{\rm cut}-H$ are all intended to ensure
that the reported forecasting skill is genuinely out of sample.
Separately, the 9-d minimum block duration used to define flare
labels (Section~\ref{sec:bb}) means this backtest targets sustained
flares of duration $\gtrsim$9 d, not all flare-like activity.
The main limitation of the present backtest is that the held-out interval
contains only two flare onsets. The lead-time values should therefore be
viewed as provisional, rather than as precise population-level numbers.
Choices such as the 365-d rolling window, the 7-d step, and the
minimum of 30 points per window are heuristic. Our held-out test data
has only two flare events, too few to properly test whether changing
these numbers improves or worsens the results. Testing this is left to
future work, once a larger multi-source backtest gives more flare
events to test with. The next step is to apply the same causal protocol to a larger sample of
bright LAT blazars, test whether the WATCH thresholds remain useful across
sources, and improve probability calibration with a larger number of
positive calibration windows. A particularly useful extension would be to
include contemporaneous optical, radio, X-ray, and polarization data,
because these observables trace different parts of the jet and may help
turn a broad WATCH state into a more selective
Target-of-Opportunity trigger. In this sense, the present result is best
viewed as a first demonstration that long-term \fermi\ light curves
contain predictive information, rather than as the final form of a
real-time blazar flare alert system.
\section{Summary and Conclusions}
\label{sec:conclusions}

We have presented a strictly causal machine-learning framework for
forecasting $\gamma$-ray blazar flares from long-term \fermi\ light
curves. The pipeline combines Bayesian-Blocks flare identification,
rolling-window variability features, separate WATCH and TRIGGER targets,
TRAIN-only calibration and threshold selection, and a fully held-out
backtest on 4FGL\,J1048.4$+$7143. The main conclusions are as follows.
\begin{enumerate}
  \item A strictly causal backtest still retains measurable predictive
  skill. Among the tested models, polynomial logistic regression (PLR)
  gives the strongest held-out WATCH performance, with $\AUC=0.884$,
  AP $=0.392$, and a block-permutation $p$-value of 0.009. This
  conclusion rests on the WATCH task; the TRIGGER result is weaker and
  remains tentative (item 4). These results suggest that the
  long-term \fermi\ light curve of 4FGL\,J1048.4$+$7143 contains useful
  pre-flare information for this source during the period considered.
  Importantly, this holds even though the model is only ever given
  information that would have been available in real time, with no
  access to future data. Extending this test to a larger sample of
  blazars is left for future work.
  \item The best held-out model is not simply the most flexible one.
  Although the random forest reaches the largest pooled out-of-fold WATCH
  AUC on the TRAIN data, PLR gives the strongest held-out WATCH
  discrimination and also a positive held-out WATCH BSS.
  This suggests that pairwise interactions between variability features
  are useful, but that greater model complexity does not automatically
  lead to better post-cutoff generalization.
  \item Operationally, the present system behaves mainly as a WATCH
  predictor rather than as a hard trigger. For the adopted PLR model, the
  held-out WATCH threshold $\tau_{\rm W}=0.06$ recovers 21 of the
  21 WATCH-positive windows, and the WATCH state appears before both
  held-out flare onsets. The last accepted WATCH windows occur 4.5 and
  2.5 d before flare onset, while the underlying WATCH-active episode
  itself begins 158.5 d before the first onset and continues, without
  interruption, through the second onset 212.5 d later.
  \item The onset-based TRIGGER task remains more challenging. The causal
  TRAIN optimisation selects a PLR trigger threshold of
  $\tau_{\rm T}=0.77$, and no valid pre-onset TRIGGER crossing is obtained
  in the held-out interval. The held-out TRIGGER AP ($=0.121$) is only about
  twice the random baseline, and with only nine positive windows its
  performance cannot yet be validated. The present data therefore support a broad
  elevated-risk WATCH state more robustly than a sparse high-confidence
  TRIGGER state.
  \item The WATCH probability scale is modestly reliable in
  absolute terms. For the adopted PLR model, the held-out WATCH BSS is
  positive ($\BSS=0.197$), so the outputs are better than a
  simple constant baseline forecast and can be interpreted as meaningful
  flare-risk probabilities, even though their most useful practical role
  remains the identification of elevated-risk states.
\end{enumerate}

The present results are encouraging but not yet definitive, because the
held-out interval contains only two flare onsets. The next steps are
therefore clear: apply the same causal protocol to a larger set of bright
LAT blazars, test whether WATCH thresholds and probability behaviour
remain stable across sources, and improve the probability calibration with
more positive calibration examples. Any such extension should also
document the causal methodology explicitly -- fitting the feature
standardization and feature selection using only TRAIN-period data,
tuning hyperparameters in a time-aware way, and recording the exact
thresholds and gap lengths used in validation. Hyperparameter choices
should also be checked with a sensitivity test: report the range of
values tried before the final value is selected -- so that the pipeline
stays fully reproducible as it is applied to more sources. A
particularly valuable extension will
be the inclusion of contemporaneous optical, radio, X-ray, and
polarization information, which may help convert a broad WATCH state into
a more selective Target-of-Opportunity trigger.

Even in its present form, the framework provides a reproducible and
leakage-resistant way to translate blazar $\gamma$-ray variability into
advance-warning states. If this behavior is confirmed on a larger sample,
causal WATCH-style alerts from \fermi\ light curves could become useful
for scheduling intensified monitoring and coordinated multi-wavelength
follow-up before the brightest phase of a flare.

\section*{Acknowledgements}
The authors thanks the \fermi\ Collaboration for making the 
Light Curve Repository publicly available.
This research made use of \textsc{astropy}, a community-developed 
core Python package for Astronomy \citep{Astropy2022}; 
\textsc{scikit-learn} \citep{Pedregosa2011}; 
\textsc{NumPy} \citep{Harris2020}; 
\textsc{SciPy} \citep{Virtanen2020}; 
\textsc{Matplotlib} \citep{Hunter2007}.

\section*{Data and Software Availability}

To facilitate reproducibility of the machine learning analysis, we provide the analysis code, the \textit{Fermi}-LAT light-curve inputs, and the training dataset through publicly accessible repositories. The complete analysis code is available in the GitHub repository \url{https://github.com/Sikandar1992/blazar-flare-forecasting} and is archived as version 1.1 on Zenodo at \url{https://doi.org/10.5281/zenodo.23978959}. The archived software release contains the code used for Bayesian-blocks flare identification, causal rolling-window feature extraction, WATCH/TRIGGER label construction, model training, calibration, threshold selection, and held-out evaluation.

The 14 \textit{Fermi}-LAT light curves used as inputs to the analysis are provided with the GitHub repository. The complete training feature dataset used for model development is archived on Zenodo at \url{https://doi.org/10.5281/zenodo.23980645} (Version 4). The deposited file, \texttt{training\_dataset\_all\_sources.csv}, contains 9004 training windows from the 14 sources, with 47 columns comprising source and temporal metadata, 42 causally constructed features, and the \texttt{Y\_watch} and \texttt{Y\_trigger} target labels. Only TRAIN windows are included in this archival dataset; TEST windows are intentionally excluded because their retrospective evaluation labels are generated separately using the full flare catalog. The deposited training dataset therefore allows readers to examine the feature representation and labels used for model development without exposing the held-out test set. The repository README provides descriptions of the input files, processing procedure, software requirements, and instructions for reproducing the analysis.

\bibliography{ref}

@ARTICLE{Abdo2010_var,
       author = {{Abdo}, A.~A. and {Ackermann}, M. and {Ajello}, M. and {Antolini}, E. and {Baldini}, L. and {Ballet}, J. and {Barbiellini}, G. and {Bastieri}, D. and {Bechtol}, K. and {Bellazzini}, R. and {Berenji}, B. and {Blandford}, R.~D. and {Bloom}, E.~D. and {Bonamente}, E. and {Borgland}, A.~W. and {Bouvier}, A. and {Bregeon}, J. and {Brez}, A. and {Brigida}, M. and {Bruel}, P. and {Buehler}, R. and {Burnett}, T.~H. and {Buson}, S. and {Caliandro}, G.~A. and {Cameron}, R.~A. and {Caraveo}, P.~A. and {Carrigan}, S. and {Casandjian}, J.~M. and {Cavazzuti}, E. and {Cecchi}, C. and {{\c{C}}elik}, {\"O}. and {Chekhtman}, A. and {Cheung}, C.~C. and {Chiang}, J. and {Ciprini}, S. and {Claus}, R. and {Cohen-Tanugi}, J. and {Cominsky}, L.~R. and {Conrad}, J. and {Costamante}, L. and {Cutini}, S. and {Dermer}, C.~D. and {de Angelis}, A. and {de Palma}, F. and {Silva}, E. do Couto e. and {Drell}, P.~S. and {Dubois}, R. and {Dumora}, D. and {Farnier}, C. and {Favuzzi}, C. and {Fegan}, S.~J. and {Focke}, W.~B. and {Fortin}, P. and {Frailis}, M. and {Fukazawa}, Y. and {Funk}, S. and {Fusco}, P. and {Gargano}, F. and {Gasparrini}, D. and {Gehrels}, N. and {Germani}, S. and {Giebels}, B. and {Giglietto}, N. and {Giommi}, P. and {Giordano}, F. and {Glanzman}, T. and {Godfrey}, G. and {Grenier}, I.~A. and {Grondin}, M.-H. and {Grove}, J.~E. and {Guiriec}, S. and {Hadasch}, D. and {Hayashida}, M. and {Hays}, E. and {Healey}, S.~E. and {Horan}, D. and {Hughes}, R.~E. and {Itoh}, R. and {J{\'o}hannesson}, G. and {Johnson}, A.~S. and {Johnson}, W.~N. and {Kamae}, T. and {Katagiri}, H. and {Kataoka}, J. and {Kawai}, N. and {Kn{\"o}dlseder}, J. and {Kuss}, M. and {Lande}, J. and {Larsson}, S. and {Latronico}, L. and {Lemoine-Goumard}, M. and {Longo}, F. and {Loparco}, F. and {Lott}, B. and {Lovellette}, M.~N. and {Lubrano}, P. and {Madejski}, G.~M. and {Makeev}, A. and {Massaro}, E. and {Mazziotta}, M.~N. and {McEnery}, J.~E. and {Michelson}, P.~F. and {Mitthumsiri}, W. and {Mizuno}, T. and {Moiseev}, A.~A. and {Monte}, C. and {Monzani}, M.~E. and {Morselli}, A. and {Moskalenko}, I.~V. and {Mueller}, M. and {Murgia}, S. and {Nolan}, P.~L. and {Norris}, J.~P. and {Nuss}, E. and {Ohno}, M. and {Ohsugi}, T. and {Omodei}, N. and {Orlando}, E. and {Ormes}, J.~F. and {Ozaki}, M. and {Panetta}, J.~H. and {Parent}, D. and {Pelassa}, V. and {Pepe}, M. and {Pesce-Rollins}, M. and {Piron}, F. and {Porter}, T.~A. and {Rain{\`o}}, S. and {Rando}, R. and {Razzano}, M. and {Reimer}, A. and {Reimer}, O. and {Ritz}, S. and {Rodriguez}, A.~Y. and {Romani}, R.~W. and {Roth}, M. and {Ryde}, F. and {Sadrozinski}, H.~F.-W. and {Sander}, A. and {Scargle}, J.~D. and {Sgr{\`o}}, C. and {Shaw}, M.~S. and {Smith}, P.~D. and {Spandre}, G. and {Spinelli}, P. and {Starck}, J.-L. and {Strickman}, M.~S. and {Suson}, D.~J. and {Takahashi}, H. and {Takahashi}, T. and {Tanaka}, T. and {Thayer}, J.~B. and {Thayer}, J.~G. and {Thompson}, D.~J. and {Tibaldo}, L. and {Torres}, D.~F. and {Tosti}, G. and {Tramacere}, A. and {Uchiyama}, Y. and {Usher}, T.~L. and {Vasileiou}, V. and {Vilchez}, N. and {Vitale}, V. and {Waite}, A.~P. and {Wallace}, E. and {Wang}, P. and {Winer}, B.~L. and {Wood}, K.~S. and {Yang}, Z. and {Ylinen}, T. and {Ziegler}, M.},
        title = "{Gamma-ray Light Curves and Variability of Bright Fermi-detected Blazars}",
      journal = {\apj},
         year = 2010,
        month = oct,
       volume = {722},
       number = {1},
        pages = {520-542},
          doi = {10.1088/0004-637X/722/1/520},
archivePrefix = {arXiv},
       eprint = {1004.0348},
 primaryClass = {astro-ph.HE},
       adsurl = {https://ui.adsabs.harvard.edu/abs/2010ApJ...722..520A}
}

@ARTICLE{Abdollahi2023,
       author = {{Abdollahi}, S. and {Ajello}, M. and {Baldini}, L. and {Ballet}, J. and {Bastieri}, D. and {Becerra Gonzalez}, J. and {Bellazzini}, R. and {Berretta}, A. and {Bissaldi}, E. and {Bonino}, R. and {Brill}, A. and {Bruel}, P. and {Burns}, E. and {Buson}, S. and {Cameron}, R.~A. and {Caputo}, R. and {Caraveo}, P.~A. and {Cibrario}, N. and {Ciprini}, S. and {Cristarella Orestano}, P. and {Crnogorcevic}, M. and {Cutini}, S. and {D'Ammando}, F. and {De Gaetano}, S. and {Digel}, S.~W. and {Di Lalla}, N. and {Di Venere}, L. and {Dom{\'\i}nguez}, A. and {Ramazani}, V. Fallah and {Fegan}, S.~J. and {Ferrara}, E.~C. and {Fiori}, A. and {Fleischhack}, H. and {Franckowiak}, A. and {Fukazawa}, Y. and {Fusco}, P. and {Gammaldi}, V. and {Gargano}, F. and {Garrappa}, S. and {Gasbarra}, C. and {Gasparrini}, D. and {Giglietto}, N. and {Giordano}, F. and {Giroletti}, M. and {Green}, D. and {Grenier}, I.~A. and {Guiriec}, S. and {Gustafsson}, M. and {Hays}, E. and {Horan}, D. and {Hou}, X. and {J{\'o}hannesson}, G. and {Kerr}, M. and {Kocevski}, D. and {Kuss}, M. and {Latronico}, L. and {Li}, J. and {Liodakis}, I. and {Longo}, F. and {Loparco}, F. and {Lorusso}, L. and {Lott}, B. and {Lovellette}, M.~N. and {Lubrano}, P. and {Maldera}, S. and {Manfreda}, A. and {Mart{\'\i}-Devesa}, G. and {Mazziotta}, M.~N. and {Mereu}, I. and {Meyer}, M. and {Michelson}, P.~F. and {Mizuno}, T. and {Monzani}, M.~E. and {Morselli}, A. and {Moskalenko}, I.~V. and {Negro}, M. and {Omodei}, N. and {Orlando}, E. and {Ormes}, J.~F. and {Paneque}, D. and {Panzarini}, G. and {Perkins}, J.~S. and {Persic}, M. and {Pesce-Rollins}, M. and {Pillera}, R. and {Porter}, T.~A. and {Principe}, G. and {Racusin}, J.~L. and {Rain{\`o}}, S. and {Rando}, R. and {Rani}, B. and {Razzano}, M. and {Razzaque}, S. and {Reimer}, A. and {Reimer}, O. and {S{\'a}nchez-Conde}, M. and {Parkinson}, P.~M. Saz and {Scargle}, Jeff and {Scotton}, L. and {Serini}, D. and {Sgr{\`o}}, C. and {Siskind}, E.~J. and {Spandre}, G. and {Spinelli}, P. and {Suson}, D.~J. and {Tajima}, H. and {Thompson}, D.~J. and {Torres}, D.~F. and {Valverde}, J. and {Venters}, T. and {Wadiasingh}, Z. and {Wagner}, S. and {Wood}, K.},
        title = "{The Fermi-LAT Lightcurve Repository}",
      journal = {\apjs},
         year = 2023,
        month = apr,
       volume = {265},
       number = {2},
          eid = {31},
        pages = {31},
          doi = {10.3847/1538-4365/acbb6a},
archivePrefix = {arXiv},
       eprint = {2301.01607},
 primaryClass = {astro-ph.HE},
       adsurl = {https://ui.adsabs.harvard.edu/abs/2023ApJS..265...31A}
}

@ARTICLE{Ackermann2011_2LAC,
       author = {{Ackermann}, M. and {Ajello}, M. and {Allafort}, A. and {Antolini}, E. and {Atwood}, W.~B. and {Axelsson}, M. and {Baldini}, L. and {Ballet}, J. and {Barbiellini}, G. and {Bastieri}, D. and {Bechtol}, K. and {Bellazzini}, R. and {Berenji}, B. and {Blandford}, R.~D. and {Bloom}, E.~D. and {Bonamente}, E. and {Borgland}, A.~W. and {Bottacini}, E. and {Bouvier}, A. and {Bregeon}, J. and {Brigida}, M. and {Bruel}, P. and {Buehler}, R. and {Burnett}, T.~H. and {Buson}, S. and {Caliandro}, G.~A. and {Cameron}, R.~A. and {Caraveo}, P.~A. and {Casandjian}, J.~M. and {Cavazzuti}, E. and {Cecchi}, C. and {Charles}, E. and {Cheung}, C.~C. and {Chiang}, J. and {Ciprini}, S. and {Claus}, R. and {Cohen-Tanugi}, J. and {Conrad}, J. and {Costamante}, L. and {Cutini}, S. and {de Angelis}, A. and {de Palma}, F. and {Dermer}, C.~D. and {Digel}, S.~W. and {Silva}, E. do Couto e. and {Drell}, P.~S. and {Dubois}, R. and {Escande}, L. and {Favuzzi}, C. and {Fegan}, S.~J. and {Ferrara}, E.~C. and {Finke}, J. and {Focke}, W.~B. and {Fortin}, P. and {Frailis}, M. and {Fukazawa}, Y. and {Funk}, S. and {Fusco}, P. and {Gargano}, F. and {Gasparrini}, D. and {Gehrels}, N. and {Germani}, S. and {Giebels}, B. and {Giglietto}, N. and {Giommi}, P. and {Giordano}, F. and {Giroletti}, M. and {Glanzman}, T. and {Godfrey}, G. and {Grenier}, I.~A. and {Grove}, J.~E. and {Guiriec}, S. and {Gustafsson}, M. and {Hadasch}, D. and {Hayashida}, M. and {Hays}, E. and {Healey}, S.~E. and {Horan}, D. and {Hou}, X. and {Hughes}, R.~E. and {Iafrate}, G. and {J{\'o}hannesson}, G. and {Johnson}, A.~S. and {Johnson}, W.~N. and {Kamae}, T. and {Katagiri}, H. and {Kataoka}, J. and {Kn{\"o}dlseder}, J. and {Kuss}, M. and {Lande}, J. and {Larsson}, S. and {Latronico}, L. and {Longo}, F. and {Loparco}, F. and {Lott}, B. and {Lovellette}, M.~N. and {Lubrano}, P. and {Madejski}, G.~M. and {Mazziotta}, M.~N. and {McConville}, W. and {McEnery}, J.~E. and {Michelson}, P.~F. and {Mitthumsiri}, W. and {Mizuno}, T. and {Moiseev}, A.~A. and {Monte}, C. and {Monzani}, M.~E. and {Moretti}, E. and {Morselli}, A. and {Moskalenko}, I.~V. and {Murgia}, S. and {Nakamori}, T. and {Naumann-Godo}, M. and {Nolan}, P.~L. and {Norris}, J.~P. and {Nuss}, E. and {Ohno}, M. and {Ohsugi}, T. and {Okumura}, A. and {Omodei}, N. and {Orienti}, M. and {Orlando}, E. and {Ormes}, J.~F. and {Ozaki}, M. and {Paneque}, D. and {Parent}, D. and {Pesce-Rollins}, M. and {Pierbattista}, M. and {Piranomonte}, S. and {Piron}, F. and {Pivato}, G. and {Porter}, T.~A. and {Rain{\`o}}, S. and {Rando}, R. and {Razzano}, M. and {Razzaque}, S. and {Reimer}, A. and {Reimer}, O. and {Ritz}, S. and {Rochester}, L.~S. and {Romani}, R.~W. and {Roth}, M. and {Sanchez}, D.~A. and {Sbarra}, C. and {Scargle}, J.~D. and {Schalk}, T.~L. and {Sgr{\`o}}, C. and {Shaw}, M.~S. and {Siskind}, E.~J. and {Spandre}, G. and {Spinelli}, P. and {Strong}, A.~W. and {Suson}, D.~J. and {Tajima}, H. and {Takahashi}, H. and {Takahashi}, T. and {Tanaka}, T. and {Thayer}, J.~G. and {Thayer}, J.~B. and {Thompson}, D.~J. and {Tibaldo}, L. and {Tinivella}, M. and {Torres}, D.~F. and {Tosti}, G. and {Troja}, E. and {Uchiyama}, Y. and {Vandenbroucke}, J. and {Vasileiou}, V. and {Vianello}, G. and {Vitale}, V. and {Waite}, A.~P. and {Wallace}, E. and {Wang}, P. and {Winer}, B.~L. and {Wood}, D.~L. and {Wood}, K.~S. and {Zimmer}, S.},
        title = "{The Second Catalog of Active Galactic Nuclei Detected by the Fermi Large Area Telescope}",
      journal = {\apj},
         year = 2011,
        month = dec,
       volume = {743},
       number = {2},
          eid = {171},
        pages = {171},
          doi = {10.1088/0004-637X/743/2/171},
archivePrefix = {arXiv},
       eprint = {1108.1420},
 primaryClass = {astro-ph.HE},
       adsurl = {https://ui.adsabs.harvard.edu/abs/2011ApJ...743..171A}
}

@ARTICLE{Ackermann2015,
       author = {{Ackermann}, M. and {Ajello}, M. and {Albert}, A. and {Atwood}, W.~B. and {Baldini}, L. and {Ballet}, J. and {Barbiellini}, G. and {Bastieri}, D. and {Becerra Gonzalez}, J. and {Bellazzini}, R. and {Bissaldi}, E. and {Blandford}, R.~D. and {Bloom}, E.~D. and {Bonino}, R. and {Bottacini}, E. and {Bregeon}, J. and {Bruel}, P. and {Buehler}, R. and {Buson}, S. and {Caliandro}, G.~A. and {Cameron}, R.~A. and {Caputo}, R. and {Caragiulo}, M. and {Caraveo}, P.~A. and {Cavazzuti}, E. and {Cecchi}, C. and {Chekhtman}, A. and {Chiang}, J. and {Chiaro}, G. and {Ciprini}, S. and {Cohen-Tanugi}, J. and {Conrad}, J. and {Cutini}, S. and {D'Ammando}, F. and {de Angelis}, A. and {de Palma}, F. and {Desiante}, R. and {Di Venere}, L. and {Dom{\'\i}nguez}, A. and {Drell}, P.~S. and {Favuzzi}, C. and {Fegan}, S.~J. and {Ferrara}, E.~C. and {Focke}, W.~B. and {Fuhrmann}, L. and {Fukazawa}, Y. and {Fusco}, P. and {Gargano}, F. and {Gasparrini}, D. and {Giglietto}, N. and {Giommi}, P. and {Giordano}, F. and {Giroletti}, M. and {Godfrey}, G. and {Green}, D. and {Grenier}, I.~A. and {Grove}, J.~E. and {Guiriec}, S. and {Harding}, A.~K. and {Hays}, E. and {Hewitt}, J.~W. and {Hill}, A.~B. and {Horan}, D. and {Jogler}, T. and {J{\'o}hannesson}, G. and {Johnson}, A.~S. and {Kamae}, T. and {Kuss}, M. and {Larsson}, S. and {Latronico}, L. and {Li}, J. and {Li}, L. and {Longo}, F. and {Loparco}, F. and {Lott}, B. and {Lovellette}, M.~N. and {Lubrano}, P. and {Magill}, J. and {Maldera}, S. and {Manfreda}, A. and {Max-Moerbeck}, W. and {Mayer}, M. and {Mazziotta}, M.~N. and {McEnery}, J.~E. and {Michelson}, P.~F. and {Mizuno}, T. and {Monzani}, M.~E. and {Morselli}, A. and {Moskalenko}, I.~V. and {Murgia}, S. and {Nuss}, E. and {Ohno}, M. and {Ohsugi}, T. and {Ojha}, R. and {Omodei}, N. and {Orlando}, E. and {Ormes}, J.~F. and {Paneque}, D. and {Pearson}, T.~J. and {Perkins}, J.~S. and {Perri}, M. and {Pesce-Rollins}, M. and {Petrosian}, V. and {Piron}, F. and {Pivato}, G. and {Porter}, T.~A. and {Rain{\`o}}, S. and {Rando}, R. and {Razzano}, M. and {Readhead}, A. and {Reimer}, A. and {Reimer}, O. and {Schulz}, A. and {Sgr{\`o}}, C. and {Siskind}, E.~J. and {Spada}, F. and {Spandre}, G. and {Spinelli}, P. and {Suson}, D.~J. and {Takahashi}, H. and {Thayer}, J.~B. and {Thompson}, D.~J. and {Tibaldo}, L. and {Torres}, D.~F. and {Tosti}, G. and {Troja}, E. and {Uchiyama}, Y. and {Vianello}, G. and {Wood}, K.~S. and {Wood}, M. and {Zimmer}, S. and {Berdyugin}, A. and {Corbet}, R.~H.~D. and {Hovatta}, T. and {Lindfors}, E. and {Nilsson}, K. and {Reinthal}, R. and {Sillanp{\"a}{\"a}}, A. and {Stamerra}, A. and {Takalo}, L.~O. and {Valtonen}, M.~J.},
        title = "{Multiwavelength Evidence for Quasi-periodic Modulation in the Gamma-Ray Blazar PG 1553+113}",
      journal = {\apjl},
         year = 2015,
        month = nov,
       volume = {813},
       number = {2},
          eid = {L41},
        pages = {L41},
          doi = {10.1088/2041-8205/813/2/L41},
archivePrefix = {arXiv},
       eprint = {1509.02063},
 primaryClass = {astro-ph.HE},
       adsurl = {https://ui.adsabs.harvard.edu/abs/2015ApJ...813L..41A}
}

@ARTICLE{Agudo2011,
       author = {{Agudo}, Iv{\'a}n and {Jorstad}, Svetlana G. and {Marscher}, Alan P. and {Larionov}, Valeri M. and {G{\'o}mez}, Jos{\'e} L. and {L{\"a}hteenm{\"a}ki}, Anne and {Gurwell}, Mark and {Smith}, Paul S. and {Wiesemeyer}, Helmut and {Thum}, Clemens and {Heidt}, Jochen and {Blinov}, Dmitriy A. and {D'Arcangelo}, Francesca D. and {Hagen-Thorn}, Vladimir A. and {Morozova}, Daria A. and {Nieppola}, Elina and {Roca-Sogorb}, Mar and {Schmidt}, Gary D. and {Taylor}, Brian and {Tornikoski}, Merja and {Troitsky}, Ivan S.},
        title = "{Location of {\ensuremath{\gamma}}-ray Flare Emission in the Jet of the BL Lacertae Object OJ287 More than 14 pc from the Central Engine}",
      journal = {\apjl},
         year = 2011,
        month = jan,
       volume = {726},
       number = {1},
          eid = {L13},
        pages = {L13},
          doi = {10.1088/2041-8205/726/1/L13},
archivePrefix = {arXiv},
       eprint = {1011.6454},
 primaryClass = {astro-ph.CO},
       adsurl = {https://ui.adsabs.harvard.edu/abs/2011ApJ...726L..13A}
}

@ARTICLE{Aharonian2007,
       author = {{Aharonian}, F. and {Akhperjanian}, A.~G. and {Bazer-Bachi}, A.~R. and {Behera}, B. and {Beilicke}, M. and {Benbow}, W. and {Berge}, D. and {Bernl{\"o}hr}, K. and {Boisson}, C. and {Bolz}, O. and {Borrel}, V. and {Boutelier}, T. and {Braun}, I. and {Brion}, E. and {Brown}, A.~M. and {B{\"u}hler}, R. and {B{\"u}sching}, I. and {Bulik}, T. and {Carrigan}, S. and {Chadwick}, P.~M. and {Clapson}, A.~C. and {Chounet}, L.-M. and {Coignet}, G. and {Cornils}, R. and {Costamante}, L. and {Degrange}, B. and {Dickinson}, H.~J. and {Djannati-Ata{\"\i}}, A. and {Domainko}, W. and {Drury}, L. O'C. and {Dubus}, G. and {Dyks}, J. and {Egberts}, K. and {Emmanoulopoulos}, D. and {Espigat}, P. and {Farnier}, C. and {Feinstein}, F. and {Fiasson}, A. and {F{\"o}rster}, A. and {Fontaine}, G. and {Funk}, Seb. and {Funk}, S. and {F{\"u}{\ss}ling}, M. and {Gallant}, Y.~A. and {Giebels}, B. and {Glicenstein}, J.~F. and {Gl{\"u}ck}, B. and {Goret}, P. and {Hadjichristidis}, C. and {Hauser}, D. and {Hauser}, M. and {Heinzelmann}, G. and {Henri}, G. and {Hermann}, G. and {Hinton}, J.~A. and {Hoffmann}, A. and {Hofmann}, W. and {Holleran}, M. and {Hoppe}, S. and {Horns}, D. and {Jacholkowska}, A. and {de Jager}, O.~C. and {Kendziorra}, E. and {Kerschhaggl}, M. and {Kh{\'e}lifi}, B. and {Komin}, Nu. and {Kosack}, K. and {Lamanna}, G. and {Latham}, I.~J. and {Le Gallou}, R. and {Lemi{\`e}re}, A. and {Lemoine-Goumard}, M. and {Lenain}, J.-P. and {Lohse}, T. and {Martin}, J.~M. and {Martineau-Huynh}, O. and {Marcowith}, A. and {Masterson}, C. and {Maurin}, G. and {McComb}, T.~J.~L. and {Moderski}, R. and {Moulin}, E. and {de Naurois}, M. and {Nedbal}, D. and {Nolan}, S.~J. and {Olive}, J.-P. and {Orford}, K.~J. and {Osborne}, J.~L. and {Ostrowski}, M. and {Panter}, M. and {Pedaletti}, G. and {Pelletier}, G. and {Petrucci}, P.-O. and {Pita}, S. and {P{\"u}hlhofer}, G. and {Punch}, M. and {Ranchon}, S. and {Raubenheimer}, B.~C. and {Raue}, M. and {Rayner}, S.~M. and {Renaud}, M. and {Ripken}, J. and {Rob}, L. and {Rolland}, L. and {Rosier-Lees}, S. and {Rowell}, G. and {Rudak}, B. and {Ruppel}, J. and {Sahakian}, V. and {Santangelo}, A. and {Saug{\'e}}, L. and {Schlenker}, S. and {Schlickeiser}, R. and {Schr{\"o}der}, R. and {Schwanke}, U. and {Schwarzburg}, S. and {Schwemmer}, S. and {Shalchi}, A. and {Sol}, H. and {Spangler}, D. and {Stawarz}, {\L}. and {Steenkamp}, R. and {Stegmann}, C. and {Superina}, G. and {Tam}, P.~H. and {Tavernet}, J.-P. and {Terrier}, R. and {van Eldik}, C. and {Vasileiadis}, G. and {Venter}, C. and {Vialle}, J.~P. and {Vincent}, P. and {Vivier}, M. and {V{\"o}lk}, H.~J. and {Volpe}, F. and {Wagner}, S.~J. and {Ward}, M. and {Zdziarski}, A.~A.},
        title = "{An Exceptional Very High Energy Gamma-Ray Flare of PKS 2155-304}",
      journal = {\apjl},
         year = 2007,
        month = aug,
       volume = {664},
       number = {2},
        pages = {L71-L74},
          doi = {10.1086/520635},
archivePrefix = {arXiv},
       eprint = {0706.0797},
 primaryClass = {astro-ph},
       adsurl = {https://ui.adsabs.harvard.edu/abs/2007ApJ...664L..71A}
}

@ARTICLE{Albert2007,
       author = {{Albert}, J. and {Aliu}, E. and {Anderhub}, H. and {Antoranz}, P. and {Armada}, A. and {Baixeras}, C. and {Barrio}, J.~A. and {Bartko}, H. and {Bastieri}, D. and {Becker}, J.~K. and {Bednarek}, W. and {Berger}, K. and {Bigongiari}, C. and {Biland}, A. and {Bock}, R.~K. and {Bordas}, P. and {Bosch-Ramon}, V. and {Bretz}, T. and {Britvitch}, I. and {Camara}, M. and {Carmona}, E. and {Chilingarian}, A. and {Coarasa}, J.~A. and {Commichau}, S. and {Contreras}, J.~L. and {Cortina}, J. and {Costado}, M.~T. and {Curtef}, V. and {Danielyan}, V. and {Dazzi}, F. and {De Angelis}, A. and {Delgado}, C. and {de los Reyes}, R. and {De Lotto}, B. and {Domingo-Santamar{\'\i}a}, E. and {Dorner}, D. and {Doro}, M. and {Errando}, M. and {Fagiolini}, M. and {Ferenc}, D. and {Fern{\'a}ndez}, E. and {Firpo}, R. and {Flix}, J. and {Fonseca}, M.~V. and {Font}, L. and {Fuchs}, M. and {Galante}, N. and {Garc{\'\i}a-L{\'o}pez}, R.~J. and {Garczarczyk}, M. and {Gaug}, M. and {Giller}, M. and {Goebel}, F. and {Hakobyan}, D. and {Hayashida}, M. and {Hengstebeck}, T. and {Herrero}, A. and {H{\"o}hne}, D. and {Hose}, J. and {Hrupec}, D. and {Hsu}, C.~C. and {Jacon}, P. and {Jogler}, T. and {Kosyra}, R. and {Kranich}, D. and {Kritzer}, R. and {Laille}, A. and {Lindfors}, E. and {Lombardi}, S. and {Longo}, F. and {L{\'o}pez}, J. and {L{\'o}pez}, M. and {Lorenz}, E. and {Majumdar}, P. and {Maneva}, G. and {Mannheim}, K. and {Mansutti}, O. and {Mariotti}, M. and {Mart{\'\i}nez}, M. and {Mazin}, D. and {Merck}, C. and {Meucci}, M. and {Meyer}, M. and {Miranda}, J.~M. and {Mirzoyan}, R. and {Mizobuchi}, S. and {Moralejo}, A. and {Nieto}, D. and {Nilsson}, K. and {Ninkovic}, J. and {O{\~n}a-Wilhelmi}, E. and {Otte}, N. and {Oya}, I. and {Paneque}, D. and {Panniello}, M. and {Paoletti}, R. and {Paredes}, J.~M. and {Pasanen}, M. and {Pascoli}, D. and {Pauss}, F. and {Pegna}, R. and {Persic}, M. and {Peruzzo}, L. and {Piccioli}, A. and {Prandini}, E. and {Puchades}, N. and {Raymers}, A. and {Rhode}, W. and {Rib{\'o}}, M. and {Rico}, J. and {Rissi}, M. and {Robert}, A. and {R{\"u}gamer}, S. and {Saggion}, A. and {Saito}, T. and {S{\'a}nchez}, A. and {Sartori}, P. and {Scalzotto}, V. and {Scapin}, V. and {Schmitt}, R. and {Schweizer}, T. and {Shayduk}, M. and {Shinozaki}, K. and {Shore}, S.~N. and {Sidro}, N. and {Sillanp{\"a}{\"a}}, A. and {Sobczynska}, D. and {Stamerra}, A. and {Stark}, L.~S. and {Takalo}, L. and {Tavecchio}, F. and {Temnikov}, P. and {Tescaro}, D. and {Teshima}, M. and {Torres}, D.~F. and {Turini}, N. and {Vankov}, H. and {Vitale}, V. and {Wagner}, R.~M. and {Wibig}, T. and {Wittek}, W. and {Zandanel}, F. and {Zanin}, R. and {Zapatero}, J.},
        title = "{Variable Very High Energy {\ensuremath{\gamma}}-Ray Emission from Markarian 501}",
      journal = {\apj},
         year = 2007,
        month = nov,
       volume = {669},
       number = {2},
        pages = {862-883},
          doi = {10.1086/521382},
archivePrefix = {arXiv},
       eprint = {astro-ph/0702008},
 primaryClass = {astro-ph},
       adsurl = {https://ui.adsabs.harvard.edu/abs/2007ApJ...669..862A}
}

@ARTICLE{Astropy2022,
       author = {{Astropy Collaboration} and {Price-Whelan}, Adrian M. and {Lim}, Pey Lian and {Earl}, Nicholas and {Starkman}, Nathaniel and {Bradley}, Larry and {Shupe}, David L. and {Patil}, Aarya A. and {Corrales}, Lia and {Brasseur}, C.~E. and {N{\"o}the}, Maximilian and {Donath}, Axel and {Tollerud}, Erik and {Morris}, Brett M. and {Ginsburg}, Adam and {Vaher}, Eero and {Weaver}, Benjamin A. and {Tocknell}, James and {Jamieson}, William and {van Kerkwijk}, Marten H. and {Robitaille}, Thomas P. and {Merry}, Bruce and {Bachetti}, Matteo and {G{\"u}nther}, H. Moritz and {Aldcroft}, Thomas L. and {Alvarado-Montes}, Jaime A. and {Archibald}, Anne M. and {B{\'o}di}, Attila and {Bapat}, Shreyas and {Barentsen}, Geert and {Baz{\'a}n}, Juanjo and {Biswas}, Manish and {Boquien}, M{\'e}d{\'e}ric and {Burke}, D.~J. and {Cara}, Daria and {Cara}, Mihai and {Conroy}, Kyle E. and {Conseil}, Simon and {Craig}, Matthew W. and {Cross}, Robert M. and {Cruz}, Kelle L. and {D'Eugenio}, Francesco and {Dencheva}, Nadia and {Devillepoix}, Hadrien A.~R. and {Dietrich}, J{\"o}rg P. and {Eigenbrot}, Arthur Davis and {Erben}, Thomas and {Ferreira}, Leonardo and {Foreman-Mackey}, Daniel and {Fox}, Ryan and {Freij}, Nabil and {Garg}, Suyog and {Geda}, Robel and {Glattly}, Lauren and {Gondhalekar}, Yash and {Gordon}, Karl D. and {Grant}, David and {Greenfield}, Perry and {Groener}, Austen M. and {Guest}, Steve and {Gurovich}, Sebastian and {Handberg}, Rasmus and {Hart}, Akeem and {Hatfield-Dodds}, Zac and {Homeier}, Derek and {Hosseinzadeh}, Griffin and {Jenness}, Tim and {Jones}, Craig K. and {Joseph}, Prajwel and {Kalmbach}, J. Bryce and {Karamehmetoglu}, Emir and {Ka{\l}uszy{\'n}ski}, Miko{\l}aj and {Kelley}, Michael S.~P. and {Kern}, Nicholas and {Kerzendorf}, Wolfgang E. and {Koch}, Eric W. and {Kulumani}, Shankar and {Lee}, Antony and {Ly}, Chun and {Ma}, Zhiyuan and {MacBride}, Conor and {Maljaars}, Jakob M. and {Muna}, Demitri and {Murphy}, N.~A. and {Norman}, Henrik and {O'Steen}, Richard and {Oman}, Kyle A. and {Pacifici}, Camilla and {Pascual}, Sergio and {Pascual-Granado}, J. and {Patil}, Rohit R. and {Perren}, Gabriel I. and {Pickering}, Timothy E. and {Rastogi}, Tanuj and {Roulston}, Benjamin R. and {Ryan}, Daniel F. and {Rykoff}, Eli S. and {Sabater}, Jose and {Sakurikar}, Parikshit and {Salgado}, Jes{\'u}s and {Sanghi}, Aniket and {Saunders}, Nicholas and {Savchenko}, Volodymyr and {Schwardt}, Ludwig and {Seifert-Eckert}, Michael and {Shih}, Albert Y. and {Jain}, Anany Shrey and {Shukla}, Gyanendra and {Sick}, Jonathan and {Simpson}, Chris and {Singanamalla}, Sudheesh and {Singer}, Leo P. and {Singhal}, Jaladh and {Sinha}, Manodeep and {Sip{\H{o}}cz}, Brigitta M. and {Spitler}, Lee R. and {Stansby}, David and {Streicher}, Ole and {{\v{S}}umak}, Jani and {Swinbank}, John D. and {Taranu}, Dan S. and {Tewary}, Nikita and {Tremblay}, Grant R. and {de Val-Borro}, Miguel and {Van Kooten}, Samuel J. and {Vasovi{\'c}}, Zlatan and {Verma}, Shresth and {de Miranda Cardoso}, Jos{\'e} Vin{\'\i}cius and {Williams}, Peter K.~G. and {Wilson}, Tom J. and {Winkel}, Benjamin and {Wood-Vasey}, W.~M. and {Xue}, Rui and {Yoachim}, Peter and {Zhang}, Chen and {Zonca}, Andrea and {Astropy Project Contributors}},
        title = "{The Astropy Project: Sustaining and Growing a Community-oriented Open-source Project and the Latest Major Release (v5.0) of the Core Package}",
      journal = {\apj},
         year = 2022,
        month = aug,
       volume = {935},
       number = {2},
          eid = {167},
        pages = {167},
          doi = {10.3847/1538-4357/ac7c74},
archivePrefix = {arXiv},
       eprint = {2206.14220},
 primaryClass = {astro-ph.IM},
       adsurl = {https://ui.adsabs.harvard.edu/abs/2022ApJ...935..167A}
}

@ARTICLE{Atwood2009,
       author = {{Atwood}, W.~B. and {Abdo}, A.~A. and {Ackermann}, M. and {Althouse}, W. and {Anderson}, B. and {Axelsson}, M. and {Baldini}, L. and {Ballet}, J. and {Band}, D.~L. and {Barbiellini}, G. and {Bartelt}, J. and {Bastieri}, D. and {Baughman}, B.~M. and {Bechtol}, K. and {B{\'e}d{\'e}r{\`e}de}, D. and {Bellardi}, F. and {Bellazzini}, R. and {Berenji}, B. and {Bignami}, G.~F. and {Bisello}, D. and {Bissaldi}, E. and {Blandford}, R.~D. and {Bloom}, E.~D. and {Bogart}, J.~R. and {Bonamente}, E. and {Bonnell}, J. and {Borgland}, A.~W. and {Bouvier}, A. and {Bregeon}, J. and {Brez}, A. and {Brigida}, M. and {Bruel}, P. and {Burnett}, T.~H. and {Busetto}, G. and {Caliandro}, G.~A. and {Cameron}, R.~A. and {Caraveo}, P.~A. and {Carius}, S. and {Carlson}, P. and {Casandjian}, J.~M. and {Cavazzuti}, E. and {Ceccanti}, M. and {Cecchi}, C. and {Charles}, E. and {Chekhtman}, A. and {Cheung}, C.~C. and {Chiang}, J. and {Chipaux}, R. and {Cillis}, A.~N. and {Ciprini}, S. and {Claus}, R. and {Cohen-Tanugi}, J. and {Condamoor}, S. and {Conrad}, J. and {Corbet}, R. and {Corucci}, L. and {Costamante}, L. and {Cutini}, S. and {Davis}, D.~S. and {Decotigny}, D. and {DeKlotz}, M. and {Dermer}, C.~D. and {de Angelis}, A. and {Digel}, S.~W. and {do Couto e Silva}, E. and {Drell}, P.~S. and {Dubois}, R. and {Dumora}, D. and {Edmonds}, Y. and {Fabiani}, D. and {Farnier}, C. and {Favuzzi}, C. and {Flath}, D.~L. and {Fleury}, P. and {Focke}, W.~B. and {Funk}, S. and {Fusco}, P. and {Gargano}, F. and {Gasparrini}, D. and {Gehrels}, N. and {Gentit}, F.-X. and {Germani}, S. and {Giebels}, B. and {Giglietto}, N. and {Giommi}, P. and {Giordano}, F. and {Glanzman}, T. and {Godfrey}, G. and {Grenier}, I.~A. and {Grondin}, M.-H. and {Grove}, J.~E. and {Guillemot}, L. and {Guiriec}, S. and {Haller}, G. and {Harding}, A.~K. and {Hart}, P.~A. and {Hays}, E. and {Healey}, S.~E. and {Hirayama}, M. and {Hjalmarsdotter}, L. and {Horn}, R. and {Hughes}, R.~E. and {J{\'o}hannesson}, G. and {Johansson}, G. and {Johnson}, A.~S. and {Johnson}, R.~P. and {Johnson}, T.~J. and {Johnson}, W.~N. and {Kamae}, T. and {Katagiri}, H. and {Kataoka}, J. and {Kavelaars}, A. and {Kawai}, N. and {Kelly}, H. and {Kerr}, M. and {Klamra}, W. and {Kn{\"o}dlseder}, J. and {Kocian}, M.~L. and {Komin}, N. and {Kuehn}, F. and {Kuss}, M. and {Landriu}, D. and {Latronico}, L. and {Lee}, B. and {Lee}, S.-H. and {Lemoine-Goumard}, M. and {Lionetto}, A.~M. and {Longo}, F. and {Loparco}, F. and {Lott}, B. and {Lovellette}, M.~N. and {Lubrano}, P. and {Madejski}, G.~M. and {Makeev}, A. and {Marangelli}, B. and {Massai}, M.~M. and {Mazziotta}, M.~N. and {McEnery}, J.~E. and {Menon}, N. and {Meurer}, C. and {Michelson}, P.~F. and {Minuti}, M. and {Mirizzi}, N. and {Mitthumsiri}, W. and {Mizuno}, T. and {Moiseev}, A.~A. and {Monte}, C. and {Monzani}, M.~E. and {Moretti}, E. and {Morselli}, A. and {Moskalenko}, I.~V. and {Murgia}, S. and {Nakamori}, T. and {Nishino}, S. and {Nolan}, P.~L. and {Norris}, J.~P. and {Nuss}, E. and {Ohno}, M. and {Ohsugi}, T. and {Omodei}, N. and {Orlando}, E. and {Ormes}, J.~F. and {Paccagnella}, A. and {Paneque}, D. and {Panetta}, J.~H. and {Parent}, D. and {Pearce}, M. and {Pepe}, M. and {Perazzo}, A. and {Pesce-Rollins}, M. and {Picozza}, P. and {Pieri}, L. and {Pinchera}, M. and {Piron}, F. and {Porter}, T.~A. and {Poupard}, L. and {Rain{\`o}}, S. and {Rando}, R. and {Rapposelli}, E. and {Razzano}, M. and {Reimer}, A. and {Reimer}, O. and {Reposeur}, T. and {Reyes}, L.~C. and {Ritz}, S. and {Rochester}, L.~S. and {Rodriguez}, A.~Y. and {Romani}, R.~W. and {Roth}, M. and {Russell}, J.~J. and {Ryde}, F. and {Sabatini}, S. and {Sadrozinski}, H.~F.-W. and {Sanchez}, D. and {Sander}, A. and {Sapozhnikov}, L. and {Parkinson}, P.~M. Saz and {Scargle}, J.~D. and {Schalk}, T.~L. and {Scolieri}, G.},
        title = "{The Large Area Telescope on the Fermi Gamma-Ray Space Telescope Mission}",
      journal = {\apj},
         year = 2009,
        month = jun,
       volume = {697},
       number = {2},
        pages = {1071-1102},
          doi = {10.1088/0004-637X/697/2/1071},
archivePrefix = {arXiv},
       eprint = {0902.1089},
 primaryClass = {astro-ph.IM},
       adsurl = {https://ui.adsabs.harvard.edu/abs/2009ApJ...697.1071A}
}

@ARTICLE{Bhatta2024,
       author = {{Bhatta}, Gopal and {Gharat}, Sarvesh and {Borthakur}, Abhimanyu and {Kumar}, Aman},
        title = "{Gamma-ray blazar classification using machine learning with advanced weight initialization and self-supervised learning techniques}",
      journal = {\mnras},
         year = 2024,
        month = feb,
       volume = {528},
       number = {1},
        pages = {976-986},
          doi = {10.1093/mnras/stae028},
archivePrefix = {arXiv},
       eprint = {2310.06095},
 primaryClass = {astro-ph.HE},
       adsurl = {https://ui.adsabs.harvard.edu/abs/2024MNRAS.528..976B}
}

@ARTICLE{Biteau2012,
       author = {{Biteau}, J. and {Giebels}, B.},
        title = "{The minijets-in-a-jet statistical model and the rms-flux correlation}",
      journal = {\aap},
         year = 2012,
        month = dec,
       volume = {548},
          eid = {A123},
        pages = {A123},
          doi = {10.1051/0004-6361/201220056},
archivePrefix = {arXiv},
       eprint = {1210.2045},
 primaryClass = {astro-ph.HE},
       adsurl = {https://ui.adsabs.harvard.edu/abs/2012A&A...548A.123B}
}

@ARTICLE{Blandford1978,
       author = {{Blandford}, R.~D. and {Rees}, M.~J.},
        title = "{Extended and compact extragalactic radio sources: interpretation and theory.}",
      journal = {\physscr},
         year = 1978,
        month = mar,
       volume = {17},
        pages = {265-274},
          doi = {10.1088/0031-8949/17/3/020},
       adsurl = {https://ui.adsabs.harvard.edu/abs/1978PhyS...17..265B}
}

@ARTICLE{Blinov2015,
       author = {{Blinov}, D. and {Pavlidou}, V. and {Papadakis}, I. and {Kiehlmann}, S. and {Panopoulou}, G. and {Liodakis}, I. and {King}, O.~G. and {Angelakis}, E. and {Balokovi{\'c}}, M. and {Das}, H. and {Feiler}, R. and {Fuhrmann}, L. and {Hovatta}, T. and {Khodade}, P. and {Kus}, A. and {Kylafis}, N. and {Mahabal}, A. and {Myserlis}, I. and {Modi}, D. and {Pazderska}, B. and {Pazderski}, E. and {Papamastorakis}, I. and {Pearson}, T.~J. and {Rajarshi}, C. and {Ramaprakash}, A. and {Reig}, P. and {Readhead}, A.~C.~S. and {Tassis}, K. and {Zensus}, J.~A.},
        title = "{RoboPol: first season rotations of optical polarization plane in blazars}",
      journal = {\mnras},
         year = 2015,
        month = oct,
       volume = {453},
       number = {2},
        pages = {1669-1683},
          doi = {10.1093/mnras/stv1723},
archivePrefix = {arXiv},
       eprint = {1505.07467},
 primaryClass = {astro-ph.HE},
       adsurl = {https://ui.adsabs.harvard.edu/abs/2015MNRAS.453.1669B}
}

@ARTICLE{Bloom1996,
       author = {{Bloom}, Steven D. and {Marscher}, Alan P.},
        title = "{An Analysis of the Synchrotron Self-Compton Model for the Multi--Wave Band Spectra of Blazars}",
      journal = {\apj},
         year = 1996,
        month = apr,
       volume = {461},
        pages = {657},
          doi = {10.1086/177092},
       adsurl = {https://ui.adsabs.harvard.edu/abs/1996ApJ...461..657B}
}

@ARTICLE{Bottcher2013,
       author = {{B{\"o}ttcher}, M. and {Reimer}, A. and {Sweeney}, K. and {Prakash}, A.},
        title = "{Leptonic and Hadronic Modeling of Fermi-detected Blazars}",
      journal = {\apj},
         year = 2013,
        month = may,
       volume = {768},
       number = {1},
          eid = {54},
        pages = {54},
          doi = {10.1088/0004-637X/768/1/54},
archivePrefix = {arXiv},
       eprint = {1304.0605},
 primaryClass = {astro-ph.HE},
       adsurl = {https://ui.adsabs.harvard.edu/abs/2013ApJ...768...54B}
}

@ARTICLE{Breiman1996,
       author = {{Breiman}, Leo},
        title = "{Bagging Predictors}",
      journal = {Machine Learning},
         year = 1996,
        month = aug,
       volume = {24},
       number = {2},
        pages = {123-140},
          doi = {10.1023/A:1018054314350},
       adsurl = {https://ui.adsabs.harvard.edu/abs/1996MLear..24..123B}
}

@ARTICLE{Breiman2001,
       author = {{Breiman}, Leo},
        title = "{Random Forests.}",
      journal = {Machine Learning},
         year = 2001,
        month = jan,
       volume = {45},
        pages = {5-32},
          doi = {10.1023/A:1010933404324},
       adsurl = {https://ui.adsabs.harvard.edu/abs/2001MachL..45....5B}
}

@ARTICLE{Chiaro2016,
       author = {{Chiaro}, G. and {Salvetti}, D. and {La Mura}, G. and {Giroletti}, M. and {Thompson}, D.~J. and {Bastieri}, D.},
        title = "{Blazar flaring patterns (B-FlaP) classifying blazar candidate of uncertain type in the third Fermi-LAT catalogue by artificial neural networks}",
      journal = {\mnras},
         year = 2016,
        month = nov,
       volume = {462},
       number = {3},
        pages = {3180-3195},
          doi = {10.1093/mnras/stw1830},
archivePrefix = {arXiv},
       eprint = {1607.07822},
 primaryClass = {astro-ph.HE},
       adsurl = {https://ui.adsabs.harvard.edu/abs/2016MNRAS.462.3180C}
}

@ARTICLE{Dermer1993,
       author = {{Dermer}, Charles D. and {Schlickeiser}, Reinhard},
        title = "{Model for the High-Energy Emission from Blazars}",
      journal = {\apj},
         year = 1993,
        month = oct,
       volume = {416},
        pages = {458},
          doi = {10.1086/173251},
       adsurl = {https://ui.adsabs.harvard.edu/abs/1993ApJ...416..458D}
}

@ARTICLE{Edelson2002,
       author = {{Edelson}, Rick and {Turner}, T.~J. and {Pounds}, Ken and {Vaughan}, Simon and {Markowitz}, Alex and {Marshall}, Herman and {Dobbie}, Paul and {Warwick}, Robert},
        title = "{X-Ray Spectral Variability and Rapid Variability of the Soft X-Ray Spectrum Seyfert 1 Galaxies Arakelian 564 and Ton S180}",
      journal = {\apj},
         year = 2002,
        month = apr,
       volume = {568},
       number = {2},
        pages = {610-626},
          doi = {10.1086/323779},
archivePrefix = {arXiv},
       eprint = {astro-ph/0108387},
 primaryClass = {astro-ph},
       adsurl = {https://ui.adsabs.harvard.edu/abs/2002ApJ...568..610E}
}

@ARTICLE{Ghisellini1993,
       author = {{Ghisellini}, G. and {Padovani}, P. and {Celotti}, A. and {Maraschi}, L.},
        title = "{Relativistic Bulk Motion in Active Galactic Nuclei}",
      journal = {\apj},
         year = 1993,
        month = apr,
       volume = {407},
        pages = {65},
          doi = {10.1086/172493},
       adsurl = {https://ui.adsabs.harvard.edu/abs/1993ApJ...407...65G}
}

@ARTICLE{Giebels2009,
       author = {{Giebels}, B. and {Degrange}, B.},
        title = "{Lognormal variability in BL Lacertae}",
      journal = {\aap},
         year = 2009,
        month = sep,
       volume = {503},
       number = {3},
        pages = {797-799},
          doi = {10.1051/0004-6361/200912303},
archivePrefix = {arXiv},
       eprint = {0907.2425},
 primaryClass = {astro-ph.CO},
       adsurl = {https://ui.adsabs.harvard.edu/abs/2009A&A...503..797G}
}

@book{Good2013,
  author = {Good, P.},
  title = {Permutation Tests: A Practical Guide to Resampling Methods for Testing Hypotheses},
  year = {2013},
  publisher = {Springer},
  address = {New York}
}

@ARTICLE{Harris2020,
       author = {{Harris}, Charles R. and {Millman}, K. Jarrod and {van der Walt}, St{\'e}fan J. and {Gommers}, Ralf and {Virtanen}, Pauli and {Cournapeau}, David and {Wieser}, Eric and {Taylor}, Julian and {Berg}, Sebastian and {Smith}, Nathaniel J. and {Kern}, Robert and {Picus}, Matti and {Hoyer}, Stephan and {van Kerkwijk}, Marten H. and {Brett}, Matthew and {Haldane}, Allan and {del R{\'\i}o}, Jaime Fern{\'a}ndez and {Wiebe}, Mark and {Peterson}, Pearu and {G{\'e}rard-Marchant}, Pierre and {Sheppard}, Kevin and {Reddy}, Tyler and {Weckesser}, Warren and {Abbasi}, Hameer and {Gohlke}, Christoph and {Oliphant}, Travis E.},
        title = "{Array programming with NumPy}",
      journal = {\nat},
         year = 2020,
        month = sep,
       volume = {585},
       number = {7825},
        pages = {357-362},
          doi = {10.1038/s41586-020-2649-2},
archivePrefix = {arXiv},
       eprint = {2006.10256},
 primaryClass = {cs.MS},
       adsurl = {https://ui.adsabs.harvard.edu/abs/2020Natur.585..357H}
}

@ARTICLE{Hayashida2015,
       author = {{Hayashida}, M. and {Nalewajko}, K. and {Madejski}, G.~M. and {Sikora}, M. and {Itoh}, R. and {Ajello}, M. and {Blandford}, R.~D. and {Buson}, S. and {Chiang}, J. and {Fukazawa}, Y. and {Furniss}, A.~K. and {Urry}, C.~M. and {Hasan}, I. and {Harrison}, F.~A. and {Alexander}, D.~M. and {Balokovi{\'c}}, M. and {Barret}, D. and {Boggs}, S.~E. and {Christensen}, F.~E. and {Craig}, W.~W. and {Forster}, K. and {Giommi}, P. and {Grefenstette}, B. and {Hailey}, C. and {Hornstrup}, A. and {Kitaguchi}, T. and {Koglin}, J.~E. and {Madsen}, K.~K. and {Mao}, P.~H. and {Miyasaka}, H. and {Mori}, K. and {Perri}, M. and {Pivovaroff}, M.~J. and {Puccetti}, S. and {Rana}, V. and {Stern}, D. and {Tagliaferri}, G. and {Westergaard}, N.~J. and {Zhang}, W.~W. and {Zoglauer}, A. and {Gurwell}, M.~A. and {Uemura}, M. and {Akitaya}, H. and {Kawabata}, K.~S. and {Kawaguchi}, K. and {Kanda}, Y. and {Moritani}, Y. and {Takaki}, K. and {Ui}, T. and {Yoshida}, M. and {Agarwal}, A. and {Gupta}, A.~C.},
        title = "{Rapid Variability of Blazar 3C 279 during Flaring States in 2013-2014 with Joint Fermi-LAT, NuSTAR, Swift, and Ground-Based Multiwavelength Observations}",
      journal = {\apj},
         year = 2015,
        month = jul,
       volume = {807},
       number = {1},
          eid = {79},
        pages = {79},
          doi = {10.1088/0004-637X/807/1/79},
archivePrefix = {arXiv},
       eprint = {1502.04699},
 primaryClass = {astro-ph.HE},
       adsurl = {https://ui.adsabs.harvard.edu/abs/2015ApJ...807...79H}
}

@ARTICLE{Hunter2007,
       author = {{Hunter}, John D.},
        title = "{Matplotlib: A 2D Graphics Environment}",
      journal = {Computing in Science and Engineering},
         year = 2007,
        month = jan,
       volume = {9},
       number = {3},
        pages = {90-95},
          doi = {10.1109/MCSE.2007.55},
       adsurl = {https://ui.adsabs.harvard.edu/abs/2007CSE.....9...90H}
}

@book{Hyndman2018,
  author = {Hyndman, R. J. and Athanasopoulos, G.},
  title = {Forecasting: Principles and Practice},
  edition = {2},
  year = {2018},
  publisher = {OTexts},
  address = {Melbourne}
}

@ARTICLE{Katarzynski2005,
       author = {{Katarzy{\'n}ski}, K. and {Ghisellini}, G. and {Tavecchio}, F. and {Maraschi}, L. and {Fossati}, G. and {Mastichiadis}, A.},
        title = "{Correlation between the TeV and X-ray emission in high-energy peaked BL Lac objects}",
      journal = {\aap},
         year = 2005,
        month = apr,
       volume = {433},
       number = {2},
        pages = {479-496},
          doi = {10.1051/0004-6361:20041556},
archivePrefix = {arXiv},
       eprint = {astro-ph/0412405},
 primaryClass = {astro-ph},
       adsurl = {https://ui.adsabs.harvard.edu/abs/2005A&A...433..479K}
}

@ARTICLE{Kiehlmann2016,
       author = {{Kiehlmann}, S. and {Savolainen}, T. and {Jorstad}, S.~G. and {Sokolovsky}, K.~V. and {Schinzel}, F.~K. and {Marscher}, A.~P. and {Larionov}, V.~M. and {Agudo}, I. and {Akitaya}, H. and {Ben{\'\i}tez}, E. and {Berdyugin}, A. and {Blinov}, D.~A. and {Bochkarev}, N.~G. and {Borman}, G.~A. and {Burenkov}, A.~N. and {Casadio}, C. and {Doroshenko}, V.~T. and {Efimova}, N.~V. and {Fukazawa}, Y. and {G{\'o}mez}, J.~L. and {Grishina}, T.~S. and {Hagen-Thorn}, V.~A. and {Heidt}, J. and {Hiriart}, D. and {Itoh}, R. and {Joshi}, M. and {Kawabata}, K.~S. and {Kimeridze}, G.~N. and {Kopatskaya}, E.~N. and {Korobtsev}, I.~V. and {Krajci}, T. and {Kurtanidze}, O.~M. and {Kurtanidze}, S.~O. and {Larionova}, E.~G. and {Larionova}, L.~V. and {Lindfors}, E. and {L{\'o}pez}, J.~M. and {McHardy}, I.~M. and {Molina}, S.~N. and {Moritani}, Y. and {Morozova}, D.~A. and {Nazarov}, S.~V. and {Nikolashvili}, M.~G. and {Nilsson}, K. and {Pulatova}, N.~G. and {Reinthal}, R. and {Sadun}, A. and {Sasada}, M. and {Savchenko}, S.~S. and {Sergeev}, S.~G. and {Sigua}, L.~A. and {Smith}, P.~S. and {Sorcia}, M. and {Spiridonova}, O.~I. and {Takaki}, K. and {Takalo}, L.~O. and {Taylor}, B. and {Troitsky}, I.~S. and {Uemura}, M. and {Ugolkova}, L.~S. and {Ui}, T. and {Yoshida}, M. and {Zensus}, J.~A. and {Zhdanova}, V.~E.},
        title = "{Polarization angle swings in blazars: The case of <ASTROBJ>3C 279</ASTROBJ>}",
      journal = {\aap},
         year = 2016,
        month = may,
       volume = {590},
          eid = {A10},
        pages = {A10},
          doi = {10.1051/0004-6361/201527725},
archivePrefix = {arXiv},
       eprint = {1603.00249},
 primaryClass = {astro-ph.HE},
       adsurl = {https://ui.adsabs.harvard.edu/abs/2016A&A...590A..10K}
}

@ARTICLE{Kovacevic2020,
       author = {{Kova{\v{c}}evi{\'c}}, M. and {Chiaro}, G. and {Cutini}, S. and {Tosti}, G.},
        title = "{Classification of blazar candidates of uncertain type from the Fermi LAT 8-yr source catalogue with an artificial neural network}",
      journal = {\mnras},
         year = 2020,
        month = apr,
       volume = {493},
       number = {2},
        pages = {1926-1935},
          doi = {10.1093/mnras/staa394},
archivePrefix = {arXiv},
       eprint = {2002.10256},
 primaryClass = {astro-ph.HE},
       adsurl = {https://ui.adsabs.harvard.edu/abs/2020MNRAS.493.1926K}
}

@article{Kunsch1989,
  author = {K{"u}nsch, H. R.},
  title = {The jackknife and the bootstrap for general stationary observations},
  year = {1989},
  journal = {The Annals of Statistics},
  volume = {17},
  pages = {1217--1241}
}

@ARTICLE{Macdonald2015,
       author = {{MacDonald}, Nicholas R. and {Marscher}, Alan P. and {Jorstad}, Svetlana G. and {Joshi}, Manasvita},
        title = "{Through the Ring of Fire: Gamma-Ray Variability in Blazars by a Moving Plasmoid Passing a Local Source of Seed Photons}",
      journal = {\apj},
         year = 2015,
        month = may,
       volume = {804},
       number = {2},
          eid = {111},
        pages = {111},
          doi = {10.1088/0004-637X/804/2/111},
archivePrefix = {arXiv},
       eprint = {1505.01239},
 primaryClass = {astro-ph.HE},
       adsurl = {https://ui.adsabs.harvard.edu/abs/2015ApJ...804..111M}
}

@ARTICLE{Mannheim1993,
       author = {{Mannheim}, K.},
        title = "{The proton blazar.}",
      journal = {\aap},
         year = 1993,
        month = mar,
       volume = {269},
        pages = {67-76},
          doi = {10.48550/arXiv.astro-ph/9302006},
archivePrefix = {arXiv},
       eprint = {astro-ph/9302006},
 primaryClass = {astro-ph},
       adsurl = {https://ui.adsabs.harvard.edu/abs/1993A&A...269...67M}
}

@ARTICLE{Maraschi1992,
       author = {{Maraschi}, L. and {Ghisellini}, G. and {Celotti}, A.},
        title = "{A Jet Model for the Gamma-Ray--emitting Blazar 3C 279}",
      journal = {\apjl},
         year = 1992,
        month = sep,
       volume = {397},
        pages = {L5},
          doi = {10.1086/186531},
       adsurl = {https://ui.adsabs.harvard.edu/abs/1992ApJ...397L...5M}
}

@ARTICLE{Marscher1985,
       author = {{Marscher}, A.~P. and {Gear}, W.~K.},
        title = "{Models for high-frequency radio outbursts in extragalactic sources, with application to the early 1983 millimeter-to-infrared flare of 3C 273.}",
      journal = {\apj},
         year = 1985,
        month = nov,
       volume = {298},
        pages = {114-127},
          doi = {10.1086/163592},
       adsurl = {https://ui.adsabs.harvard.edu/abs/1985ApJ...298..114M}
}

@ARTICLE{Marscher2010,
       author = {{Marscher}, Alan P. and {Jorstad}, Svetlana G. and {Larionov}, Valeri M. and {Aller}, Margo F. and {Aller}, Hugh D. and {L{\"a}hteenm{\"a}ki}, Anne and {Agudo}, Iv{\'a}n and {Smith}, Paul S. and {Gurwell}, Mark and {Hagen-Thorn}, Vladimir A. and {Konstantinova}, Tatiana S. and {Larionova}, Elena G. and {Larionova}, Liudmila V. and {Melnichuk}, Daria A. and {Blinov}, Dmitry A. and {Kopatskaya}, Evgenia N. and {Troitsky}, Ivan S. and {Tornikoski}, Merja and {Hovatta}, Talvikki and {Schmidt}, Gary D. and {D'Arcangelo}, Francesca D. and {Bhattarai}, Dipesh and {Taylor}, Brian and {Olmstead}, Alice R. and {Manne-Nicholas}, Emily and {Roca-Sogorb}, Mar and {G{\'o}mez}, Jos{\'e} L. and {McHardy}, Ian M. and {Kurtanidze}, Omar and {Nikolashvili}, Maria G. and {Kimeridze}, Givi N. and {Sigua}, Lorand A.},
        title = "{Probing the Inner Jet of the Quasar PKS 1510-089 with Multi-Waveband Monitoring During Strong Gamma-Ray Activity}",
      journal = {\apjl},
         year = 2010,
        month = feb,
       volume = {710},
       number = {2},
        pages = {L126-L131},
          doi = {10.1088/2041-8205/710/2/L126},
archivePrefix = {arXiv},
       eprint = {1001.2574},
 primaryClass = {astro-ph.CO},
       adsurl = {https://ui.adsabs.harvard.edu/abs/2010ApJ...710L.126M}
}

@ARTICLE{Marscher2014,
       author = {{Marscher}, Alan P.},
        title = "{Turbulent, Extreme Multi-zone Model for Simulating Flux and Polarization Variability in Blazars}",
      journal = {\apj},
         year = 2014,
        month = jan,
       volume = {780},
       number = {1},
          eid = {87},
        pages = {87},
          doi = {10.1088/0004-637X/780/1/87},
archivePrefix = {arXiv},
       eprint = {1311.7665},
 primaryClass = {astro-ph.HE},
       adsurl = {https://ui.adsabs.harvard.edu/abs/2014ApJ...780...87M}
}

@ARTICLE{Mucke2003,
       author = {{M{\"u}cke}, A. and {Protheroe}, R.~J. and {Engel}, R. and {Rachen}, J.~P. and {Stanev}, T.},
        title = "{BL Lac objects in the synchrotron proton blazar model}",
      journal = {Astroparticle Physics},
         year = 2003,
        month = mar,
       volume = {18},
       number = {6},
        pages = {593-613},
          doi = {10.1016/S0927-6505(02)00185-8},
archivePrefix = {arXiv},
       eprint = {astro-ph/0206164},
 primaryClass = {astro-ph},
       adsurl = {https://ui.adsabs.harvard.edu/abs/2003APh....18..593M}
}

@ARTICLE{Nalewajko2013,
       author = {{Nalewajko}, Krzysztof},
        title = "{The brightest gamma-ray flares of blazars}",
      journal = {\mnras},
         year = 2013,
        month = apr,
       volume = {430},
       number = {2},
        pages = {1324-1333},
          doi = {10.1093/mnras/sts711},
archivePrefix = {arXiv},
       eprint = {1211.0274},
 primaryClass = {astro-ph.HE},
       adsurl = {https://ui.adsabs.harvard.edu/abs/2013MNRAS.430.1324N}
}

@ARTICLE{Ojha2013_ATel4941,
       author = {{Ojha}, Roopesh and {Carpenter}, Bryce and {Dutka}, Michael},
        title = "{Fermi LAT detection of a gamma-ray flare from FSRQ S5 1044+71}",
      journal = {The Astronomer's Telegram},
         year = 2013,
        month = apr,
       volume = {4941},
        pages = {1},
       adsurl = {https://ui.adsabs.harvard.edu/abs/2013ATel.4941....1O}
}

@ARTICLE{Ojha2017_ATel9928,
       author = {{Ojha}, Roopesh and {Carpen}, Bryce},
        title = "{Fermi LAT detection of an increase in gamma-ray activity of the FSRQ S5 1044+71}",
      journal = {The Astronomer's Telegram},
         year = 2017,
        month = jan,
       volume = {9928},
        pages = {1},
       adsurl = {https://ui.adsabs.harvard.edu/abs/2017ATel.9928....1O}
}

@ARTICLE{Padovani1995,
       author = {{Padovani}, Paolo and {Giommi}, Paolo},
        title = "{The Connection between X-Ray-- and Radio-selected BL Lacertae Objects}",
      journal = {\apj},
         year = 1995,
        month = may,
       volume = {444},
        pages = {567},
          doi = {10.1086/175631},
archivePrefix = {arXiv},
       eprint = {astro-ph/9412073},
 primaryClass = {astro-ph},
       adsurl = {https://ui.adsabs.harvard.edu/abs/1995ApJ...444..567P}
}

@ARTICLE{Pedregosa2011,
       author = {{Pedregosa}, Fabian and {Varoquaux}, Ga{\"e}l and {Gramfort}, Alexandre and {Michel}, Vincent and {Thirion}, Bertrand and {Grisel}, Olivier and {Blondel}, Mathieu and {M{\"u}ller}, Andreas and {Nothman}, Joel and {Louppe}, Gilles and {Prettenhofer}, Peter and {Weiss}, Ron and {Dubourg}, Vincent and {Vanderplas}, Jake and {Passos}, Alexandre and {Cournapeau}, David and {Brucher}, Matthieu and {Perrot}, Matthieu and {Duchesnay}, {\'E}douard},
        title = "{Scikit-learn: Machine Learning in Python}",
      journal = {Journal of Machine Learning Research},
         year = 2011,
        month = oct,
       volume = {12},
        pages = {2825-2830},
          doi = {10.48550/arXiv.1201.0490},
archivePrefix = {arXiv},
       eprint = {1201.0490},
 primaryClass = {cs.LG},
       adsurl = {https://ui.adsabs.harvard.edu/abs/2011JMLR...12.2825P}
}

@article{Phipson2010,
  author = {Phipson, B. and Smyth, G. K.},
  title = {Permutation P-values Should Never Be Zero: Calculating Exact P-values When Permutations Are Randomly Drawn},
  year = {2010},
  journal = {Statistical Applications in Genetics and Molecular Biology},
  volume = {9},
  pages = {Article 39}
}

@incollection{Platt1999,
  author = {Platt, J. C.},
  title = {Probabilistic Outputs for Support Vector Machines and Comparisons to Regularized Likelihood Methods},
  booktitle = {Advances in Large Margin Classifiers},
  editor = {Smola, A. J. and Bartlett, P. and Sch{"o}lkopf, B. and Schuurmans, D.},
  year = {1999},
  publisher = {MIT Press},
  address = {Cambridge, MA},
  pages = {61--74}
}

@ARTICLE{Polatidis1995,
       author = {{Polatidis}, A.~G. and {Wilkinson}, P.~N. and {Xu}, W. and {Readhead}, A.~C.~S. and {Pearson}, T.~J. and {Taylor}, G.~B. and {Vermeulen}, R.~C.},
        title = "{The First Caltech--Jodrell Bank VLBI Survey. I. lambda = 18 Centimeter Observations of 87 Sources}",
      journal = {\apjs},
         year = 1995,
        month = may,
       volume = {98},
        pages = {1},
          doi = {10.1086/192152},
       adsurl = {https://ui.adsabs.harvard.edu/abs/1995ApJS...98....1P}
}

@ARTICLE{Politis1994,
       author = {{Politis}, Dimitris N. and {Romano}, Joseph P.},
        title = "{The Stationary Bootstrap}",
      journal = {Journal of the American Statistical Association},
         year = 1994,
        month = dec,
       volume = {89},
       number = {428},
        pages = {1303-1313},
          doi = {10.1080/01621459.1994.10476870},
       adsurl = {https://ui.adsabs.harvard.edu/abs/1994JASA...89.1303P}
}

@incollection{Politis1992,
  author    = {Politis, Dimitris N. and Romano, Joseph P.},
  title     = {A Circular Block-Resampling Procedure for Stationary Data},
  booktitle = {Exploring the Limits of Bootstrap},
  editor    = {LePage, R. and Billard, L.},
  publisher = {Wiley},
  address   = {New York},
  year      = {1992},
  pages     = {263--270}
}

@ARTICLE{Raiteri2017,
       author = {{Raiteri}, C.~M. and {Villata}, M. and {Acosta-Pulido}, J.~A. and {Agudo}, I. and {Arkharov}, A.~A. and {Bachev}, R. and {Baida}, G.~V. and {Ben{\'\i}tez}, E. and {Borman}, G.~A. and {Boschin}, W. and {Bozhilov}, V. and {Butuzova}, M.~S. and {Calcidese}, P. and {Carnerero}, M.~I. and {Carosati}, D. and {Casadio}, C. and {Castro-Segura}, N. and {Chen}, W.-P. and {Damljanovic}, G. and {D'Ammando}, F. and {di Paola}, A. and {Echevarr{\'\i}a}, J. and {Efimova}, N.~V. and {Ehgamberdiev}, Sh. A. and {Espinosa}, C. and {Fuentes}, A. and {Giunta}, A. and {G{\'o}mez}, J.~L. and {Grishina}, T.~S. and {Gurwell}, M.~A. and {Hiriart}, D. and {Jermak}, H. and {Jordan}, B. and {Jorstad}, S.~G. and {Joshi}, M. and {Kopatskaya}, E.~N. and {Kuratov}, K. and {Kurtanidze}, O.~M. and {Kurtanidze}, S.~O. and {L{\"a}hteenm{\"a}ki}, A. and {Larionov}, V.~M. and {Larionova}, E.~G. and {Larionova}, L.~V. and {L{\'a}zaro}, C. and {Lin}, C.~S. and {Malmrose}, M.~P. and {Marscher}, A.~P. and {Matsumoto}, K. and {McBreen}, B. and {Michel}, R. and {Mihov}, B. and {Minev}, M. and {Mirzaqulov}, D.~O. and {Mokrushina}, A.~A. and {Molina}, S.~N. and {Moody}, J.~W. and {Morozova}, D.~A. and {Nazarov}, S.~V. and {Nikolashvili}, M.~G. and {Ohlert}, J.~M. and {Okhmat}, D.~N. and {Ovcharov}, E. and {Pinna}, F. and {Polakis}, T.~A. and {Protasio}, C. and {Pursimo}, T. and {Redondo-Lorenzo}, F.~J. and {Rizzi}, N. and {Rodriguez-Coira}, G. and {Sadakane}, K. and {Sadun}, A.~C. and {Samal}, M.~R. and {Savchenko}, S.~S. and {Semkov}, E. and {Skiff}, B.~A. and {Slavcheva-Mihova}, L. and {Smith}, P.~S. and {Steele}, I.~A. and {Strigachev}, A. and {Tammi}, J. and {Thum}, C. and {Tornikoski}, M. and {Troitskaya}, Yu. V. and {Troitsky}, I.~S. and {Vasilyev}, A.~A. and {Vince}, O.},
        title = "{Blazar spectral variability as explained by a twisted inhomogeneous jet}",
      journal = {\nat},
         year = 2017,
        month = dec,
       volume = {552},
       number = {7685},
        pages = {374-377},
          doi = {10.1038/nature24623},
archivePrefix = {arXiv},
       eprint = {1712.02098},
 primaryClass = {astro-ph.HE},
       adsurl = {https://ui.adsabs.harvard.edu/abs/2017Natur.552..374R}
}

@ARTICLE{Sahakyan2023,
       author = {{Sahakyan}, N. and {Vardanyan}, V. and {Khachatryan}, M.},
        title = "{Gradient boosting decision trees classification of blazars of uncertain type in the fourth Fermi-LAT catalogue}",
      journal = {\mnras},
         year = 2023,
        month = feb,
       volume = {519},
       number = {2},
        pages = {3000-3010},
          doi = {10.1093/mnras/stac3701},
archivePrefix = {arXiv},
       eprint = {2212.06614},
 primaryClass = {astro-ph.HE},
       adsurl = {https://ui.adsabs.harvard.edu/abs/2023MNRAS.519.3000S}
}

@ARTICLE{Saito2015,
       author = {{Saito}, Takaya and {Rehmsmeier}, Marc},
        title = "{The Precision-Recall Plot Is More Informative than the ROC Plot When Evaluating Binary Classifiers on Imbalanced Datasets}",
      journal = {PLoS ONE},
         year = 2015,
        month = mar,
       volume = {10},
       number = {3},
        pages = {e0118432},
          doi = {10.1371/journal.pone.0118432},
       adsurl = {https://ui.adsabs.harvard.edu/abs/2015PLoSO..1018432S}
}

@ARTICLE{Scargle2013,
       author = {{Scargle}, Jeffrey D. and {Norris}, Jay P. and {Jackson}, Brad and {Chiang}, James},
        title = "{Studies in Astronomical Time Series Analysis. VI. Bayesian Block Representations}",
      journal = {\apj},
         year = 2013,
        month = feb,
       volume = {764},
       number = {2},
          eid = {167},
        pages = {167},
          doi = {10.1088/0004-637X/764/2/167},
archivePrefix = {arXiv},
       eprint = {1207.5578},
 primaryClass = {astro-ph.IM},
       adsurl = {https://ui.adsabs.harvard.edu/abs/2013ApJ...764..167S}
}

@ARTICLE{Shah2018_lognormal,
       author = {{Shah}, Zahir and {Mankuzhiyil}, Nijil and {Sinha}, Atreyee and {Misra}, Ranjeev and {Sahayanathan}, Sunder and {Iqbal}, Naseer},
        title = "{Log-normal flux distribution of bright Fermi blazars}",
      journal = {Research in Astronomy and Astrophysics},
         year = 2018,
        month = oct,
       volume = {18},
       number = {11},
          eid = {141},
        pages = {141},
          doi = {10.1088/1674-4527/18/11/141},
archivePrefix = {arXiv},
       eprint = {1805.04675},
 primaryClass = {astro-ph.HE},
       adsurl = {https://ui.adsabs.harvard.edu/abs/2018RAA....18..141S}
}

@ARTICLE{2017MNRAS.470.3283S,
       author = {{Shah}, Zahir and {Sahayanathan}, S. and {Mankuzhiyil}, Nijil and {Kushwaha}, Pankaj and {Misra}, Ranjeev and {Iqbal}, Naseer},
        title = "{Clues on high-energy emission mechanism from blazar 3C 454.3 during 2015 August flare}",
      journal = {\mnras},
         year = 2017,
        month = sep,
       volume = {470},
       number = {3},
        pages = {3283-3299},
          doi = {10.1093/mnras/stx1194},
archivePrefix = {arXiv},
       eprint = {1705.06185},
 primaryClass = {astro-ph.HE},
       adsurl = {https://ui.adsabs.harvard.edu/abs/2017MNRAS.470.3283S}
}

@ARTICLE{Sikora1994,
       author = {{Sikora}, Marek and {Begelman}, Mitchell C. and {Rees}, Martin J.},
        title = "{Comptonization of Diffuse Ambient Radiation by a Relativistic Jet: The Source of Gamma Rays from Blazars?}",
      journal = {\apj},
         year = 1994,
        month = jan,
       volume = {421},
        pages = {153},
          doi = {10.1086/173633},
       adsurl = {https://ui.adsabs.harvard.edu/abs/1994ApJ...421..153S}
}

@ARTICLE{Sironi2015,
       author = {{Sironi}, Lorenzo and {Petropoulou}, Maria and {Giannios}, Dimitrios},
        title = "{Relativistic jets shine through shocks or magnetic reconnection?}",
      journal = {\mnras},
         year = 2015,
        month = jun,
       volume = {450},
       number = {1},
        pages = {183-191},
          doi = {10.1093/mnras/stv641},
archivePrefix = {arXiv},
       eprint = {1502.01021},
 primaryClass = {astro-ph.HE},
       adsurl = {https://ui.adsabs.harvard.edu/abs/2015MNRAS.450..183S}
}

@ARTICLE{Spada2001,
       author = {{Spada}, Maddalena and {Ghisellini}, Gabriele and {Lazzati}, Davide and {Celotti}, Annalisa},
        title = "{Internal shocks in the jets of radio-loud quasars}",
      journal = {\mnras},
         year = 2001,
        month = aug,
       volume = {325},
       number = {4},
        pages = {1559-1570},
          doi = {10.1046/j.1365-8711.2001.04557.x},
archivePrefix = {arXiv},
       eprint = {astro-ph/0103424},
 primaryClass = {astro-ph},
       adsurl = {https://ui.adsabs.harvard.edu/abs/2001MNRAS.325.1559S}
}

@ARTICLE{Stickel1991,
       author = {{Stickel}, M. and {Padovani}, P. and {Urry}, C.~M. and {Fried}, J.~W. and {Kuehr}, H.},
        title = "{The Complete Sample of 1 Jansky BL Lacertae Objects. I. Summary Properties}",
      journal = {\apj},
         year = 1991,
        month = jun,
       volume = {374},
        pages = {431},
          doi = {10.1086/170133},
       adsurl = {https://ui.adsabs.harvard.edu/abs/1991ApJ...374..431S}
}

@ARTICLE{Tolamatti2023,
       author = {{Tolamatti}, A. and {Singh}, K.~K. and {Yadav}, K.~K.},
        title = "{Classification of the Fermi-LAT blazar candidates of uncertain type using extreme gradient boosting}",
      journal = {\mnras},
         year = 2023,
        month = aug,
       volume = {523},
       number = {4},
        pages = {5341-5352},
          doi = {10.1093/mnras/stad1826},
archivePrefix = {arXiv},
       eprint = {2306.15393},
 primaryClass = {astro-ph.HE},
       adsurl = {https://ui.adsabs.harvard.edu/abs/2023MNRAS.523.5341T}
}

@ARTICLE{Urry1995,
       author = {{Urry}, C. Megan and {Padovani}, Paolo},
        title = "{Unified Schemes for Radio-Loud Active Galactic Nuclei}",
      journal = {\pasp},
         year = 1995,
        month = sep,
       volume = {107},
        pages = {803},
          doi = {10.1086/133630},
archivePrefix = {arXiv},
       eprint = {astro-ph/9506063},
 primaryClass = {astro-ph},
       adsurl = {https://ui.adsabs.harvard.edu/abs/1995PASP..107..803U}
}

@ARTICLE{Vaughan2003,
       author = {{Vaughan}, S. and {Edelson}, R. and {Warwick}, R.~S. and {Uttley}, P.},
        title = "{On characterizing the variability properties of X-ray light curves from active galaxies}",
      journal = {\mnras},
         year = 2003,
        month = nov,
       volume = {345},
       number = {4},
        pages = {1271-1284},
          doi = {10.1046/j.1365-2966.2003.07042.x},
archivePrefix = {arXiv},
       eprint = {astro-ph/0307420},
 primaryClass = {astro-ph},
       adsurl = {https://ui.adsabs.harvard.edu/abs/2003MNRAS.345.1271V}
}

@ARTICLE{Virtanen2020,
       author = {{Virtanen}, Pauli and {Gommers}, Ralf and {Oliphant}, Travis E. and {Haberland}, Matt and {Reddy}, Tyler and {Cournapeau}, David and {Burovski}, Evgeni and {Peterson}, Pearu and {Weckesser}, Warren and {Bright}, Jonathan and {van der Walt}, St{\'e}fan J. and {Brett}, Matthew and {Wilson}, Joshua and {Millman}, K. Jarrod and {Mayorov}, Nikolay and {Nelson}, Andrew R.~J. and {Jones}, Eric and {Kern}, Robert and {Larson}, Eric and {Carey}, C.~J. and {Polat}, {\.I}lhan and {Feng}, Yu and {Moore}, Eric W. and {VanderPlas}, Jake and {Laxalde}, Denis and {Perktold}, Josef and {Cimrman}, Robert and {Henriksen}, Ian and {Quintero}, E.~A. and {Harris}, Charles R. and {Archibald}, Anne M. and {Ribeiro}, Ant{\^o}nio H. and {Pedregosa}, Fabian and {van Mulbregt}, Paul and {SciPy 1.  0 Contributors}},
        title = "{SciPy 1.0: fundamental algorithms for scientific computing in Python}",
      journal = {Nature Medicine},
         year = 2020,
        month = feb,
       volume = {17},
        pages = {261-272},
          doi = {10.1038/s41592-019-0686-2},
archivePrefix = {arXiv},
       eprint = {1907.10121},
 primaryClass = {cs.MS},
       adsurl = {https://ui.adsabs.harvard.edu/abs/2020NatMe..17..261V}
}

@book{dePrado2018,
  author = {de Prado, M. L.},
  title = {Advances in Financial Machine Learning},
  year = {2018},
  publisher = {Wiley}
}

@ARTICLE{2024ApJ...977..111A,
       author = {{Akbar}, Sikandar and {Shah}, Zahir and {Misra}, Ranjeev and {Iqbal}, Naseer},
        title = "{Insights into the Long-term Flaring Events of Blazar PKS 0805-07: A Multiwavelength Analysis Over the Period of 2009{\textendash}2023}",
      journal = {\apj},
         year = 2024,
        month = dec,
       volume = {977},
       number = {1},
          eid = {111},
        pages = {111},
          doi = {10.3847/1538-4357/ad8ddb},
archivePrefix = {arXiv},
       eprint = {2410.23181},
 primaryClass = {astro-ph.HE},
       adsurl = {https://ui.adsabs.harvard.edu/abs/2024ApJ...977..111A}
}

@ARTICLE{2025MNRAS.539.2185M,
       author = {{Malik}, Zahoor and {Akbar}, Sikandar and {Shah}, Zahir and {Misra}, Ranjeev and {Dar}, Athar A. and {Manzoor}, Aaqib and {Ahanger}, Sajad and {Nazir}, Zeeshan and {Iqbal}, Naseer and {Rubab}, Seemin and {Tantry}, Javaid},
        title = "{Statistical Insights into flux and photon index distributions of VHE FSRQs from Fermi-LAT observations}",
      journal = {\mnras},
         year = 2025,
        month = may,
       volume = {539},
       number = {3},
        pages = {2185-2201},
          doi = {10.1093/mnras/staf620},
archivePrefix = {arXiv},
       eprint = {2505.23338},
 primaryClass = {astro-ph.HE},
       adsurl = {https://ui.adsabs.harvard.edu/abs/2025MNRAS.539.2185M}
}

@article{61tz-jk8c,
  title = {Comprehensive variability analysis of blazars using Fermi light curves across multiple timescales},
  author = {Shah, Zahir and Dar, Athar A. and Akbar, Sikandar and Peer, Anjum and Malik, Zahoor and Manzoor, Aaqib and Ahanger, Sajad and Tantry, Javaid and Nazir, Zeeshan and Bose, Debanjan and Magray, Mushtaq},
  journal = {Phys. Rev. D},
  volume = {111},
  issue = {12},
  pages = {123052},
  numpages = {17},
  year = {2025},
  month = {Jun},
  publisher = {American Physical Society},
  doi = {10.1103/61tz-jk8c},
  url = {https://link.aps.org/doi/10.1103/61tz-jk8c}
}

@ARTICLE{2025PhRvD.112f3061A,
       author = {{Akbar}, Sikandar and {Shah}, Zahir and {Misra}, Ranjeev and {Boked}, Sajad and {Iqbal}, Naseer},
        title = "{Indication for dual periodic signatures in PKS 0805-07 from multitechnique time series analysis}",
      journal = {\prd},
         year = 2025,
        month = sep,
       volume = {112},
       number = {6},
          eid = {063061},
        pages = {063061},
          doi = {10.1103/zxgv-fzv5},
archivePrefix = {arXiv},
       eprint = {2506.09602},
 primaryClass = {astro-ph.HE},
       adsurl = {https://ui.adsabs.harvard.edu/abs/2025PhRvD.112f3061A}
}

@article{AKBAR2026100608,
title = {A multi-technique search for year-scale γ-ray quasi-periodic modulation in the high-redshift FSRQ PKS 2052-47},
journal = {Journal of High Energy Astrophysics},
volume = {53},
pages = {100608},
year = {2026},
issn = {2214-4048},
doi = {https://doi.org/10.1016/j.jheap.2026.100608},
url = {https://www.sciencedirect.com/science/article/pii/S2214404826000601},
author = {Sikandar Akbar}
}

@ARTICLE{2014ApJ...796...61K,
       author = {{Kushwaha}, Pankaj and {Singh}, K.~P. and {Sahayanathan}, Sunder},
        title = "{Brightest Fermi-LAT Flares of PKS 1222+216: Implications on Emission and Acceleration Processes}",
      journal = {\apj},
         year = 2014,
        month = nov,
       volume = {796},
       number = {1},
          eid = {61},
        pages = {61},
          doi = {10.1088/0004-637X/796/1/61},
archivePrefix = {arXiv},
       eprint = {1409.8201},
 primaryClass = {astro-ph.HE},
       adsurl = {https://ui.adsabs.harvard.edu/abs/2014ApJ...796...61K}
}

@ARTICLE{Yoshida2023,
       author = {{Yoshida}, Kenji and {Petropoulou}, Maria and {Murase}, Kohta and {Oikonomou}, Foteini},
        title = "{Flare Duty Cycle of Gamma-Ray Blazars and Implications for High-Energy Neutrino Emission}",
      journal = {\apj},
         year = 2023,
       volume = {954},
          doi = {10.3847/1538-4357/acea74},
archivePrefix = {arXiv},
       eprint = {2210.10011},
       adsurl = {https://ui.adsabs.harvard.edu/abs/2023ApJ...954...36Y}
}

@ARTICLE{Kouch2026,
       author = {{Kouch}, Pouya M. and {Lindfors}, Elina and {Hovatta}, Talvikki and {Liodakis}, Ioannis and {Koljonen}, Karri I.~I. and {Paggi}, Alessandro and {Nilsson}, Kari and {Jormanainen}, Jenni and {Fallah Ramazani}, Vandad and {Kankkunen}, Sofia and {Wierda}, Folkert and {Wagner}, Sarah M. and {Graham}, Matthew J.},
        title = "{CAZ catalog and optical light curves of 7918 blazar-selected active galactic nuclei}",
      journal = {\aap},
         year = 2026,
       volume = {708},
        pages = {A382},
archivePrefix = {arXiv},
       eprint = {2510.16584}
}

@ARTICLE{Peduzzi1996,
       author = {{Peduzzi}, P. and {Concato}, J. and {Kemper}, E. and {Holford}, T.~R. and {Feinstein}, A.~R.},
        title = "{A simulation study of the number of events per variable in logistic regression analysis}",
      journal = {Journal of Clinical Epidemiology},
         year = 1996,
       volume = {49},
       number = {12},
        pages = {1373--1379},
          doi = {10.1016/S0895-4356(96)00236-3}
}

@INPROCEEDINGS{Oshiro2012,
       author = {{Oshiro}, T.~M. and {Perez}, P.~S. and {Baranauskas}, J.~A.},
        title = "{How Many Trees in a Random Forest?}",
    booktitle = {Machine Learning and Data Mining in Pattern Recognition (MLDM 2012)},
         year = 2012,
        pages = {154--168},
          doi = {10.1007/978-3-642-31537-4_13}
}

@ARTICLE{Probst2018,
       author = {{Probst}, P. and {Boulesteix}, A.-L.},
        title = "{To tune or not to tune the number of trees in random forest?}",
      journal = {Journal of Machine Learning Research},
         year = 2018,
       volume = {18},
       number = {181},
        pages = {1--18}
}

@ARTICLE{2020ApJ...891..120B,
       author = {{Bhatta}, Gopal and {Dhital}, Niraj},
        title = "{The Nature of {\ensuremath{\gamma}}-Ray Variability in Blazars}",
      journal = {\apj},
         year = 2020,
        month = mar,
       volume = {891},
       number = {2},
          eid = {120},
        pages = {120},
          doi = {10.3847/1538-4357/ab7455},
archivePrefix = {arXiv},
       eprint = {1911.08198},
 primaryClass = {astro-ph.HE},
       adsurl = {https://ui.adsabs.harvard.edu/abs/2020ApJ...891..120B}
}

@ARTICLE{2025ApJ...980..141S,
       author = {{Stolte}, Hermann and {Sinapius}, Jonas and {Sadeh}, Iftach and {Pueschel}, Elisa and {Weidlich}, Matthias and {Berge}, David},
        title = "{Early Detection of Multiwavelength Blazar Variability}",
      journal = {\apj},
         year = 2025,
        month = feb,
       volume = {980},
       number = {1},
          eid = {141},
        pages = {141},
          doi = {10.3847/1538-4357/ad960c},
archivePrefix = {arXiv},
       eprint = {2411.10140},
 primaryClass = {astro-ph.HE},
       adsurl = {https://ui.adsabs.harvard.edu/abs/2025ApJ...980..141S}
}

@ARTICLE{2021APh...13202625F,
       author = {{Fidor}, Tomasz and {Sitarek}, Julian},
        title = "{Assessing the capability of random forest to predict the evolution of enhanced gamma-ray states of active galactic nuclei}",
      journal = {Astroparticle Physics},
         year = 2021,
        month = nov,
       volume = {132},
          eid = {102625},
        pages = {102625},
          doi = {10.1016/j.astropartphys.2021.102625},
archivePrefix = {arXiv},
       eprint = {2107.04571},
 primaryClass = {astro-ph.IM},
       adsurl = {https://ui.adsabs.harvard.edu/abs/2021APh...13202625F}
}
\bibliographystyle{aasjournal}

\end{document}